\documentclass[11pt,a4paper]{article}
\pdfoutput=1
\usepackage{jcappub}

\newcommand{\dB}{\mathbf{d}}
\newcommand{\nB}{\mathbf{n}}
\newcommand{\pB}{\mathbf{p}}

\newcommand{\uB}{\mathbf{u}}
\newcommand{\yB}{\mathbf{y}}

\newcommand{\bm}[1]{\mbox{\boldmath{$#1$}}}
\newcommand{\bms}[1]{\mbox{{\tiny\boldmath{$#1$}}}}

\notoc

\begin{document}

\title{Search for features in the spectrum of primordial 
  perturbations using Planck and other datasets}
\author[a]{Paul Hunt}
\affiliation[a]{Theoretical Physics, Ludwig Maxmillians University,
Theresienstrasse 37, 80333 Munich, Germany}
\author[b,c]{\& Subir Sarkar}
\affiliation[b]{Rudolf Peierls Centre for Theoretical Physics, University
  of Oxford, 1 Keble Road, Oxford OX1 3NP, UK} 
\affiliation[c]{Niels Bohr International Academy, Copenhagen University, Blegdamsvej
  17, 2100 Copenhagen \O, Denmark}
\emailAdd{Paul.Hunt@lmu.de}
\emailAdd{s.sarkar@physics.ox.ac.uk}
 
\abstract{We reconstruct the power spectrum of primordial curvature
  perturbations by applying a well-validated non-parametric technique
  employing Tikhonov regularisation to the first data release from the
  Planck satellite. To improve the reconstruction on small spatial
  scales we include data from the ground-based ACT and SPT
  experiments, the WiggleZ galaxy redshift survey, the CFHTLenS
  tomographic weak lensing survey, and spectral analysis of the
  Lyman-$\alpha$ forest. The reconstructed scalar spectrum (assuming
  the standard $\Lambda$CDM cosmology) is not scale-free but has an
  infrared cutoff at $k \lesssim 5 \times 10^{-4}\; \mathrm{Mpc}^{-1}$
  and several $(2-3)\sigma$ features, of which two at wavenumber
  $k/\mathrm{Mpc}^{-1} \sim$ 0.0018 and 0.057 had been seen already in
  WMAP data. A higher significance feature at
  $k\sim 0.12\, \mathrm{Mpc}^{-1}$ is indicated by Planck data, but
  may be sensitive to the systematic uncertainty around multipole
  $\ell \sim 1800$ in the 217x217 GHz cross-spectrum.  In any case
  accounting for the `look elsewhere' effect decreases its global
  significance to $\sim2\sigma$.}

\date{\today}

\keywords{cosmic microwave background, cosmological parameters,
  cosmology: large-scale structure of universe, inflation,
  primordial curvature perturbation}
  
\arxivnumber{1510.03338}
\maketitle

\section{Introduction \label{intro}}

Detailed knowledge of the primordial curvature perturbation is
essential in order to elucidate the physical mechanism which generated
it. This is widely believed to be an early quasi-de Sitter phase of
exponentially fast expansion (inflation), usually assumed to be driven
by a scalar field whose `slow-roll' to the minimum of its potential
generates a close to power-law spectrum of curvature perturbations
(with small logarithmic corrections called `running').

A power-law spectrum is usually \emph{assumed} when extracting
cosmological parameters from observations of the cosmic microwave
background (CMB) and large-scale structure (LSS) in the universe. The
actual primordial power spectrum (PPS) cannot in fact be directly
extracted from the data. This is because relevant cosmological
observables are given by a \emph{convolution} of the primordial
perturbations with a smoothing kernel which depends on both the
assumed world model and the assumed matter content of the
universe. Moreover the deconvolution problem is ill-conditioned so a
regularisation scheme must be employed to control error propagation
\cite{tikhonov}.
  
We have demonstrated in some detail \cite{Hunt:2013bha} that `Tikhonov
regularisation' can reconstruct the primordial spectrum from multiple
cosmological data sets and provide reliable estimates of both its
uncertainty and resolution. Using Monte Carlo simulations we
investigated several methods for selecting the regularisation
parameter and found that generalised cross-validation and Mallow's
$C_p$ method give optimal results. We applied our inversion procedure
to data from the Wilkinson Microwave Anisotropy Probe (WMAP), other
ground-based small angular scale CMB experiments, and the Sloan
Digital Sky Survey (SDSS). The reconstructed spectrum (assuming the
standard $\Lambda$CDM cosmology) was found to have an infrared cutoff
at $k \lesssim 5 \times 10^{-4}\; \mathrm{Mpc}^{-1}$ (due to the
anomalously low CMB quadrupole) and several features with
$\sim 2 \sigma$ significance at $k/\mathrm{Mpc}^{-1} \sim$
0.0013--0.0023, 0.036--0.040 and 0.051--0.056, reflecting the `WMAP
glitches' \cite{Hunt:2013bha}. We noted that more accurate data, such
as from the Planck satellite, would be required to test whether these
features are indeed real. 
 
In this paper we apply our method to the first data release from the
Planck satellite \cite{Ade:2013kta}, and ground-based experiments such
as Atacama Cosmology Telescope (ACT) \cite{Das:2013zf} and South Pole
Telescope (SPT) \cite{Story:2012wx}, as well as the WiggleZ galaxy
redshift survey \cite{Parkinson:2012vd}, analysis of the Canada-France
Hawaii Telescope Lensing Survey (CFHTLenS) \cite{Kilbinger:2012qz},
and spectral analysis of the Lyman-$\alpha$ forest
\cite{Viel:2004bf}. Note that the Planck collaboration has estimated
cosmological parameters from their data by \emph{assuming} a power-law
PPS, with possible running included \cite{Ade:2013zuv}. Several
authors have adopted a more general parameterisation of the PPS and
used Monte Carlo Markov Chain (MCMC) analysis to simultaneously
estimate the PPS and the background cosmological parameters. However
the relative crudity of the modelling means that the resolution of the
estimated PPS is limited. Up to 4 tilted wavenumber bins with variable
locations were used in \cite{Hazra:2013nca}, while
\cite{Aslanyan:2014mqa} employed up to 5 movable `knots' with linear
and cubic spline interpolation and assessed the Bayesian evidence for
each additional knot. The Planck team applied a similar procedure to
their second data release for up to 8 movable knots with linear
interpolation \cite{Ade:2015lrj}. A cubic spline PPS with 20 fixed
knots was applied in analysis of the Planck, ACT, SPT and BOSS CMASS
data in \cite{dePutter:2014hza}, and the same method was implemented
with 12 fixed knots for the second Planck data release
\cite{Ade:2015lrj}. The PPS has also been modelled by a 12 fixed knot
cubic Hermite polynomial and estimated from CMB and WiggleZ data,
together with measurements of $\sigma_8$ from CFHTLenS and the Planck
Sunyaev-Zeldovich catalogue \cite{Gariazzo:2014dla}. In
\cite{Abazajian:2014tqa} the Planck data was used to constrain a
linear spline PPS with 1 movable knot, while a 3 fixed knot cubic
spline was used in \cite{Hu:2014aua}.

There have been far fewer \emph{non-}parametric approaches. The Planck
team have most recently used a penalised likelihood inversion method
involving a B-spline for the PPS \cite{Ade:2015lrj}; a similar scheme
with a 485 knot cubic spline had been used earlier for their first
data release \cite{Planck:2013jfk}.  Another example is the inversion
of Planck data with Richardson-Lucy deconvolution
\cite{Hazra:2014jwa}. An attractive method called {PRISM} which uses a
`sparsity' prior on features in the PPS in a wavelet basis to
regularise the inverse problem was developed in \cite{Paykari:2014cna}
and has been subsequently applied to Planck data
\cite{Lanusse:2014sra}.

In the alternative approach we follow here the background parameters
are held fixed, which permits the deconvolution of the smoothing
kernel relating the observables to the PPS in linear cosmological
perturbation theory. We have refined our earlier method and now use
the logarithm of the power spectra in the reconstruction which ensures
that the recovered spectra are positive, and allows us to set priors
on the slope of the spectra. Moreover we correct for gravitational
lensing of the CMB which is important on the small scales probed by
the latest experiments. Our method features a `regularisation
parameter' $\lambda$ that balances the influence on the solution of
prior information with that of the observed data. By studying the
trade-off between the resolution and stability of the recovered PPS we
find $\lambda=$ 400 \emph{and} 20000 to be suitable values for the
regularisation parameter (see Fig.\ref{fig:tradeoff},
Appendix~\ref{error}). We have no criterion for choosing between them,
so present results for both values.

We confirm that all the features we identified previously
\cite{Hunt:2013bha} in WMAP data are also present in the Planck data
at $\gtrsim2\sigma$ confidence.  Moreover there is a new feature at
$k/\mathrm{Mpc}^{-1} \sim$ 0.12--0.15 at $4\sigma$ confidence for
$\lambda=400$ ($2.9\sigma$ for $\lambda=20000$), even \emph{after} we
take out the 217x217 GHz data from Planck. We did so following the
suggestion \cite{Spergel:2013rxa} that there are residual systematics
in this particular channel, which was confirmed by the Planck
collaboration in their updated paper \cite{Ade:2013zuv}. This both
illustrates the problems in reliably identifying features, but it also
makes more compelling the need for further detailed studies. Reliable
detection of even \emph{one} feature in the spectrum would immediately
rule out \emph{all} slow-roll models of inflation. Hence this is a key
probe of inflation, complementary to searches for non-gaussianity and
gravitational waves (see \cite{Chluba:2015bqa} and extensive
references therein to inflationary models which generate features in
the PPS).

\section{Inversion method \label{method}}

\subsection{Tikhonov regularisation \label{tikreg}}

Let us assume there are $N$ available cosmological data sets, each
with $N_\mathbb{Z}$ data points $\mathrm{d}_a^{(\mathbb{Z})}$, from
which we wish to estimate the PPS. Here the subscript runs from $1$ to
$N_\mathbb{Z}$ and the superscript $\mathbb{Z}$ denotes the data
set. In a flat (or open) universe, the points of many data sets are
related to the power spectrum $\mathcal{P_\zeta} \left(k\right)$ of
the curvature perturbation $\mathcal{\zeta}$ \cite{Lyth:2009zz} by
\begin{equation}
  \mathrm{d}_a^{(\mathbb{Z})} = \int^{\infty}_0
  \mathcal{K}_a^{(\mathbb{Z})}\left(\bm{\theta},k\right)\mathcal{P_\zeta}
  \left(k\right)\,\mathrm{d}k + \mathrm{n}_a^{(\mathbb{Z})}.
\label{int1}
\end{equation}
Here the integral kernels $\mathcal{K}_a^{(\mathbb{Z})}$ depend on the
background cosmological parameters $\bm{\theta}$, and the noise
vectors $\mathrm{n}_a^{(\mathbb{Z})}$ have zero mean and covariance
matrices
$N_{ab}^{(\mathbb{Z})}\equiv\langle\mathrm{n}_a^{(\mathbb{Z})}
\mathrm{n}_b^{(\mathbb{Z})} \rangle$.
In what follows we also include in $\bm{\theta}$ extraneous `nuisance'
parameters associated with the likelihood functions of the data sets,
such as calibration parameters or the parameters describing the CMB
foregrounds.  We assume an estimate $\hat{\bm{\theta}}$ of the
background and nuisance parameters exists which is \emph{independent}
of the $N$ data sets, and has a zero mean uncertainty $\uB$, with
elements $\mathrm{u}_\alpha$.  Then
$\langle \mathrm{u}_\alpha \mathrm{n}_a^{(\mathbb{Z})}\rangle=0$ for
all elements of the uncertainty and noise vectors as these are
uncorrelated by assumption. The covariance matrix for the estimated
background parameters is just
$\mathsf{U}\equiv\langle \uB \uB^\mathrm{T}\rangle$, where
$^\mathrm{T}$ signifies the matrix transpose. Given our estimate of
the background and nuisance parameter set $\hat{\bm{\theta}}$, the
goal is to obtain an estimate
$\hat{\mathcal{P}}_\mathcal{\zeta}\left(k\right)$ of the PPS from the
data sets.

The PPS is approximated as a piecewise function given by a sum of
$N_j$ basis functions $\phi_i\left(k\right)$, weighted by coefficients
$\mathrm{p}_i$:
\begin{equation}
 \mathcal{P_\zeta}\left(k\right) = \sum_{i=1}^{N_j}\mathrm{p}_i\phi_i\left(k\right).
\label{basis}
\end{equation}
For a grid of wavenumbers $\left\{k_i\right\}$ the basis functions are
defined as
\begin{equation}
\label{phidef}
 \phi_i\left(k\right) \equiv \left\{\begin{array}{ll}
 1,& k_{i}<k\leq k_{i+1},  \\
 0,&  \mbox{elsewhere.}
 \end{array}\right.
\end{equation}
We use a logarithmically spaced grid between
$k_1=7\times 10^{-6}\; \mathrm{Mpc}^{-1}$ and
$k_{N_j+1}=30\; \mathrm{Mpc}^{-1}$ with $N_j=2500$. Substituting
Eq.(\ref{basis}) into Eq.(\ref{int1}) gives
\begin{equation}
 \mathrm{d}_a^{(\mathbb{Z})} = \sum_i
 W_{ai}^{(\mathbb{Z})}\left(\bm{\theta}\right)\mathrm{p}_i +
 \mathrm{n}_a^{(\mathbb{Z})},
\label{rel}   
\end{equation}
where the $N_\mathbb{Z}\times N_j$ matrices
$W_{ai}^{(\mathbb{Z})}\left(\bm{\theta}\right)$ depend on the
background parameters:
\begin{equation}
\label{rel1}
 W_{ai}^{(\mathbb{Z})}\left(\bm{\theta}\right) = \int^{k_{i+1}}_{k_i}
 \mathcal{K}_a^{(\mathbb{Z})}\left(\bm{\theta},k\right)\,
 \mathrm{d}k.   
\end{equation}

As discussed in \cite{Hunt:2013bha} solving Eq.(\ref{int1}) for the
PPS is an ill-posed inverse problem and the matrices
$W_{ai}^{(\mathbb{Z})}$ are ill-conditioned. Consequently na\"ive
attempts to determine the PPS by maximising the likelihood function
$\mathcal{L}\left(\pB,\bm{\theta}|\dB\right)$ of the data $\dB$ given
$\pB$ and \bm{\theta} produce ill-behaved spectra with wild irregular
oscillations. To overcome this, Tikhonov regularisation
\cite{tikhonov} uses a penalty function
$\mathrm{R}\left(\mathbf{p}\right)$ which takes on large values for
unphysical spectra. Then the likelihood is maximised subject to the
constraint that the penalty function at most equals a certain value
$R_0$:
\begin{equation}
\max_\mathbf{p} \mathcal{L}\left(\pB,\bm{\theta}|\dB\right)\quad
\mbox{subject to} \quad \mathrm{R}\left(\mathbf{p}\right)\le \mathrm{R}_0.  
\end{equation}  
Rather than working directly with $\mathbf{p}$ we use instead
$\mathbf{y}$ with elements $y_i=\ln p_i$ in order to enforce the
positivity constraint on the recovered PPS. Thus the estimated PPS is
given by
\begin{equation}
 \hat{\mathbf{y}}\left(\dB,\hat{\bm{\theta}},\lambda\right) =
 \min_\yB\,Q\left(\yB,\dB,\hat{\bm{\theta}},\lambda\right),
\end{equation}
where
\begin{equation}
\label{minq}
 Q\left(\yB,\dB,\hat{\bm{\theta}},\lambda\right)\equiv  
 L\left(\yB,\hat{\bm{\theta}},\dB\right) + \lambda \mathrm{R}\left(\yB\right).
\end{equation}
Here
$L\left(\pB,\bm{\theta},\dB\right)\equiv-2\ln\mathcal{L}
\left(\pB,\bm{\theta}|\dB\right)$
and the regularisation parameter $\lambda$ acts as a
Lagrange multiplier.

Since to a first approximation the PPS is a power-law with a constant
spectral index
$n_\mathrm{s}-1=\mathrm{d}\ln \mathcal{P_\zeta}/\mathrm{d}\ln k$ \cite{Lyth:2009zz}, we use
the penalty function
\begin{eqnarray}
R\left(\yB\right)& =& \sum_{i=1}^{N_k-1}
\left[y_{i+1}-y_i-\left(n_\mathrm{s}-1\right) \Delta \ln k \right]^2,\\
& \propto & \int \left(\frac{\mathrm{d}\ln \mathcal{P_\zeta}}{\mathrm{d}\ln k}-n_\mathrm{s}+1\right)^2 \,\mathrm{d}\ln k,
\end{eqnarray}
where $\Delta \ln k$ is the logarithmic separation of the
$\left\{k_i\right\}$ wavenumber grid.  Using the
$\left(N_k-1\right)\times N_k$ first difference matrix $\mathsf{L}$
and the $N_k\times N_k$ matrix $\mathsf{\Gamma}$ given by
\begin{equation}
\label{firstl}
\mathsf{L} =
\left(\begin{array}{cccccc} -1 & 1 & & & & \\ & -1 & 1 & & & \\ & 
  & \ddots & \ddots & & \\ & & & -1 & 1 & \\ & & & & -1 & 1
\end{array}\right),
\qquad
\mathsf{\Gamma}\equiv\mathsf{L}^\mathrm{T} \mathsf{L} =
\left(\begin{array}{ccccc} 1 & -1 & & & \\ -1 & 2 & -1 & & \\ & \ddots
  & \ddots & \ddots & \\ & & -1 & 2 & -1 \\ & & & -1 & 1
\end{array}\right),
\end{equation}
together with the $N_k-1$ vector $\bm{\eta}$ with elements
$\eta_i\equiv \left(n_\mathrm{s}-1\right)\Delta \ln k$ the penalty
function can be written as
\begin{equation}
R\left(\yB\right)=\yB^\mathrm{T} \mathsf{\Gamma} \yB -2 \bm{\eta}^\mathrm{T} \mathsf{L} \yB + \bm{\eta}^\mathrm{T} \bm{\eta}.
\label{scalpen}
\end{equation}
It penalises large excursions and conservatively smooths the estimated
PPS towards a power-law of amplitude set by the data and a `prior'
spectral index $n_\mathrm{s}$. The penalty function determines the way
in which the recovered PPS is smoothed, while the regularisation
parameter controls the amount of smoothing. Thus $\hat{\mathbf{y}}$
depends on $\lambda$ and $R$, and both must be chosen
carefully to give sensible results. The pseudo Newton-Raphson
algorithm of \cite{Hunt:2013bha} is again employed to minimise $Q$ and
estimate the PPS.

In our previous work \cite{Hunt:2013bha} following tests using mock
data we performed reconstructions with $\lambda=100$ and
$\lambda=5000$. We desire comparable results in this paper, but are
now working with the \emph{logarithmic} elements
$\mathrm{y}_i=\ln \mathrm{p}_i$ instead of $\mathrm{p}_i$ in the
inversion.  For data sets with Gaussian likelihood functions,
$\partial\hat{\mathrm{y}}_i/\partial \mathrm{d}_{a}^{(\mathbb{Z})}
\propto \left[\sum_{\mathbb{Z},a,b}
  \mathrm{p_i}\,W^{(\mathbb{Z})}_{ia}
  \left(N^{(\mathbb{Z})}\right)_{ab}^{-1} W^{(\mathbb{Z})}_{jb}
  \mathrm{p_j} + \lambda \Gamma_{ij} \right]^{-1}$
in the logarithm-based reconstruction, whereas
$\partial\hat{\mathrm{p}}_i/\partial \mathrm{d}_{a}^{(\mathbb{Z})}
\propto \left[\sum_{\mathbb{Z},a,b} W^{(\mathbb{Z})}_{ia}
  \left(N^{(\mathbb{Z})}\right)_{ab}^{-1} W^{(\mathbb{Z})}_{jb} +
  \lambda \Gamma_{ij} \right]^{-1}$
in the non-logarithmic case. As a result the regularisation parameter
must be a factor of $\mathrm{p}_i^2\simeq 4$ (in units of $10^{-9}$)
larger for a logarthmic reconstruction to approximate a
non-logarithmic one. Hence we now use $\lambda=$ 400 and 20000. As
shown in Fig.\ref{fig:tradeoff} of Appendix~\ref{error} these values
provide a close to \emph{optimal} compromise between the resolution
and the variance of the reconstruction. The higher value in general
yields smoother spectra, however we have no rationale for choosing one
value over the other so present our results using both values.

\subsection{CMB lensing \label{lensing}}

After last scattering CMB photons are gravitationally deflected by
large scale structure.  This CMB lensing smooths the acoustic peaks of
the temperature and polarisation angular power spectra, and also
generates B-mode polarisation on small scales. ACT, SPT and Planck
have all detected CMB lensing at high significance assuming a
power-law PPS \cite{Das:2011ak,vanEngelen:2012va,Ade:2013tyw}. Since
lensing changes the TT spectrum by around $20\%$ at $\ell=3000$ it
\emph{must} be taken into account in order to obtain an accurate PPS
reconstruction.

The deflection angle equals the gradient of the lensing potential
$\psi$, which is given by a weighted integral of the gravitational
potential along the line of sight. The power spectrum of the lensing
potential
$\mathrm{s}_\ell^\psi=\ell\left(\ell+1\right) C_\ell^\psi/2\pi$ can be
written as \cite{Lewis:2006fu}
\begin{equation}
  \mathrm{s}_\ell^\psi = \int^{\infty}_0
  \mathcal{K}_\ell^\psi\left(\bm{\theta},k\right)\mathcal{P_\zeta}
  \left(k\right)\,\mathrm{d}k.
\end{equation}
We include the effects of nonlinear structure formation in the kernel
$\mathcal{K}_\ell^\psi$ using the \texttt{Halofit}
\cite{Smith:2002dz} fitting formula for the nonlinear matter power
spectrum in the same way as \cite{Challinor:2005jy}, but applied with
a fixed fiducial PPS. Substituting Eq.(\ref{basis}) into the above
equation gives
$\mathrm{s}_\ell^\psi=\sum_i W_{\ell i}^\psi \mathrm{p}_i$. Two
quantities which characterise the statistical properties of the
deflection angle are
\begin{eqnarray} 
\sigma^2\left(r\right) & = & \sum_\ell \frac{\ell^2}{\ell+1} \left[1-J_0\left(\ell r\right)\right] \mathrm{s}_\ell^\psi, \\
C_{\mathrm{gl},2}\left(r\right) & = & \sum_\ell \frac{\ell^2}{\ell+1} J_2\left(\ell r\right) \mathrm{s}_\ell^\psi, 
\end{eqnarray} 
where $J_0$ and $J_2$ are Bessel functions. 

\begin{figure}[h!]
\includegraphics[width=0.5\columnwidth,trim = 32mm 171mm 23mm 15mm, clip]{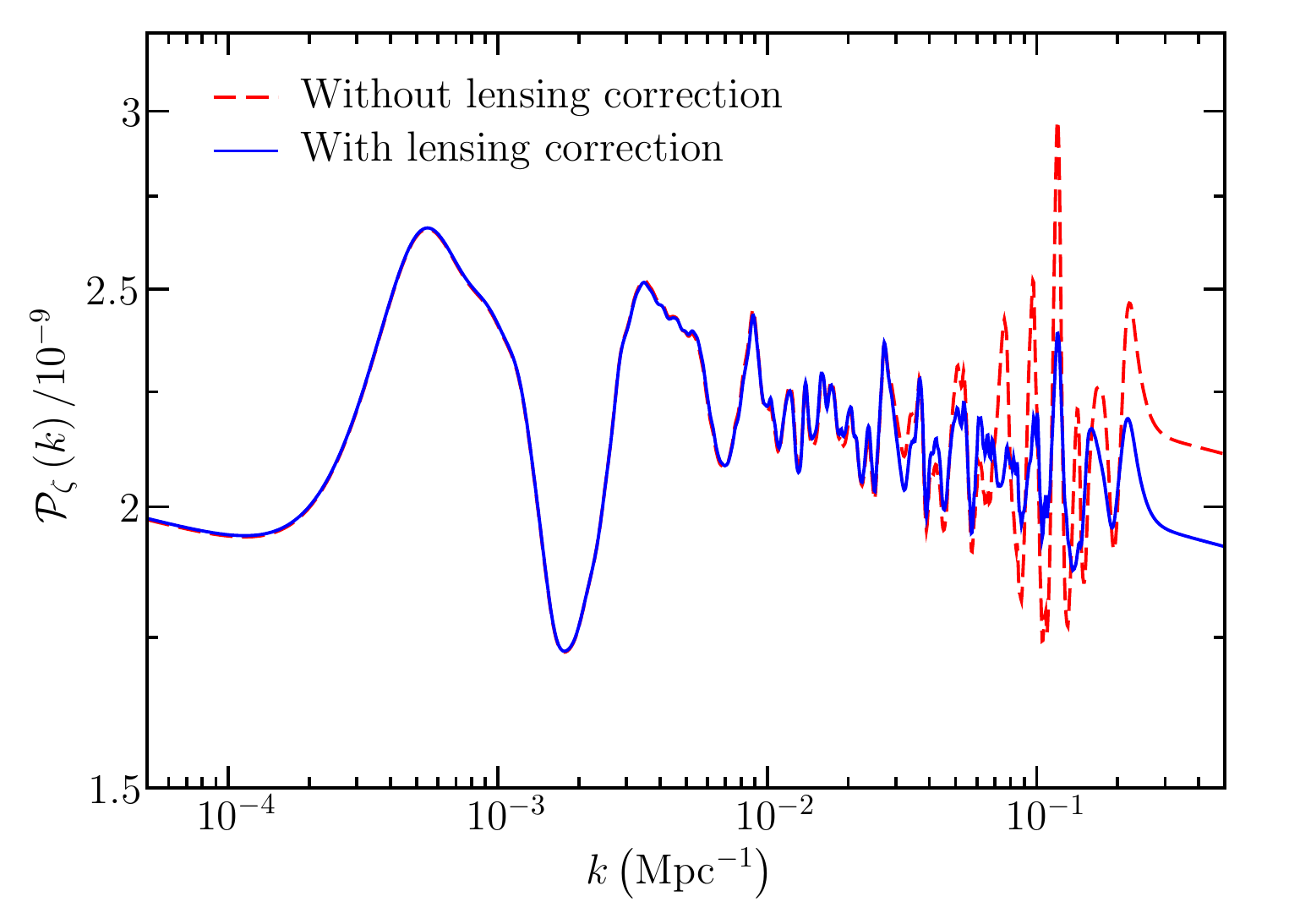}
\includegraphics[width=0.5\columnwidth,trim = 32mm 171mm 23mm 15mm, clip]{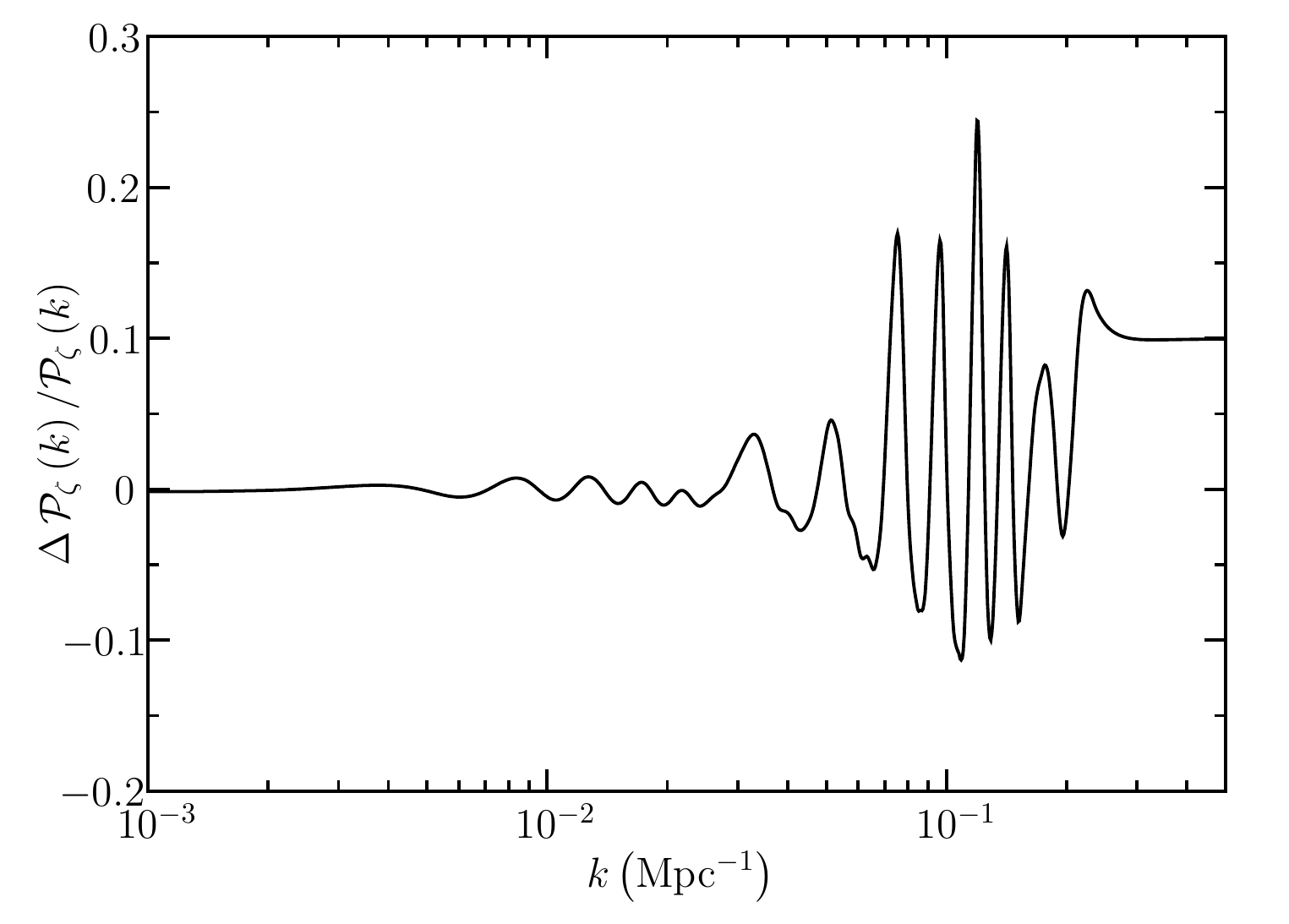}
\caption{Left: Comparison of the spectra recovered with $\lambda=400$
  from the Planck, WMAP-9 polarisation, ACT and SPT data, with and
  without correcting for gravitational lensing. Right: The fractional
  change
  $\left(\mathcal{P}_\zeta^\mathrm{No-Lens}\left(k\right)-\mathcal{P_\zeta}\left(k\right)\right)/\mathcal{P_\zeta}\left(k\right)$
  in the recovered PPS due to the lensing correction.}
\label{fig:lensing}
\end{figure}

The predicted CMB angular power spectra
$\ell\left(\ell+1\right) C_\ell^I/2\pi$ where
$I \in
\left\{\mathrm{TT},\;\mathrm{TE},\;\mathrm{EE},\;\mathrm{BB}\right\}$
are denoted $\mathrm{s}_\ell^I$.  The unlensed scalar temperature
power spectrum is
$\mathrm{s}^\mathrm{TT}_{\ell,\mathrm{s}}=\sum_i W_{\ell
  i,\mathrm{s}}^\mathrm{TT} \mathrm{p}_i$,
where the scalar matrix $W_{\ell i,\mathrm{s}}^\mathrm{TT}$ is
calculated from the scalar temperature integral kernel as in
Eq.(\ref{rel1}).  In the Boltzmann code {\tt CAMB} used for this work
the total lensed temperature power spectrum is
$\mathrm{s}_\ell^\mathrm{TT}=\tilde{\mathrm{s}}_{\ell,\mathrm{s}}^\mathrm{TT}+\mathrm{s}_{\ell,\mathrm{t}}^\mathrm{TT}$.
Here the lensed scalar TT spectrum
$\tilde{\mathrm{s}}_{\ell,\mathrm{s}}^\mathrm{TT}$ is related to the
unlensed spectrum by
$\tilde{\mathrm{s}}_{\ell,\mathrm{s}}^\mathrm{TT}=\sum_{\ell^\prime}
\mathcal{W}_{\ell\ell^\prime}^\mathrm{TT}
\mathrm{s}_{\ell^\prime,\mathrm{s}}^\mathrm{TT}$
in the perturbative `flat-sky' approximation of
\cite{Seljak:1995ve,Zaldarriaga:1998ar}, where
\begin{equation}
\mathcal{W}_{\ell\ell^\prime}^\mathrm{TT} = \frac{\ell\left(\ell+1\right)}{\ell^\prime+1} \int^\pi_0 e^{-\ell^{\prime 2} \sigma^2\left(r\right) /2}
J_0\left(\ell r\right) \left[J_0\left(\ell^\prime r\right) + \frac{\ell^{\prime 2}}{2} C_{\mathrm{gl},2}\left(r\right) J_2\left(\ell^\prime r\right) \right]
r\,\mathrm{d}r.  
\end{equation}
To obtain $\hat{\yB}$ by minimising $Q$ we need the derivative
\begin{equation}
\frac{\partial L_\mathrm{CMB}}{\partial y_i}=\sum_{\ell \ell^\prime} \frac{\partial L_\mathrm{CMB}}{\partial \mathrm{s}_\ell^\mathrm{TT}}
\left(\mathcal{W}_{\ell\ell^\prime}^\mathrm{TT} W_{\ell^\prime i,\mathrm{s}}^\mathrm{TT} 
+\sum_{\ell^{\prime\prime}}\frac{\partial \mathcal{W}_{\ell\ell^{\prime\prime}}^\mathrm{TT}}{\partial \mathrm{s}^\psi_{\ell^\prime}} 
\mathrm{s}_{\ell^{\prime\prime},\mathrm{s}}^\mathrm{TT} W^\psi_{\ell^\prime i} \right)\mathrm{p}_i.
\end{equation}
Here $L_\mathrm{CMB}$ is the sum of the ACT, SPT and Planck likelihood
functions, and
\begin{eqnarray}
\frac{\partial \mathcal{W}_{\ell\ell^\prime}^\mathrm{TT}}{\partial \mathrm{s}^\psi_{\ell^{\prime\prime}}} & = & 
\frac{\ell\left(\ell+1\right)\ell^{\prime 2} \ell^{\prime\prime 2}}{2\left(\ell^\prime+1\right)\left(\ell^{\prime\prime}+1\right)}
\int^\pi_0 e^{-\ell^{\prime 2} \sigma^2\left(r\right) /2} J_0\left(\ell r\right) 
\bigg\{J_2\left(\ell^{\prime\prime} r \right) J_0\left(\ell^\prime r \right) \bigg. \nonumber \\ 
& & \left. - \left[1-J_0\left(\ell^{\prime\prime} r\right)\right]  
\left[J_0\left(\ell^\prime r\right) + \frac{\ell^{\prime 2}}{2} C_{\mathrm{gl},2}\left(r\right) J_2\left(\ell^\prime r\right) \right]\right\}
r\,\mathrm{d}r.
\end{eqnarray}
The derivative $\partial L_\mathrm{CMB}/\partial \mathrm{s}_\ell^\mathrm{TT}$ is a
function of $\tilde{\mathrm{s}}_{\ell,\mathrm{s}}^\mathrm{TT}$, which
is calculated from $\mathrm{s}_{\ell,\mathrm{s}}^\mathrm{TT}$ at each
iteration of the Newton-Raphson minimisation algorithm using the more
accurate but complicated curved-sky correlation function method of
\cite{Challinor:2005jy} as implemented in {\tt CAMB}. The effect of
the lensing correction is shown in Fig.\,\ref{fig:lensing}. Since
lensing smooths the acoustic peaks, neglecting it means that fitting
the data requires spurious oscillatory features in the recovered PPS
on small scales. Including the lensing correction \emph{removes} these
spurious features.

Lensing of the EE and TE spectra is neglected as it has a negligible
effect for the data sets considered here. Thus
$\mathrm{s}_\ell^\mathrm{EE}=\mathrm{s}_{\ell,\mathrm{s}}^\mathrm{EE}+\mathrm{s}_{\ell,\mathrm{t}}^\mathrm{EE}$
where
$\mathrm{s}^\mathrm{EE}_{\ell,\mathrm{s}}=\sum_i W_{\ell
  i,\mathrm{s}}^\mathrm{EE} \mathrm{p}_i$
and
$\mathrm{s}^\mathrm{EE}_{\ell,\mathrm{t}}=\sum_i W_{\ell
  i,\mathrm{t}}^\mathrm{EE} \mathrm{q}_i$,
and similarly for the TE spectrum. 

\section{Results}
We choose a standard $\Lambda$CDM model when performing the
reconstructions. The background cosmological and foregound parameter
values, which are quite consistent with those obtained by the Planck
team \cite{Ade:2013zuv} are listed in Table \ref{table1}.

\begin{table}[b!]
\begin{center}
\footnotesize{
\begin{tabular}{|l|l|l|l|l|l|}
\hline
Parameter type & Parameter  & Value & Parameter type & Parameter & Value \\ \hline
Cosmological   & $\Omega_\mathrm{b} h^2$ & 0.02240 & ACT + SPT  & $A_{148}^\mathrm{PS,\;ACT}$      & 11.93      \\  
               & $\Omega_\mathrm{c} h^2$ & 0.1145  &            & $A_{218}^\mathrm{PS,\;ACT}$      & 84.7      \\  
               & $H_0$     & 69.6      &                        & $A_{95}^\mathrm{PS,\;SPT}$       & 8.00       \\  
               & $\tau$     &  0.077     &                      & $A_{150}^\mathrm{PS,\;SPT}$      & 10.51      \\ \cline{1-3} 
Planck         & $A^\mathrm{PS}_{100}$   & 223      &           & $A_{220}^\mathrm{PS,\;SPT}$      & 84.3      \\  
               & $A^\mathrm{PS}_{143}$   & 76      &            & $r^\mathrm{PS}_{95\times 150}$   & 0.924      \\ 
               & $A^\mathrm{PS}_{217}$   & 61      &            & $r^\mathrm{PS}_{95\times 220}$   & 0.751      \\  
               & $A^\mathrm{CIB}_{143}$  & 3.40      &          & $r^\mathrm{PS}_{150\times 220}$  & 0.926      \\  
               & $A^\mathrm{CIB}_{217}$  & 50      &            & $A^\mathrm{ACTs}_\mathrm{dust}$  & 0.40      \\  
               & $A^\mathrm{tSZ}_{143}$  & 4.99      &          & $A^\mathrm{ACTe}_\mathrm{dust}$  & 0.80      \\  
               & $r^\mathrm{PS}_{143\times 217}$  & 0.849 &     & $y^\mathrm{ACTs}_{148}$          & 0.9913      \\  
               & $r^\mathrm{CIB}_{143\times 217}$  & 1.000 &    & $y^\mathrm{ACTs}_{218}$          & 1.002       \\  
               & $\gamma^\mathrm{CIB}$  & 0.548      &          & $y^\mathrm{ACTe}_{148}$          & 0.9873      \\  
               & $c_{100}$  & 1.00068      &                    & $y^\mathrm{ACTe}_{218}$          & 0.961      \\  
               & $c_{217}$  & 1.00005      &                    & $y^\mathrm{SPT}_{95}$            & 0.9848      \\  
               & $\xi^{\mathrm{tSZ}\times\mathrm{CIB}}$  & 0.000  & & $y^\mathrm{SPT}_{150}$       & 0.9845      \\
               & $A^\mathrm{tSZ}_{143}$  & 4.99      &                & $y^\mathrm{SPT}_{220}$          & 1.0173      \\  \cline{4-6}
               & $A^\mathrm{kSZ}$  & 0.717      &  WiggleZ       & $b$     & 1.00069      \\  \cline{4-6}
               & $\beta_1^{100\times 100}$      & 0.710      &  Lyman-$\alpha$              &  $A$    & 0.545      \\ \hline

\end{tabular}}
\caption{Parameter values used when performing the
  reconstructions. The cosmological parameters were obtained by a fit
  to data combination IV, assuming a power-law spectrum.}
\label{table1} 
\end{center} 
\end{table}

The inversion method is applied to the following 5 combinations of data
sets:
\begin{itemize}
\item Data combination I: Planck + WMAP-9 polarisation 
\item Data combination II: Combination I + ACT + SPT
\item Data combination III: Combination II + WiggleZ + galaxy clusters
\item Data combination IV: Combination III + CFHTLenS weak lensing +
  VHS Lyman-$\alpha$ data
\end{itemize}
The data sets and their likelihood functions are discussed in
Appendix \ref{dsets}. Throughout we use a prior of
$n_\mathrm{s}=0.969$ in the scalar penalty function (\ref{scalpen}),
corresponding to the best-fit power-law PPS to data combination IV.

\begin{figure}[tbh]
\includegraphics*[angle=0,width=0.5\columnwidth,trim = 32mm 171mm 23mm
  15mm, clip]{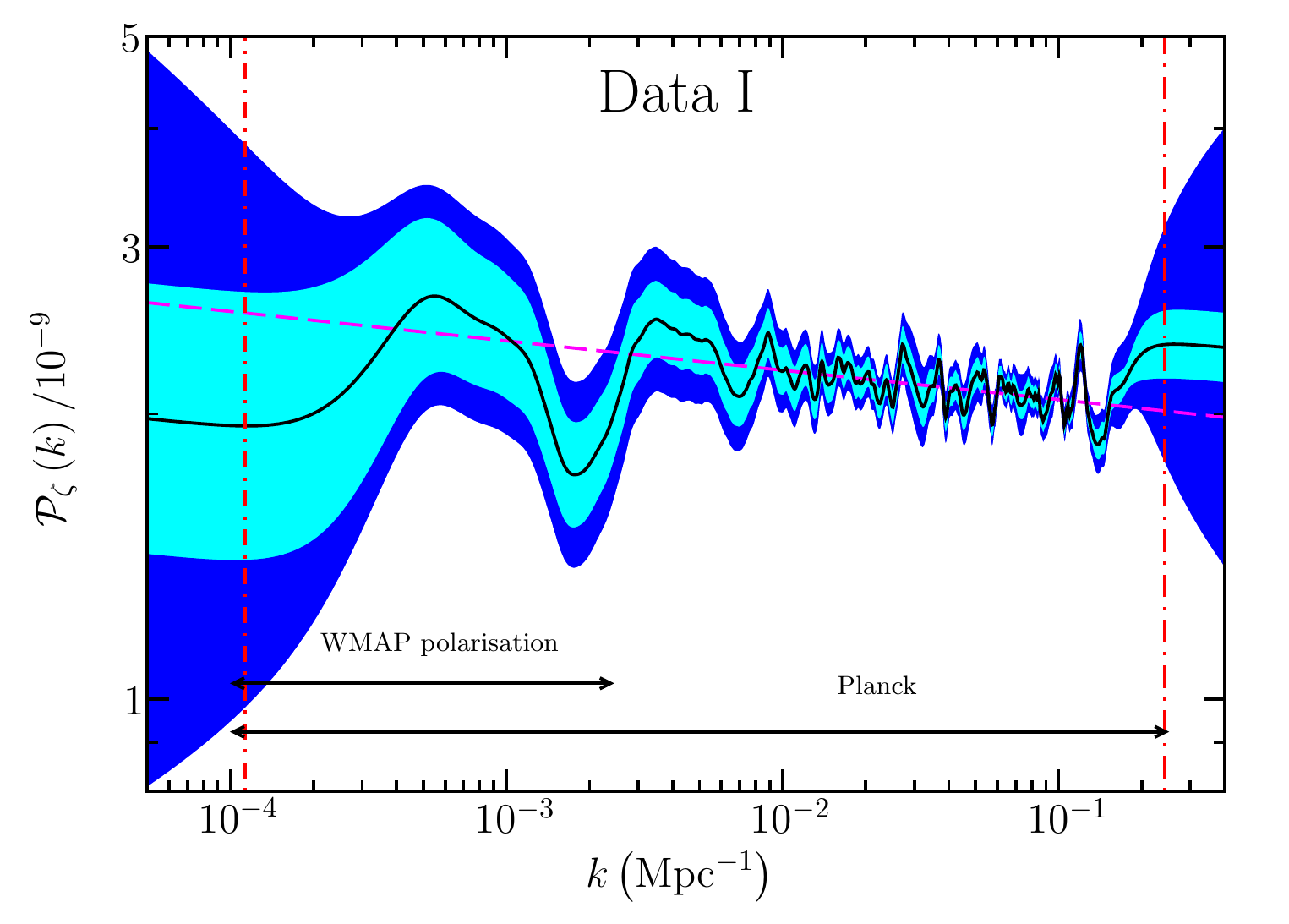}
\includegraphics*[angle=0,width=0.5\columnwidth,trim = 32mm 171mm 23mm
  15mm, clip]{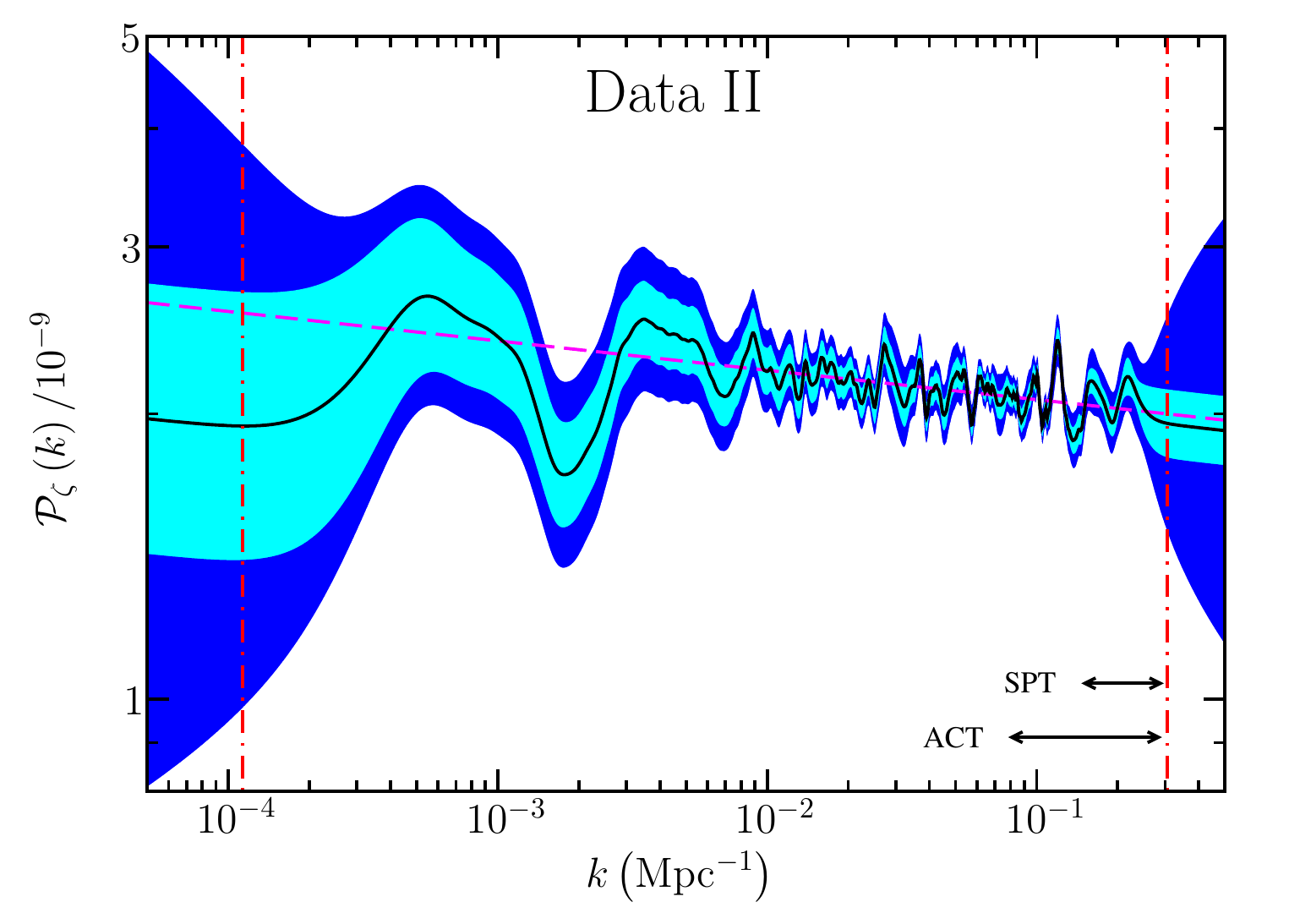}
\includegraphics*[angle=0,width=0.5\columnwidth,trim = 32mm 171mm 23mm
  15mm, clip]{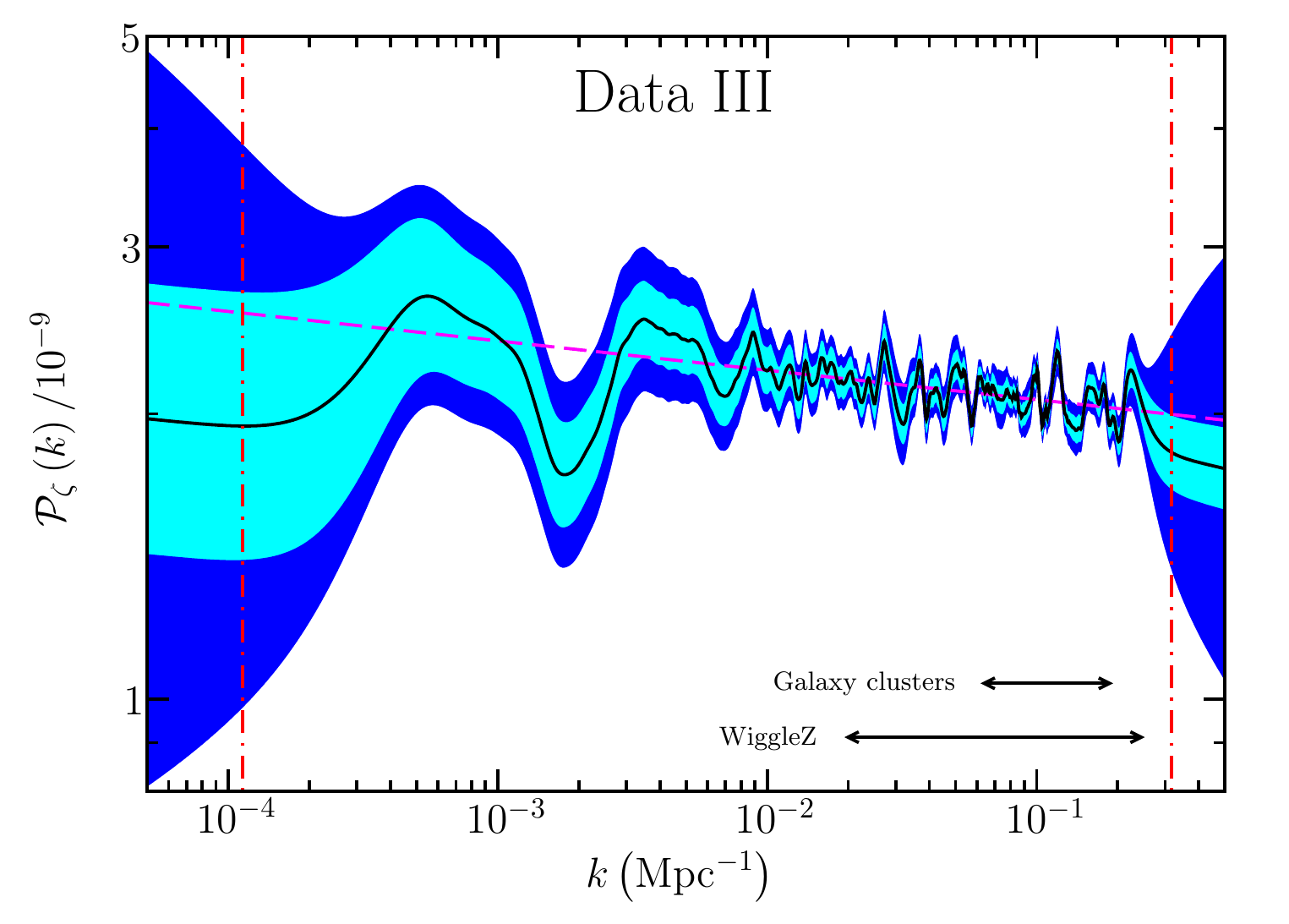}
\includegraphics*[angle=0,width=0.5\columnwidth,trim = 32mm 171mm 23mm
  15mm, clip]{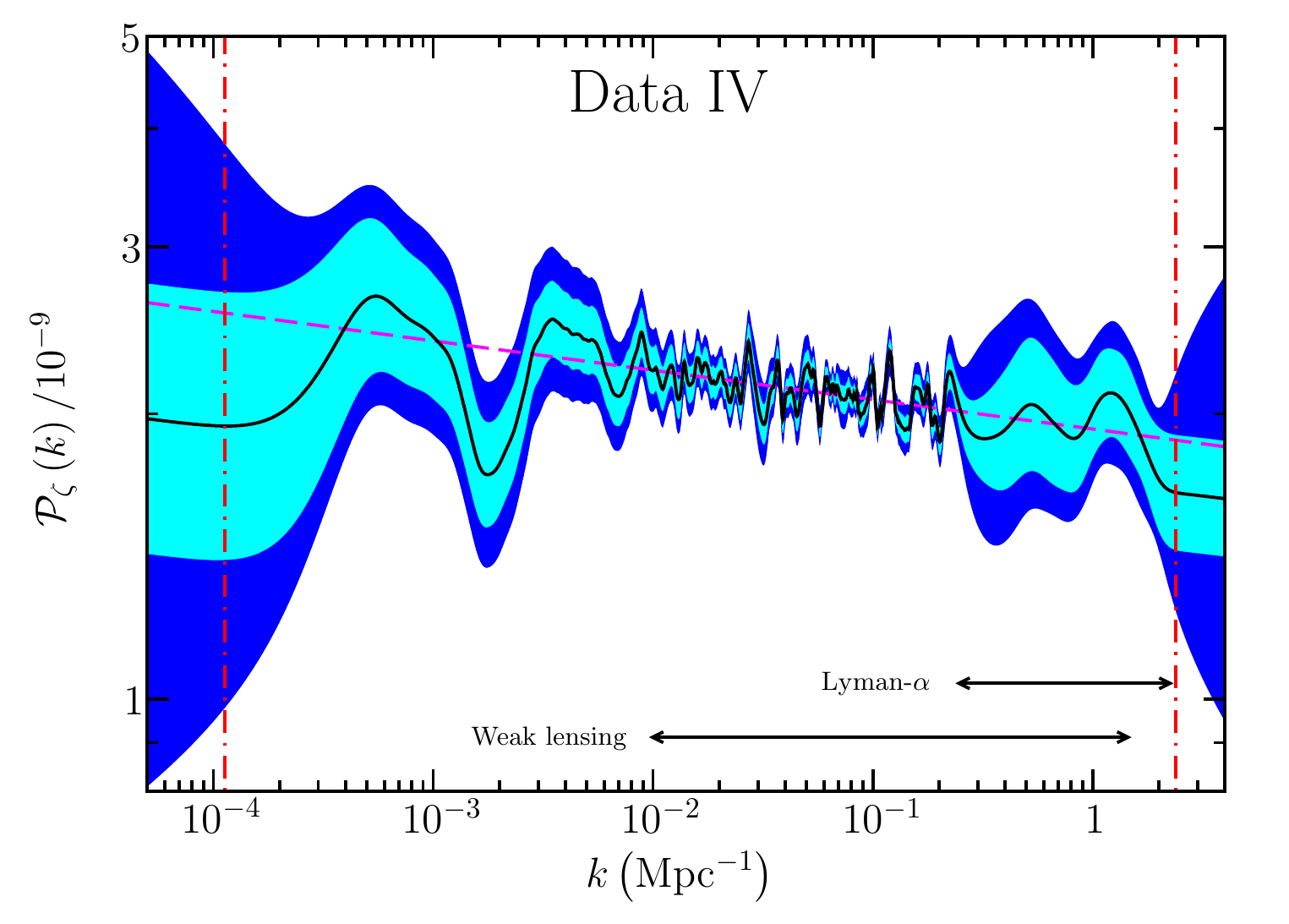}
  \caption{Primordial power spectra recovered from data combinations I
    to IV involving the Planck, WMAP-9 polarisation, ACT, SPT,
    WiggleZ, galaxy clustering, CFTHLenS and Lyman-$\alpha$ data, with
    $\lambda=400$, compared to the best-fit power-law spectrum (slope
    $n_\mathrm{s}=0.969$, magenta dashed line).  In all panels the
    central black line is the reconstruction adopting this
    $n_\mathrm{s}$ value as a prior, and the dark band is the
    $1\sigma$ error given by the square root of the diagonal elements
    of the Bayesian covariance matrix $\mathsf{\Pi}$ (\ref{sigb}),
    while the smaller light band is similarly obtained from the
    frequentist covariance matrix $\mathsf{\Sigma}_\mathrm{F}$
    (\ref{sigmaf}). Vertical lines delineate the wavenumber range
    covered by the resolution kernels (see Fig.\ref{fig:resol1}) over
    which the reconstruction is faithful to the true PPS, while the
    horizontal lines indicate the wavenumber range over which specific
    datasets have most impact.}
\label{fig:ppsplot1}
\end{figure}

\begin{figure}[tbh]
\includegraphics*[angle=0,width=0.5\columnwidth,trim = 32mm 171mm 23mm
  15mm, clip]{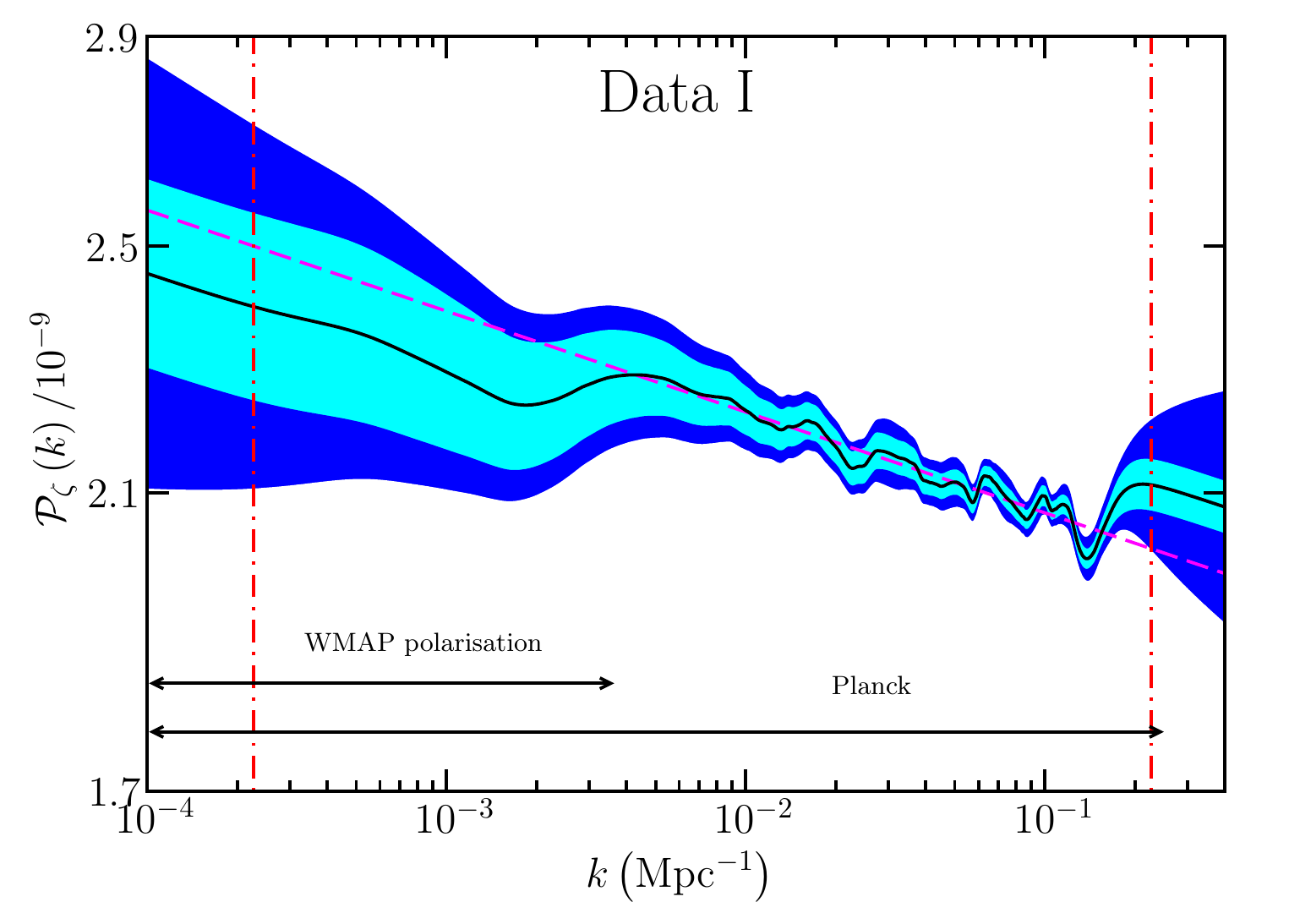}
\includegraphics*[angle=0,width=0.5\columnwidth,trim = 32mm 171mm 23mm
  15mm, clip]{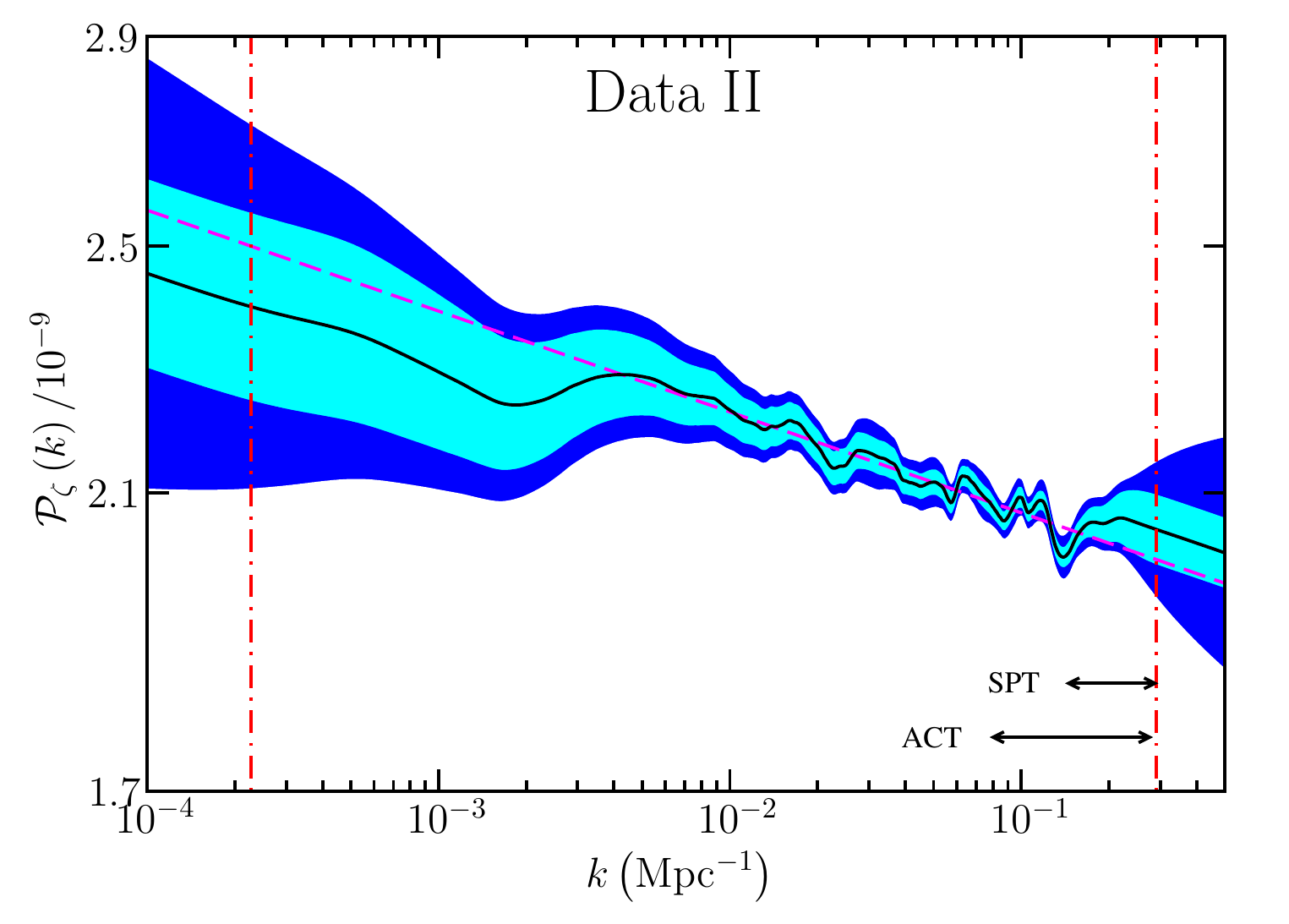}
\includegraphics*[angle=0,width=0.5\columnwidth,trim = 32mm 171mm 23mm
  15mm, clip]{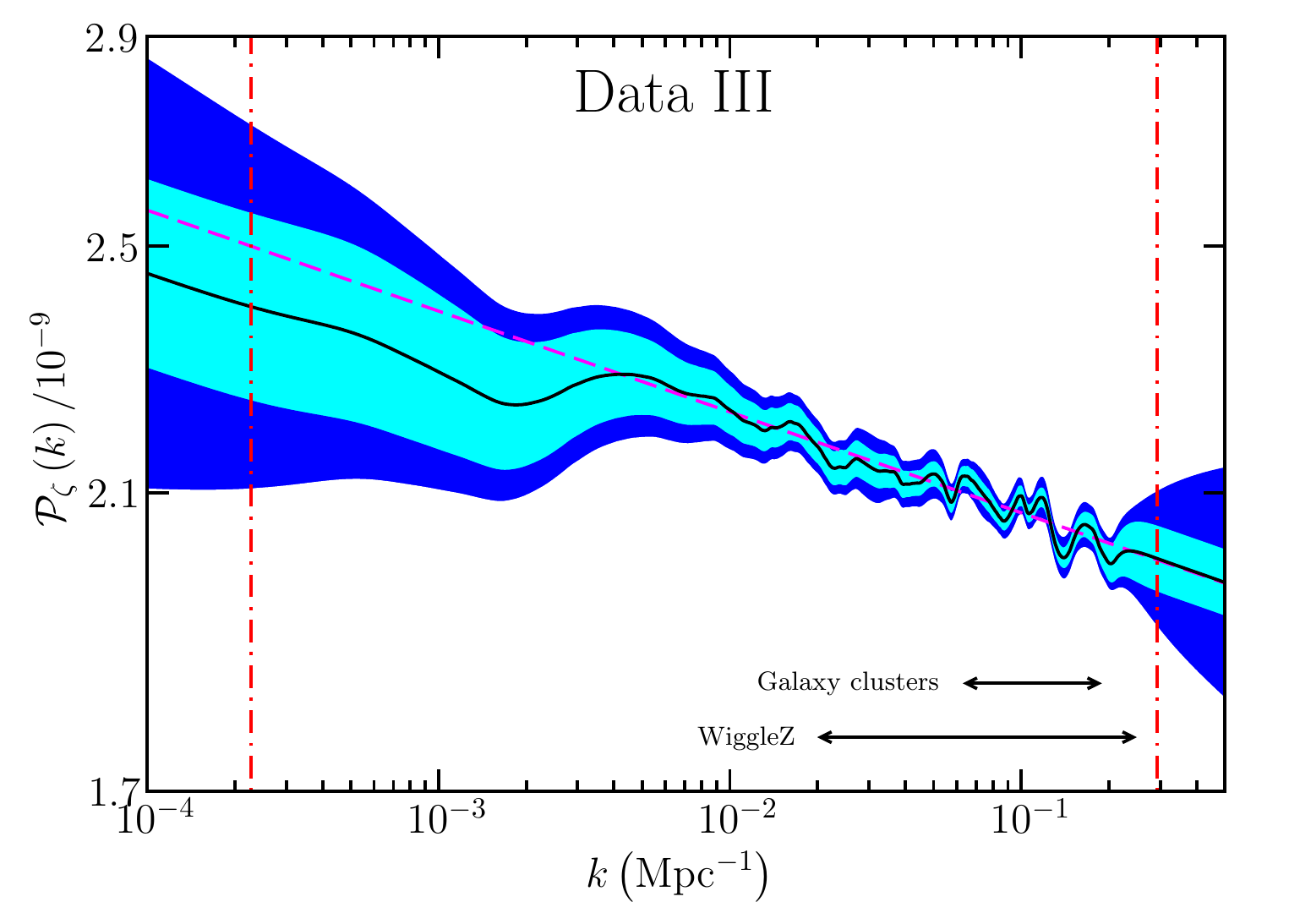}
\includegraphics*[angle=0,width=0.5\columnwidth,trim = 32mm 171mm 23mm
  15mm, clip]{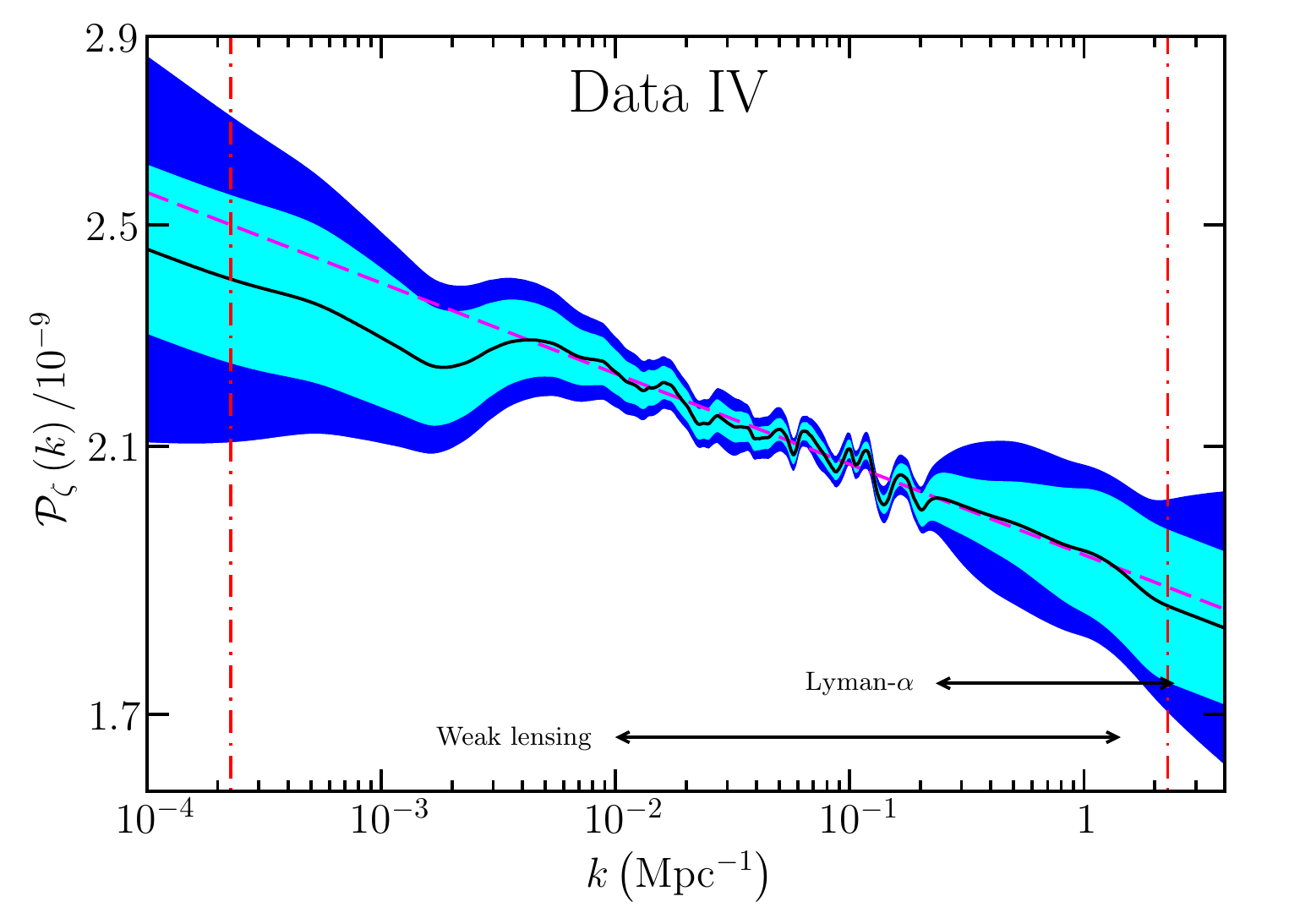}
\caption{As Fig.\ref{fig:ppsplot1} but for $\lambda=20000$.}
\label{fig:ppsplot2}
\end{figure}

The scalar PPS found from the Planck and WMAP-9 polarisation data
exhibits a number of interesting deviations from a power-law, as shown
in Figs.\ref{fig:ppsplot1} and \ref{fig:ppsplot2}. For
$k\lesssim 0.03\;\mathrm{Mpc}^{-1}$ the PPS is similar to that
recovered from the WMAP-9 temperature angular power spectrum in
\cite{Hunt:2013bha}. As in the earlier reconstruction, there is a
cutoff on large scales from the low TT quadrupole followed by dips at
$k\simeq 0.0018$, $0.0070$ and $0.013\;\mathrm{Mpc}^{-1}$ due to a
deficit in power around the $\ell\simeq 22$, $90$ and 180 multipoles
of the Planck TT spectrum, and peaks at $k\simeq 0.0034$ and
$0.0088\;\mathrm{Mpc}^{-1}$ due to the excess power around the
$\ell\simeq 40$ and 120 multipoles. A peak at
$k\simeq 0.027\;\mathrm{Mpc}^{-1}$ and dips at $k\simeq 0.032$ and
$0.039\;\mathrm{Mpc}^{-1}$ correspond to the excess around
$\ell\simeq 370$ and the deficits around $\ell\simeq 410$ and 540
respectively. A deficit around $\ell\simeq 800$ causes a dip at
$k\simeq 0.057\;\mathrm{Mpc}^{-1}$. Peaks at $k\simeq 0.10$ and
$0.12\;\mathrm{Mpc}^{-1}$ and a dip at
$k\simeq 0.105\;\mathrm{Mpc}^{-1}$ originate from the excesses around
$\ell\simeq 1350$ and 1600 and the deficit around $\ell\simeq 1450$
respectively.  We \emph{exclude} the $217\times217$ GHz cross-spectrum
for $1700<\ell<1860$ because these multipoles are known to be
contaminated by electromagnetic interference from the 4K Joule-Thomson
cryogenic cooler, as discussed in Appendix \ref{planckdat}. Hence the
dip found at $k\simeq 0.14\;\mathrm{Mpc}^{-1}$ arises from the deficit
around $\ell\simeq 1800$ in the $143\times217$ GHz cross-spectrum
\emph{alone} and is presumably uncontaminated.

Adding the ACT and SPT observations improves the reconstruction over
the range $0.08 \lesssim k\lesssim 0.28\;\mathrm{Mpc}^{-1}$.  A
deficit around $\ell\simeq 2450$ introduces a dip at
$k\simeq 0.19\;\mathrm{Mpc}^{-1}$. Including the WiggleZ and galaxy
cluster data, which cover
$0.02 \lesssim k\lesssim 0.25\;\mathrm{Mpc}^{-1}$, deepens the dips at
$k\simeq 0.032\;\mathrm{Mpc}^{-1}$ and
$k\simeq 0.19\;\mathrm{Mpc}^{-1}$. The reconstruction is extended to
smaller scales by the weak lensing and Lyman-$\alpha$ measurements,
which together span $0.01 \lesssim k\lesssim 2.0\;\mathrm{Mpc}^{-1}$.
The spectra in Fig.\ref{fig:ppsplot2} recovered with $\lambda=20000$
are suppressed for $k\lesssim 0.003\;\mathrm{Mpc}^{-1}$ due to the
deficit at $\ell\simeq 22$. The dip at
$k\simeq 0.14\;\mathrm{Mpc}^{-1}$ is the most significant.

\begin{figure}[tbh]
\begin{center}  
\includegraphics*[angle=0,width=0.63\columnwidth]{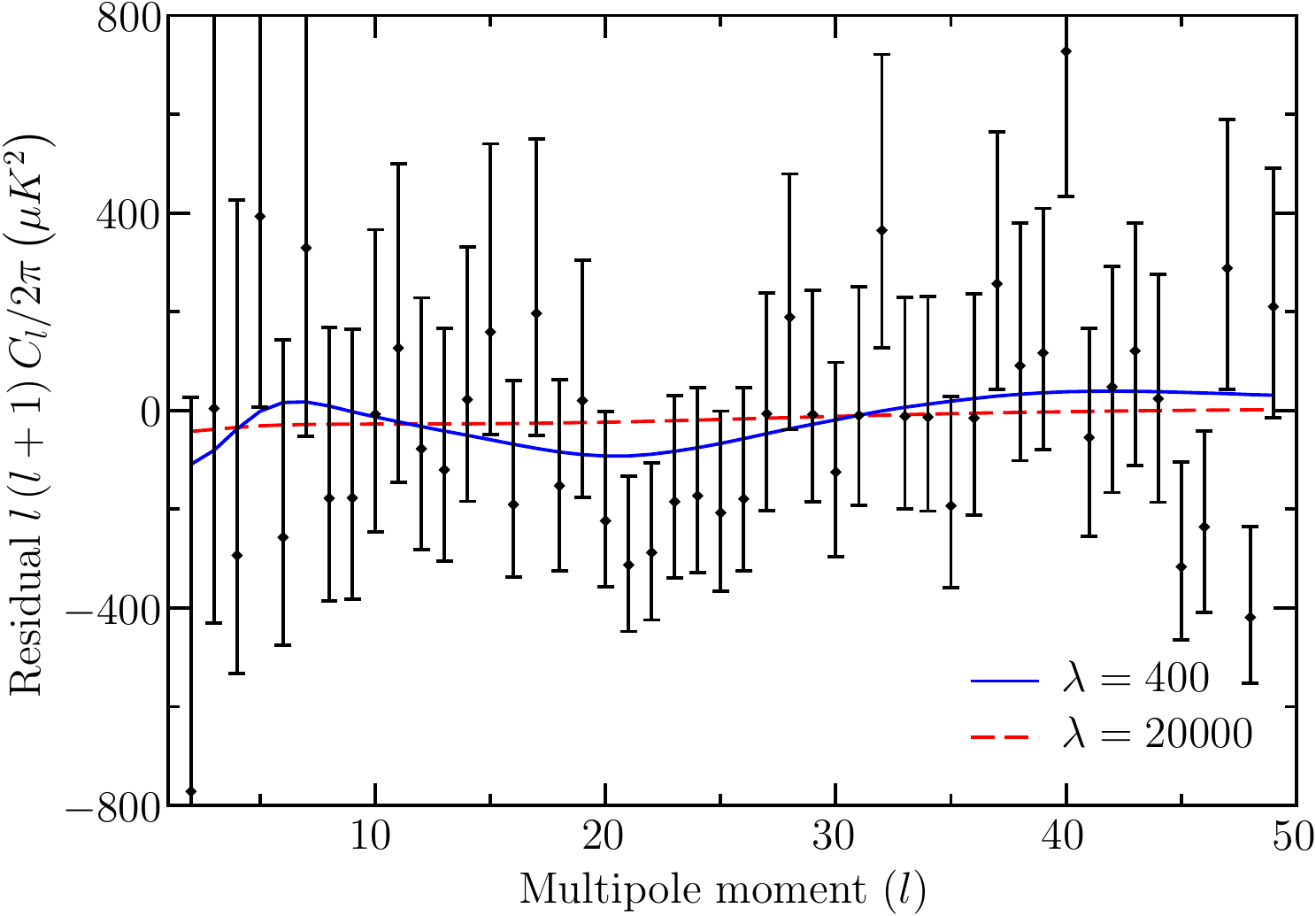}
\end{center}
\caption{Fits to the $\ell<50$ Planck TT data (residuals) of
  primordial power spectra recovered with $\lambda=400$ ((full blue
  line) and $20000$ (dashed red line) from the Planck, WMAP-9
  polarisation, ACT, SPT, WiggleZ, galaxy clustering, CFTHLenS and
  Lyman-$\alpha$ data (combination IV).  The residuals are obtained by
  subtracting off the TT spectrum of the best-fit $n_\mathrm{s}=0.969$
  power-law spectrum.}
\label{fig:cmblo}
\end{figure}

\begin{figure}[tbh]
\includegraphics*[angle=0,width=0.5\columnwidth,trim = 32mm 171mm 23mm
  15mm, clip]{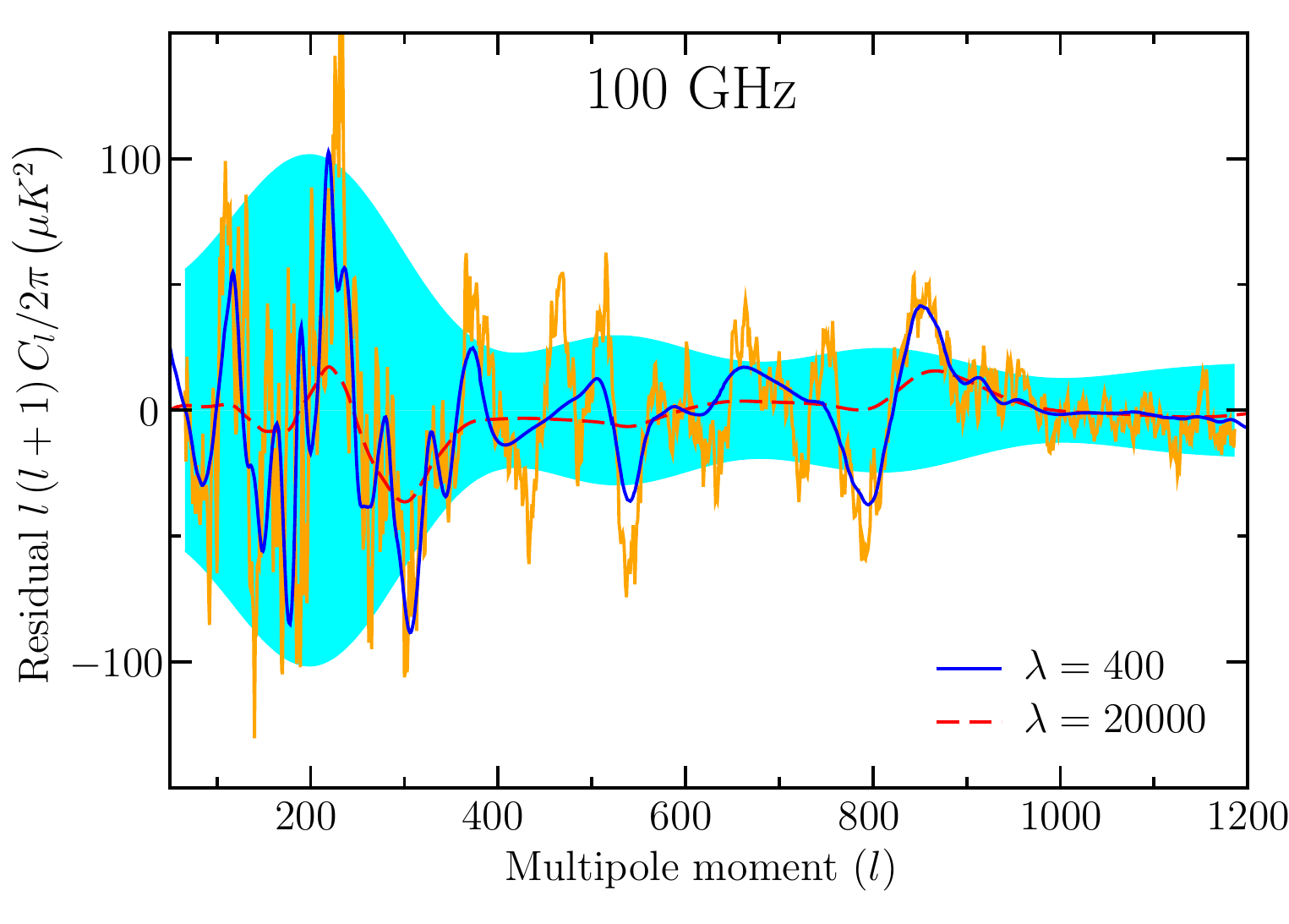}
\includegraphics*[angle=0,width=0.5\columnwidth,trim = 32mm 171mm 23mm
  15mm, clip]{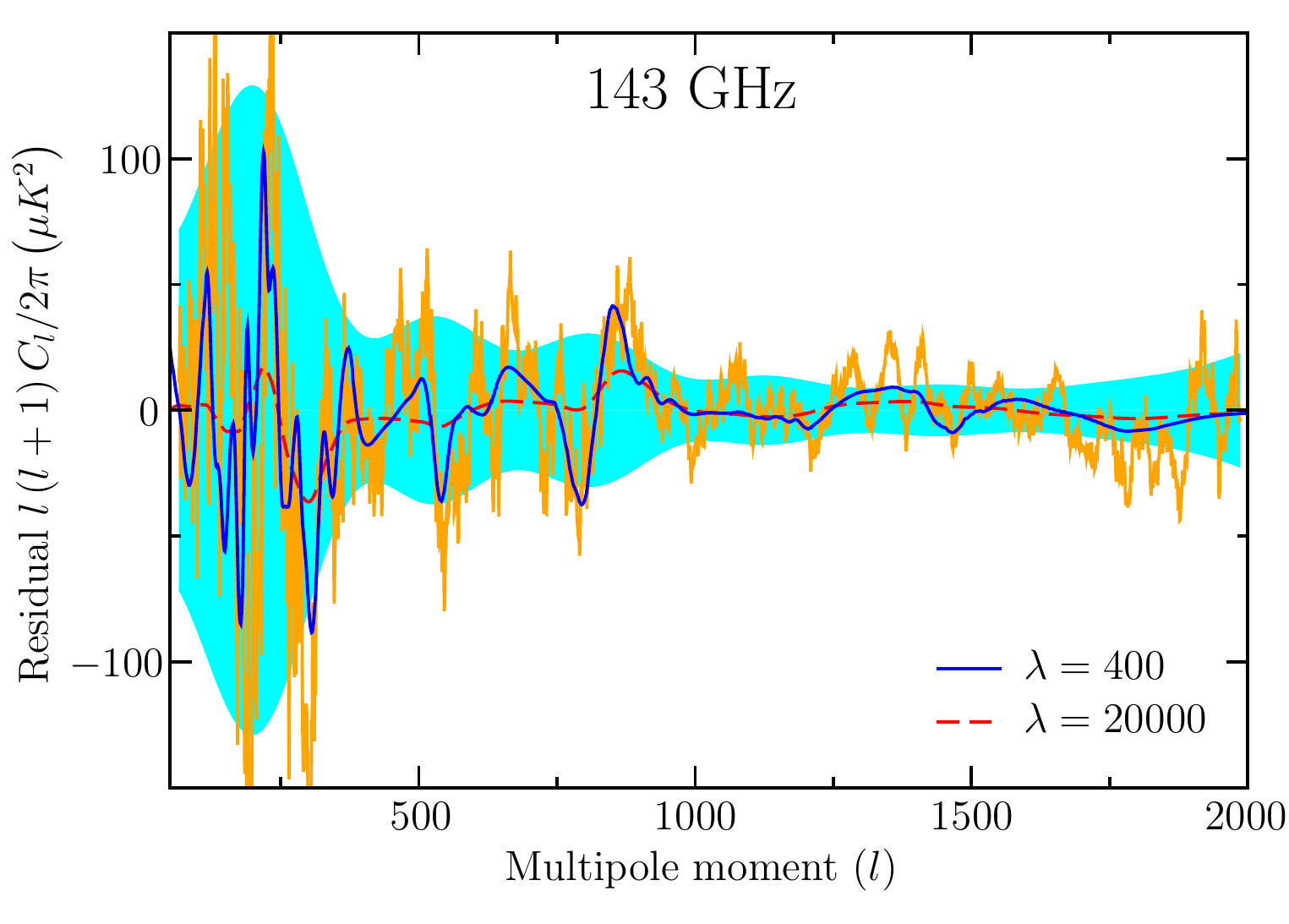}
\includegraphics*[angle=0,width=0.5\columnwidth,trim = 32mm 171mm 23mm
  15mm, clip]{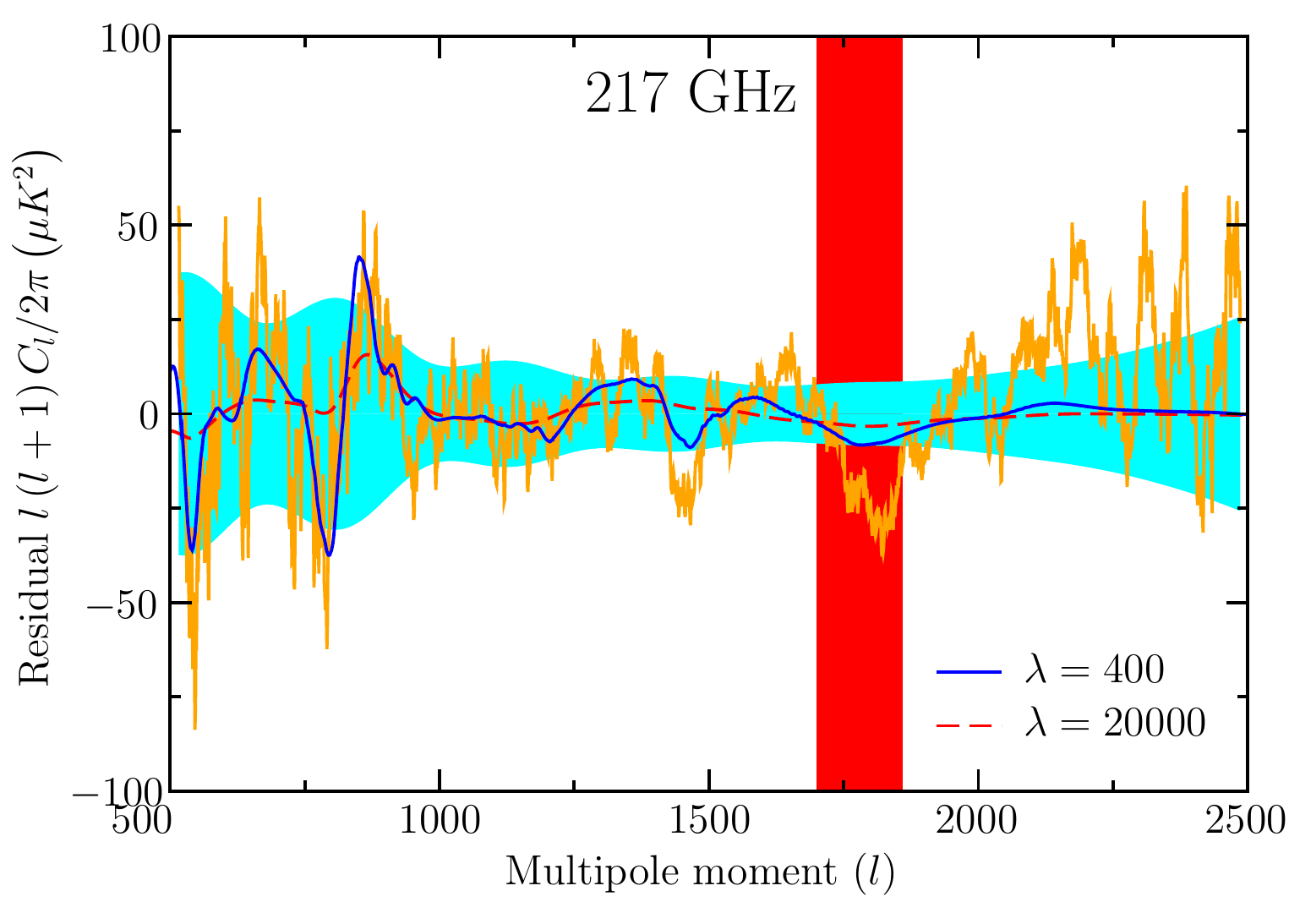}
\includegraphics*[angle=0,width=0.5\columnwidth,trim = 32mm 171mm 23mm
  15mm, clip]{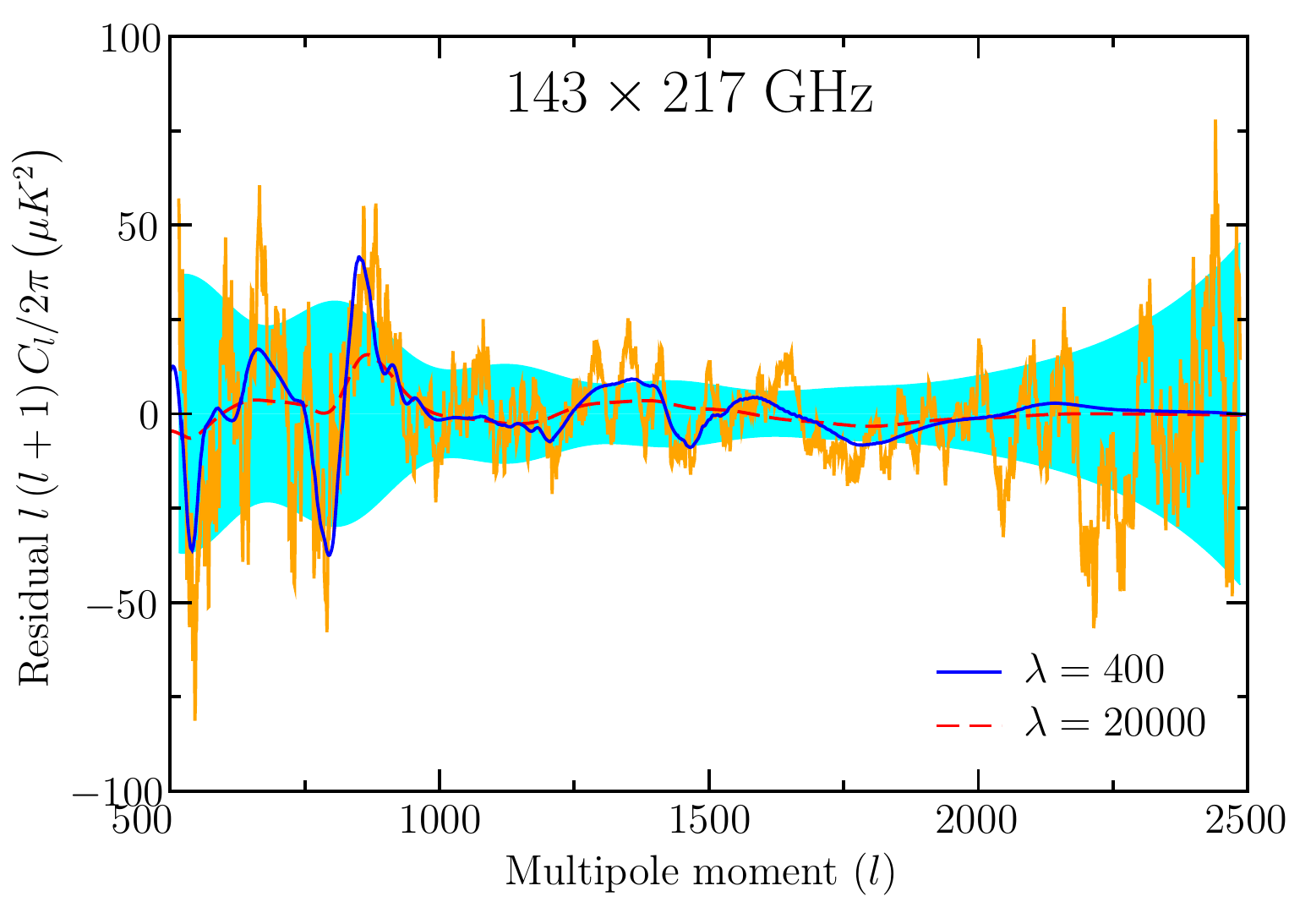}
  \caption{Comparison of residuals for the $\ell=31$ running average
    of the Planck data (orange line) with the residuals corresponding
    to the $\lambda=400$ (full blue line) and $\lambda=20000$ (red
    dashed line) reconstructions from the Planck, WMAP-9 polarisation,
    ACT, SPT, WiggleZ, galaxy clustering, CFTHLenS and Lyman-$\alpha$
    data (combination IV). The band indicates the $1\sigma$ scatter of
    the $\ell=31$ running average data, calculated from the Planck
    covariance matrix. Top left: 100 GHz data. Top right: 143 GHz
    data.  Bottom left: 217 GHz data (the vertical strip indicates the
    unused $1700<\ell<1860$ multipoles). Bottom right: $143\times217$
    GHz data.}
\label{runningaverage}
\end{figure}

\begin{figure}[tbh]
\begin{center}  
\includegraphics*[angle=0,width=0.5\columnwidth]{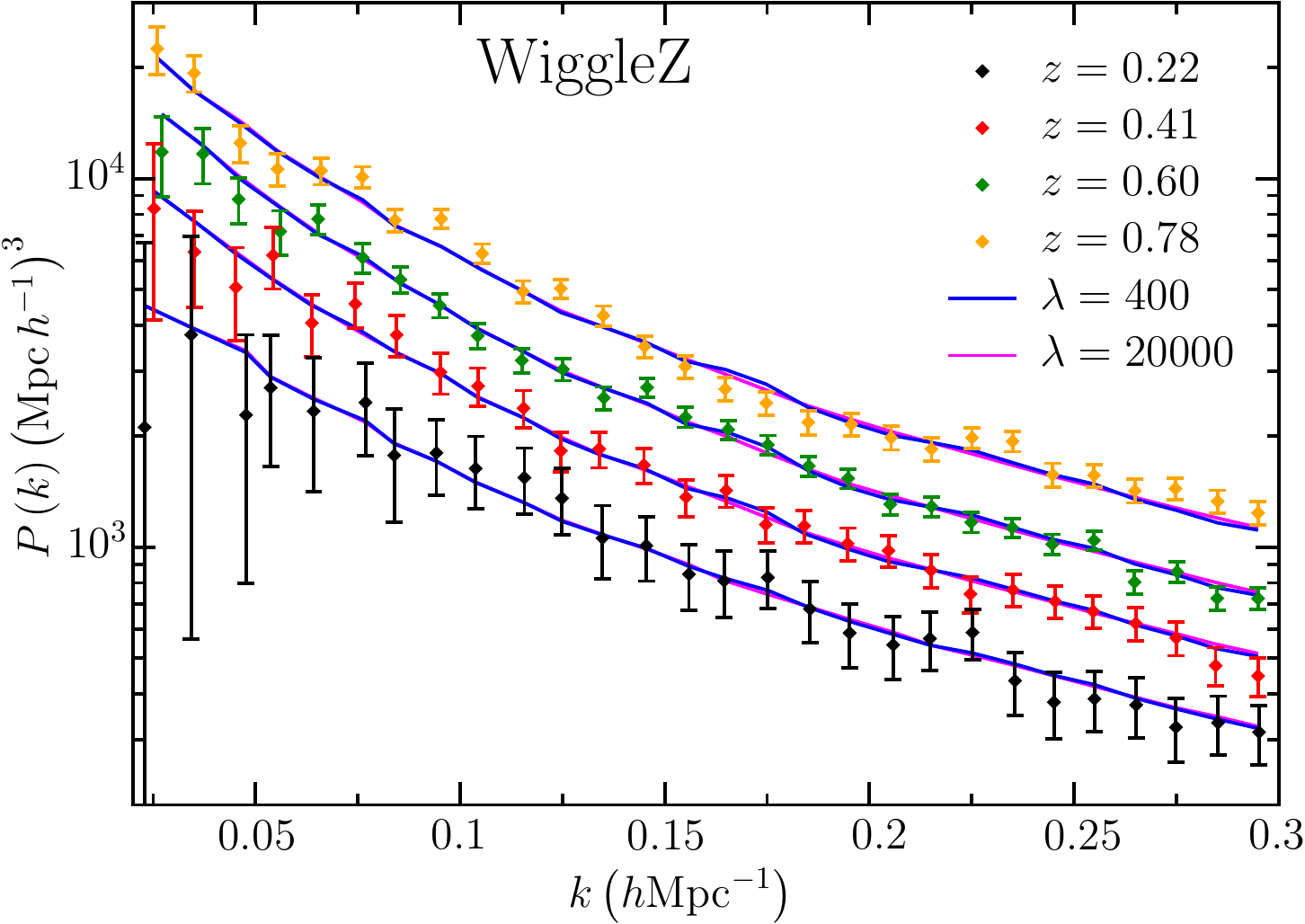}
\end{center}
\caption{Fit to WiggleZ data (in 4 redshift bins) of power spectra
  recovered with $\lambda=$ 400 and 20000 from the Planck, WMAP-9
  polarisation, ACT, SPT, WiggleZ, galaxy clustering, CFTHLenS and
  Lyman-$\alpha$ data (combination IV), convolved with the WiggleZ
  window functions.  Both data and theoretical predictions have been
  averaged over the 7 sky regions and shifted vertically for clarity
  (by a factor of 0.4, 0.6, 0.8 and 1.2 respectively for $z=$ 0.22,
  0.41, 0.60 and 0.78).}
\label{fig:wigglez}
\end{figure}

\begin{figure}[tbh]
\includegraphics*[angle=0,width=0.5\columnwidth,trim = 32mm 171mm 23mm
  15mm, clip]{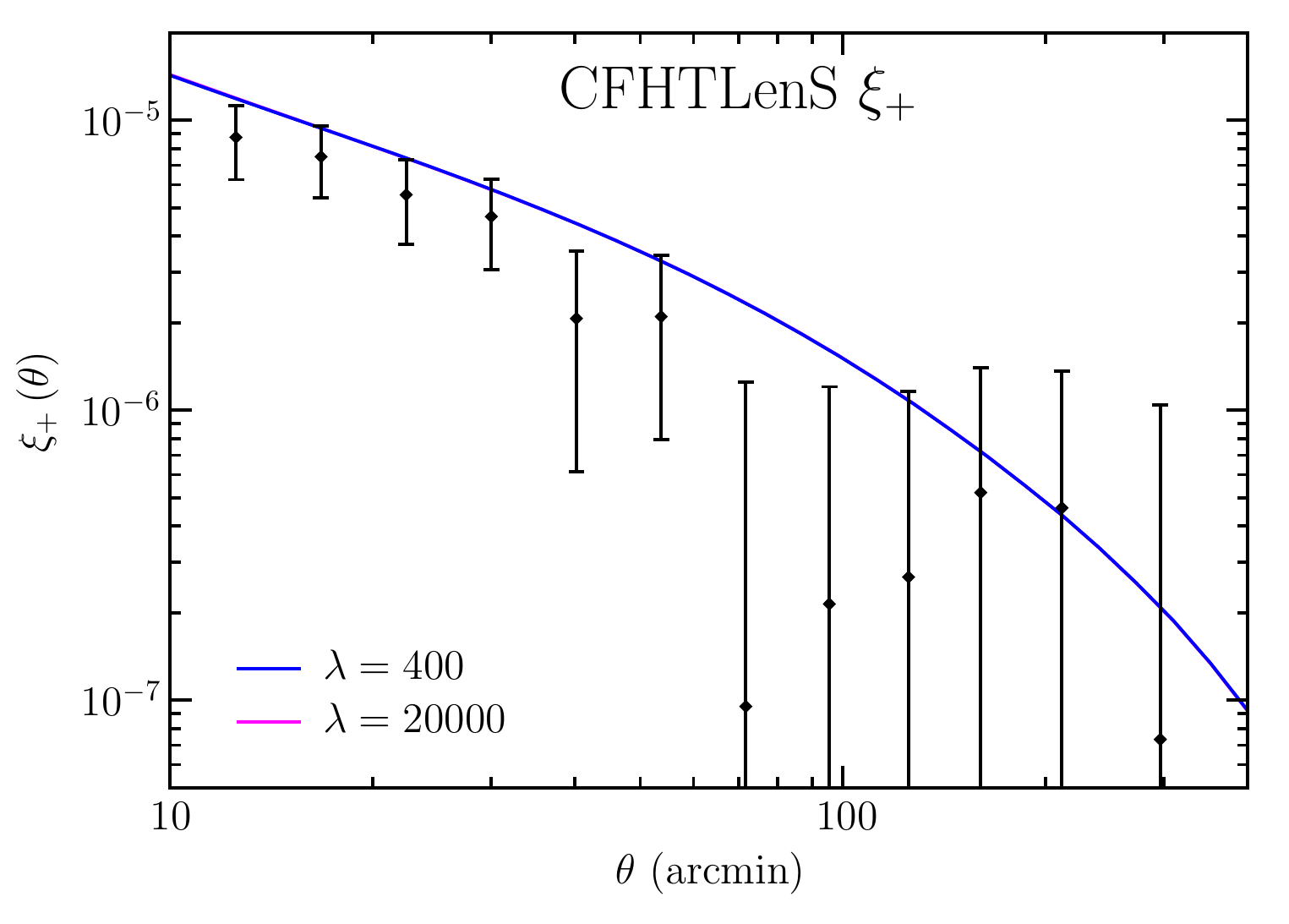}
\includegraphics*[angle=0,width=0.5\columnwidth,trim = 32mm 171mm 23mm
  15mm, clip]{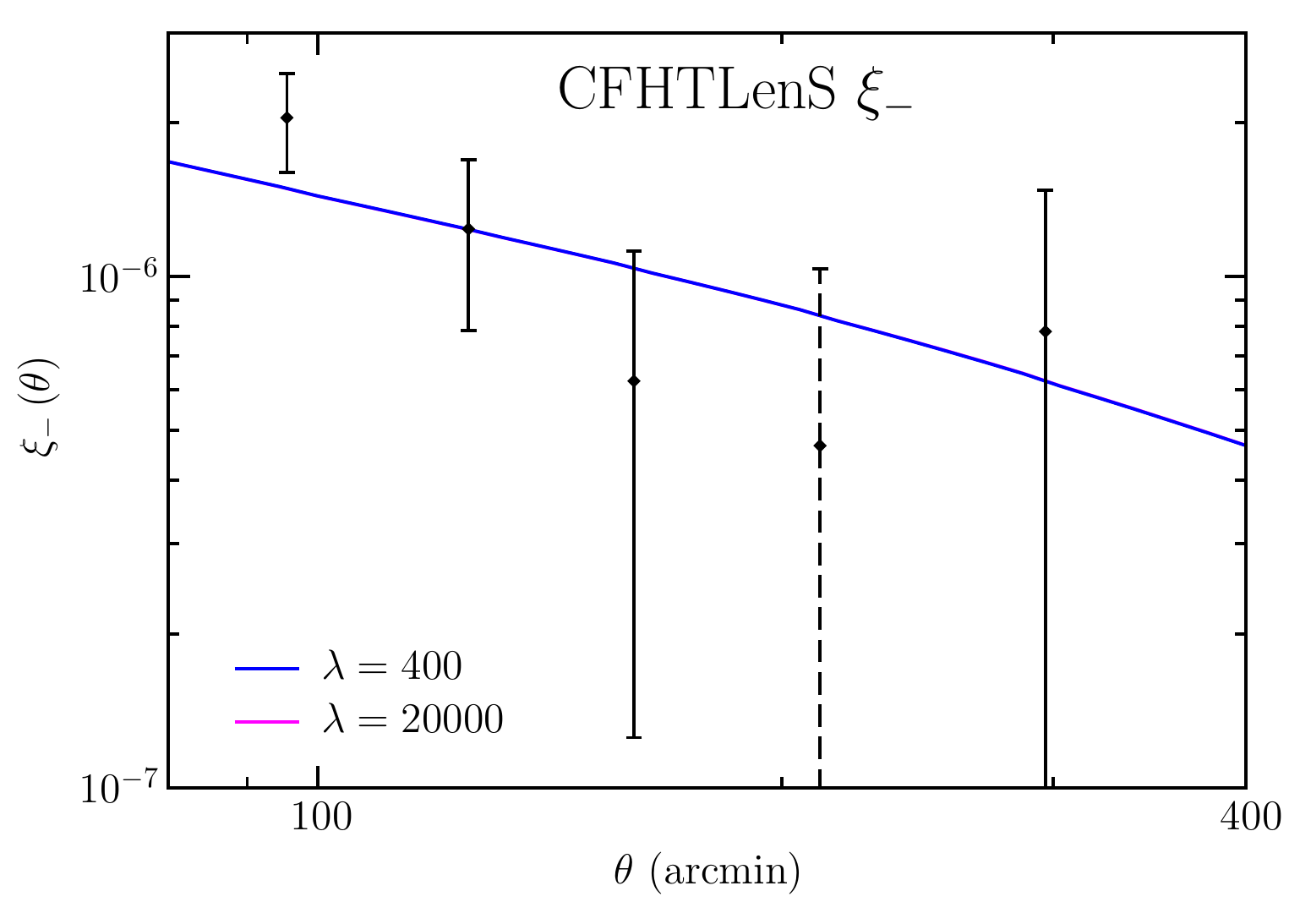}
\includegraphics*[angle=0,width=0.5\columnwidth,trim = 32mm 171mm 23mm
  15mm, clip]{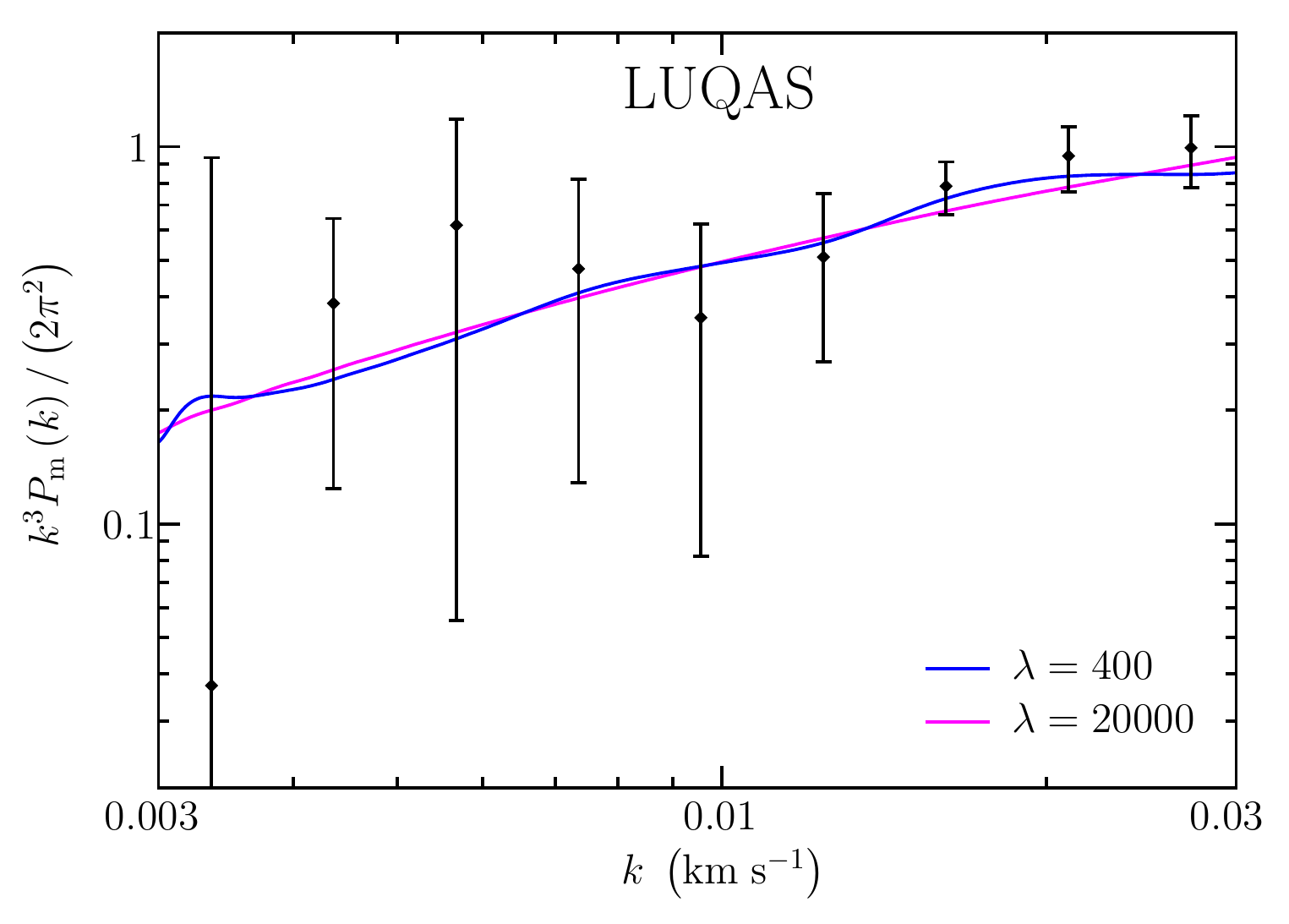}
\includegraphics*[angle=0,width=0.5\columnwidth,trim = 32mm 171mm 23mm
  15mm, clip]{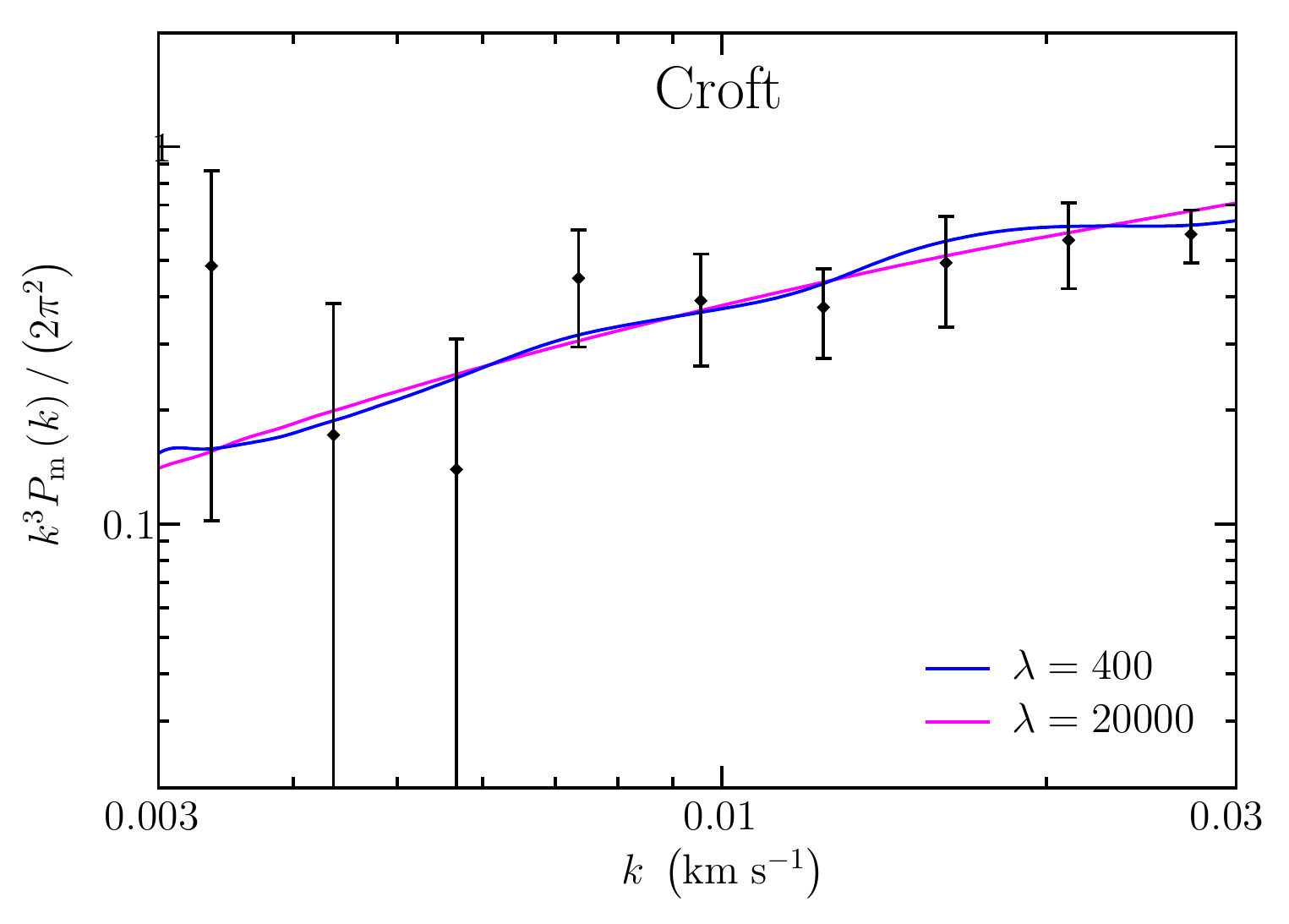}
  \caption{Fit to CFHTLenS shear correlation data
    $\xi_+\left(\theta\right)$ (top left) and
    $\xi_-\left(\theta\right)$ (top right), and Lyman-$\alpha$ data
    from LUQAS (bottom left) and Croft (bottom right), of power
    spectra recovered with $\lambda=$ 400 and 20000 from the Planck,
    WMAP-9 polarisation, ACT, SPT, WiggleZ, galaxy clustering,
    CFTHLenS and Lyman-$\alpha$ data (combination IV). The
    $\xi_-\left(\theta\right)$ measurement at $\theta=212'$ is
    \emph{negative} (indicated by a dashed error bar).}
\label{fig:wkllya}
\end{figure}

The Planck, ACT, SPT and WiggleZ data derived from the recovered
spectra are compared to the measured data in Fig.\ref{fig:cmblo} to
Fig.\ref{fig:wigglez}.  The `running average' is defined over $n$ data
points ($n$ odd) as:
$\hat{\mathrm{d}}_a^{(\mathbb{Z})}\equiv \frac{1}{n}
\sum_{b=-\left(n-1\right)/2}^{\left(n-1\right)/2}
\mathrm{d}_{a+b}^{(\mathbb{Z})}$.
The residuals after subtracting the TT spectrum of the best-fit
$n_\mathrm{s}=0.969$ PPS from that of the reconstructed spectra are
shown in Fig.\ref{runningaverage}, together with the $\ell=31$ running
average of the Planck data residuals.  In each case the predicted data
match the measurements well.  However, the predicted CFHTLenS
$\xi_+\left(\theta\right)$ shear correlation is systematically
\emph{higher} than the data, as seen in Fig.\ref{fig:wkllya}. This is
consistent with the tension between CFHTLenS and Planck for a
power-law PPS that has been reported in the literature, the cause of
which is an open question at present
\cite{Ade:2013zuv,Battye:2013xqa,MacCrann:2014wfa,Ade:2015xua,Raveri:2015maa}.
Planck is known to favour a slightly higher value of $\sigma_8$ than
galaxy cluster abundance observations \cite{Ade:2013lmv,Hu:2014qma}.
Here for $\lambda=500$ the galaxy cluster parameter is
$\Sigma_8\equiv \sigma_8
\left(\Omega_\mathrm{m}/0.27\right)^{0.3}=0.809$,
while $\Sigma_8=0.808$ for $\lambda=20000$. This is higher than, but
not inconsistent with, the value $\Sigma_8=0.797\pm 0.05$ obtained by
the cluster abundance observations listed in Appendix
\ref{clusterdat}.

\begin{figure}[tbh]
\includegraphics*[angle=0,width=0.5\columnwidth,trim = 32mm 171mm 23mm
  15mm, clip]{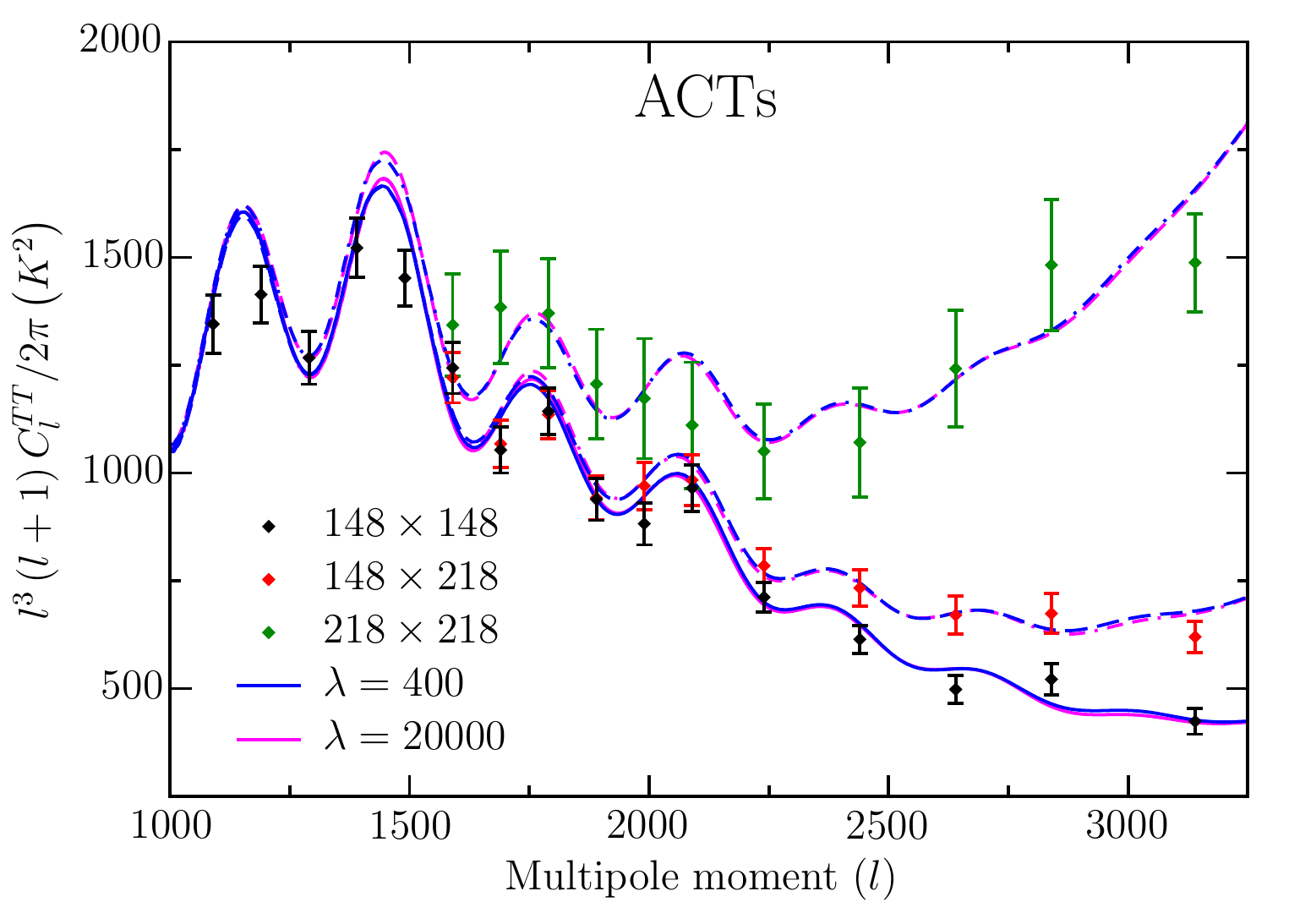}
\includegraphics*[angle=0,width=0.5\columnwidth,trim = 32mm 171mm 23mm
  15mm, clip]{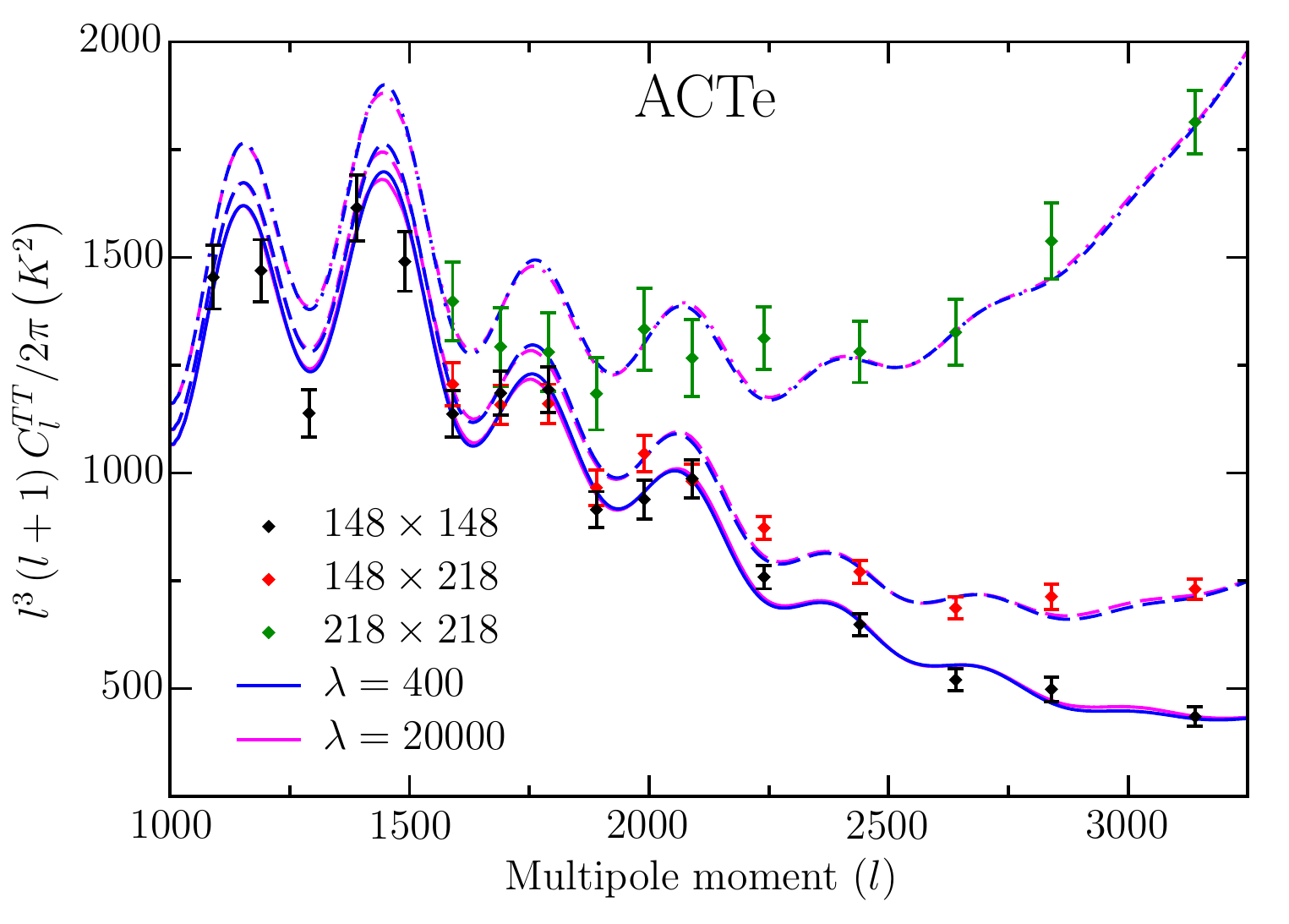}
\includegraphics*[angle=0,width=0.5\columnwidth,trim = 32mm 171mm 23mm
  15mm, clip]{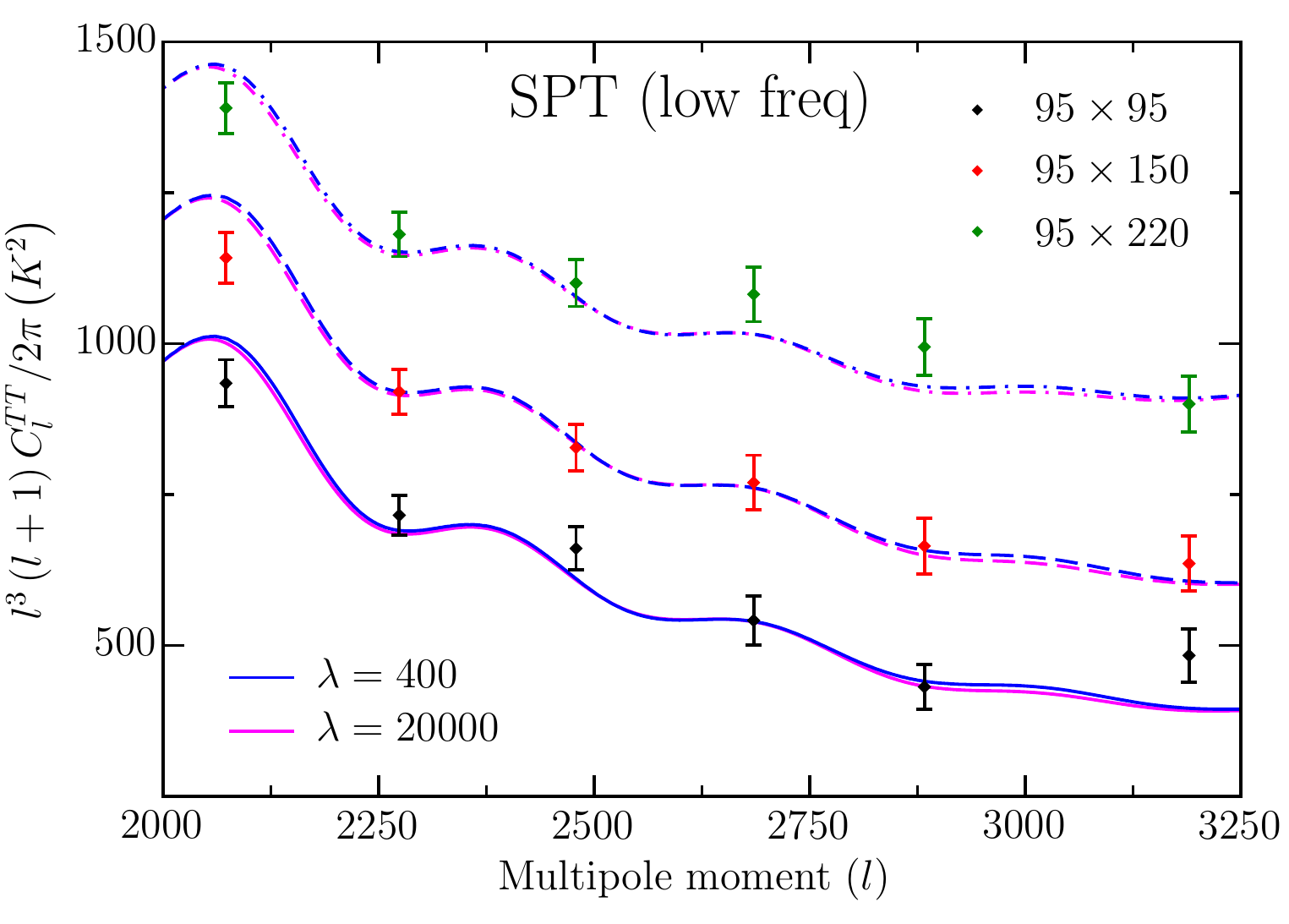}
\includegraphics*[angle=0,width=0.5\columnwidth,trim = 32mm 171mm 23mm
  15mm, clip]{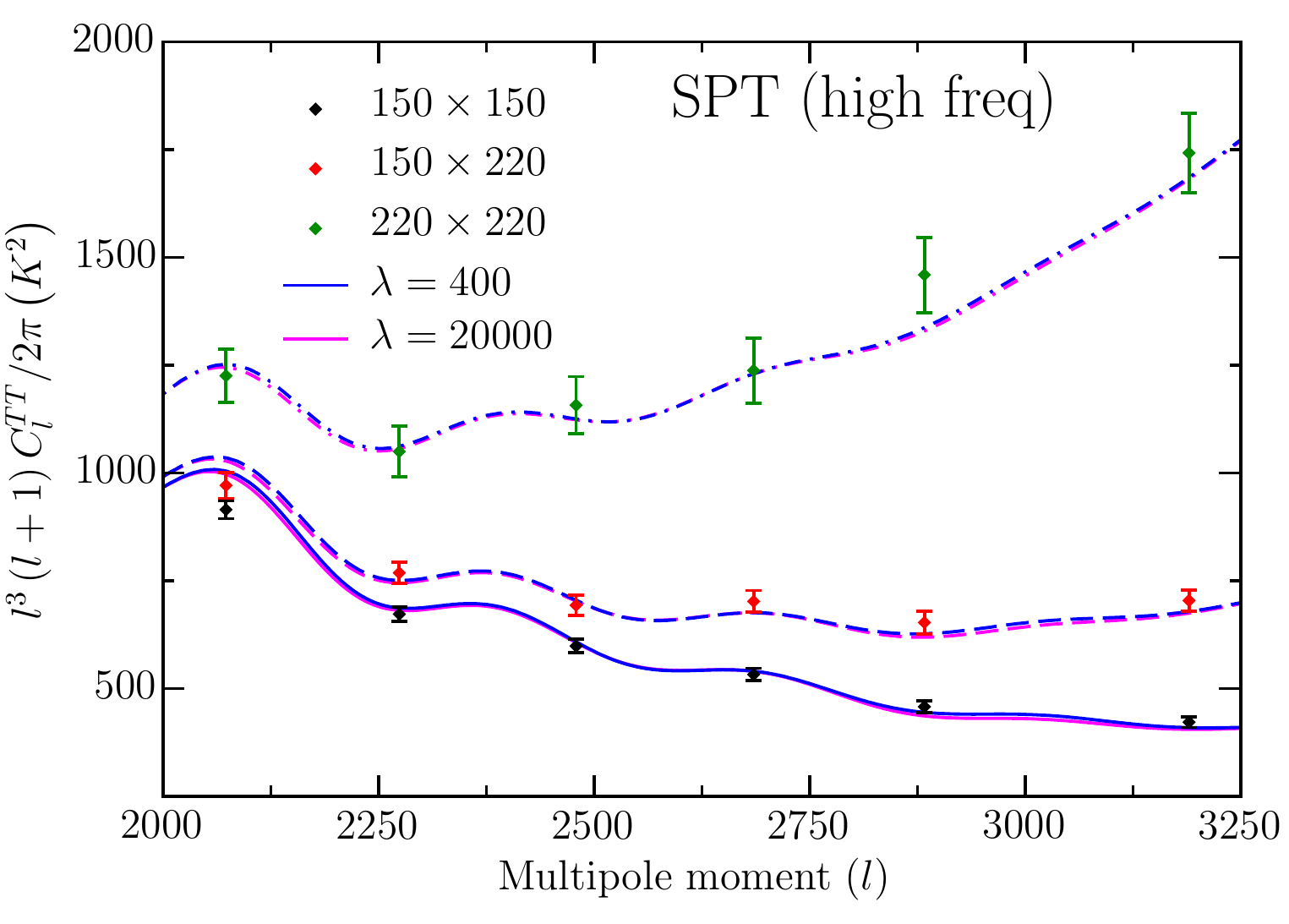}
  \caption{Fit to data from ACTs (top left), ACTe (top right), 95,
    $95\times150$, $95\times 220$ GHz SPT (bottom left) and 150,
    $150\times220$, 220 GHz SPT (bottom right) of power spectra
    recovered with $\lambda=$ 400 and 20000 from the Planck, WMAP-9
    polarisation, ACT, SPT, WiggleZ, galaxy clustering, CFTHLenS and
    Lyman-$\alpha$ data (combination IV). The $95\times150$ and
    $95\times 220$ GHz SPT spectra have been shifted vertically
    (respectively by 250 and 500 $K^2$) for clarity.}
\end{figure}

The VHS Lyman-$\alpha$ data is consistent with the CMB, WiggleZ and
galaxy cluster data for a calibration parameter $A=0.54$, a
$1.9\sigma$ deviation from the expected value of unity. This is in
agreement with \cite{Hoi:2007sf} which found that the VHS data for
$A=1$ is approximately a factor of 2 higher than expected from the
WMAP-3 results.

The large-scale cutoff, $k\simeq 0.0018\;\mathrm{Mpc}^{-1}$ dip and
$k\simeq 0.0034\;\mathrm{Mpc}^{-1}$ peak have been observed in
model-independent PPS estimates since the release of the WMAP-1 data
\cite{Chluba:2015bqa}. Our data combination I reconstruction with
$\lambda=400$ is clearly consistent with the PPS found from the
$50\leq \ell \leq 2500$ data by the Planck team using their penalised
likelihood method \cite{Planck:2013jfk}. Similar features can be seen
at e.g. $k\simeq 0.027$, $0.057$, $0.12$ and
$0.14\;\mathrm{Mpc}^{-1}$.  The latter three fluctuations were also
emphasised in the Richardson-Lucy deconvolution study
\cite{Hazra:2014jwa}.

\subsection{Background and nuisance parameter errors \label{bgpe}}

To demonstrate how errors in the background and nuisance parameters
affect the recovered PPS, we calculate the covariance matrix
$\mathsf{\Sigma}_\mathrm{P}$ (\ref{sigmap}) using the error matrix
\begin{eqnarray}
\label{bkgderror}
\mathsf{U} & = & \text{diag}\left[(0.012\,\omega_\mathrm{b})^2,
(0.022\,\omega_\mathrm{c})^2,(0.018\,h)^2,
(0.15\,\tau)^2, (0.03\,b)^2, (0.53\,A)^2, \right. \nonumber \\
& & (0.061\,A_{220}^\mathrm{PS,\;SPT})^2,
(0.026\,r^\mathrm{PS}_{150\times 220})^2, (0.2\,r^\mathrm{PS}_{95\times 220})^2,
(0.11\,r^\mathrm{PS}_{95\times 150})^2, (0.056\,A_{148}^\mathrm{PS,\;ACT})^2, \nonumber \\
& & (0.24\,A^\mathrm{PS}_{100})^2,
(0.089\,r^\mathrm{PS}_{143\times 217})^2, (0.06\,A_{218}^\mathrm{PS,\;ACT})^2,
(0.051\,A_{150}^\mathrm{PS,\;SPT})^2, (0.1\,A^\mathrm{CIB}_{217})^2, \nonumber \\
& &  (0.59\,A^\mathrm{tSZ}_{143})^2].
\end{eqnarray}
Here some selected nuisance parameters associated with the CMB
foregrounds $\mathrm{f}_\ell^I$ defined in Appendix~\ref{dsets} are
included. These errors correspond to the uncertainties in the
parameter values obtained from the Planck, ACT, SPT and WiggleZ data
assuming a power-law PPS.~\footnote{Although our analysis assumes that
  the data sets used to estimate the background and nuisance
  parameters are different from those used to recover the PPS, the
  parameter error matrix is used merely as an example.}
 
\begin{figure}[tbh]
\includegraphics*[angle=0,width=0.5\columnwidth,trim = 32mm 171mm 23mm
  15mm, clip]{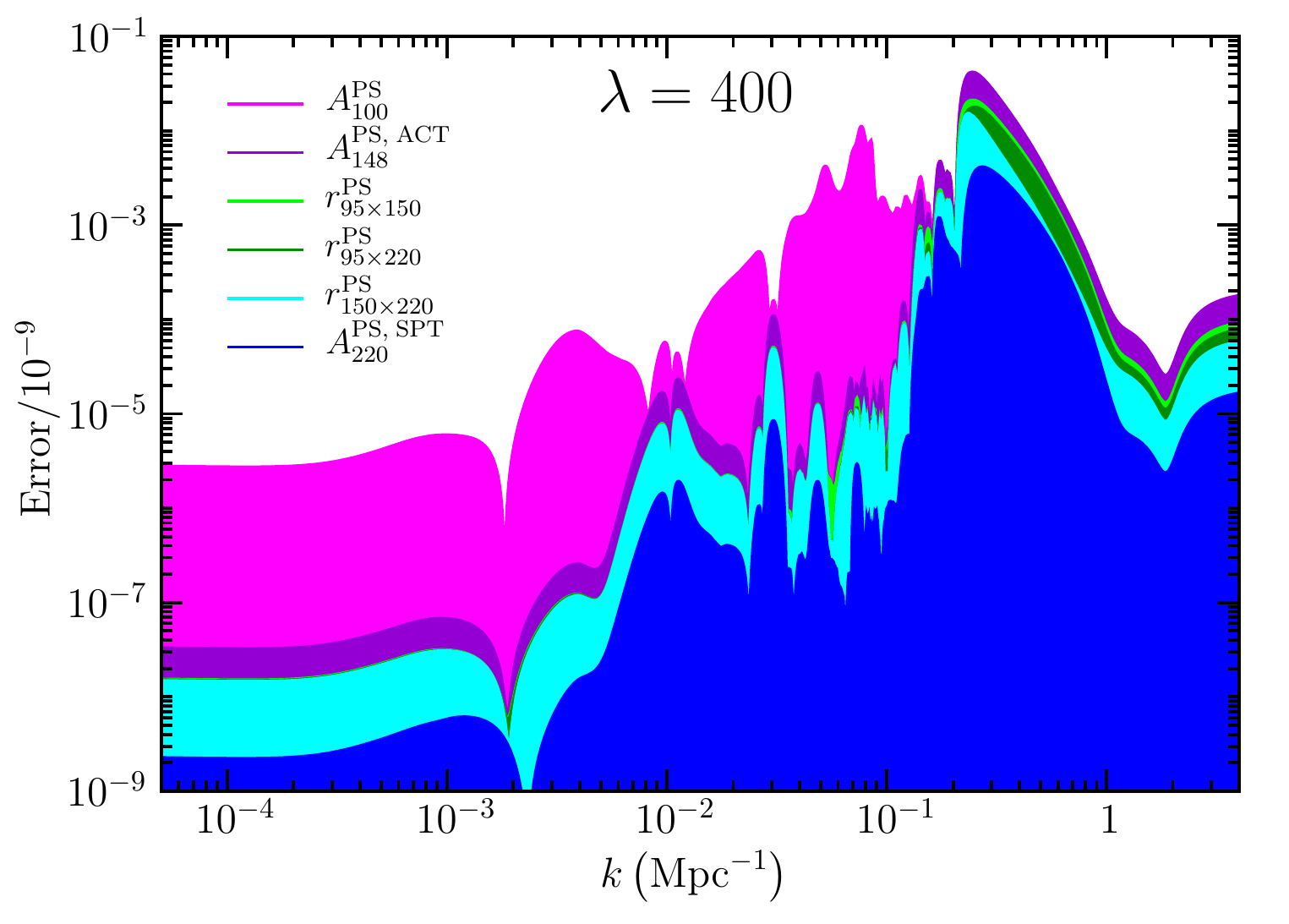}
\includegraphics*[angle=0,width=0.5\columnwidth,trim = 32mm 171mm 23mm
  15mm, clip]{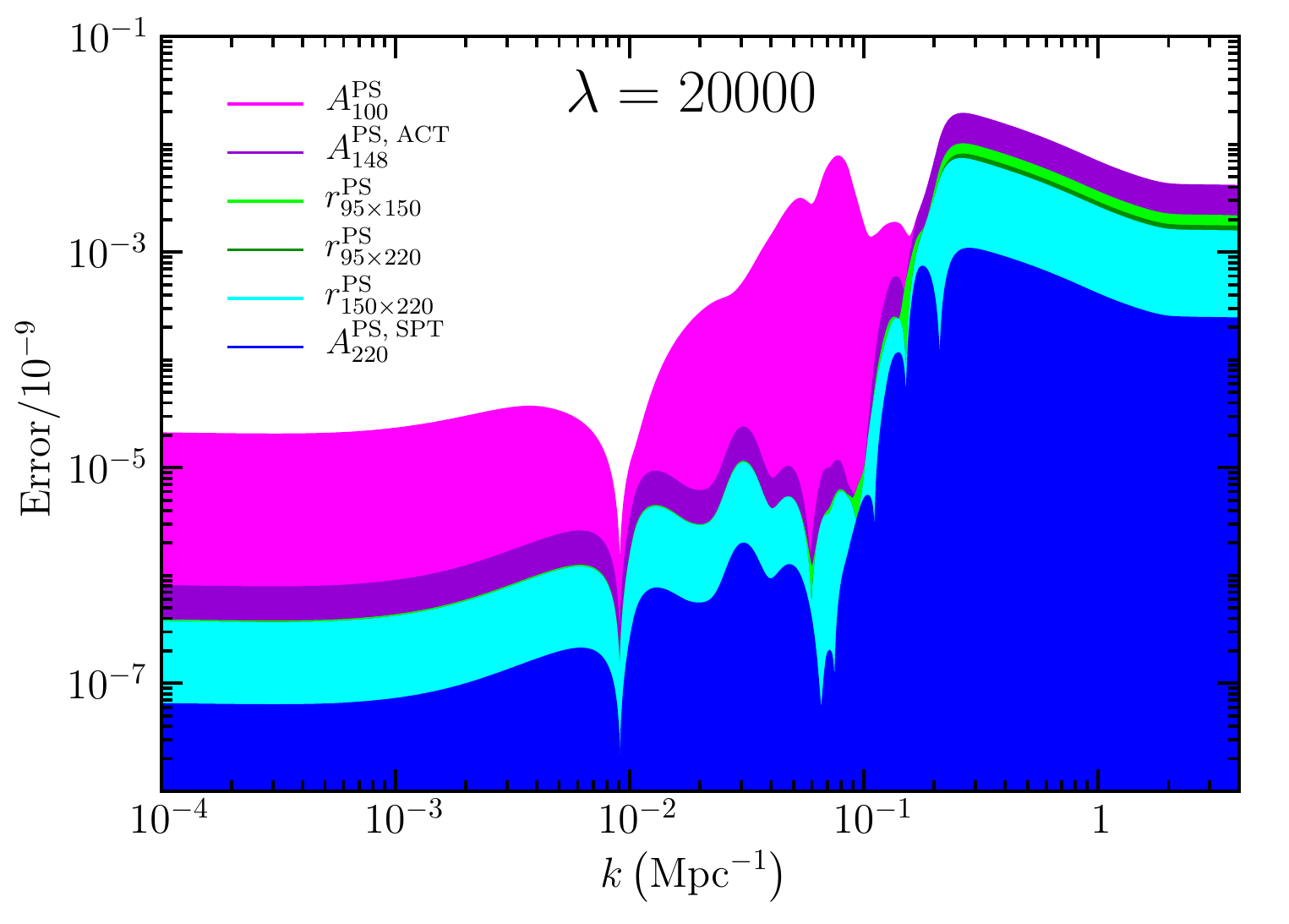}
\includegraphics*[angle=0,width=0.5\columnwidth,trim = 32mm 171mm 23mm
  15mm, clip]{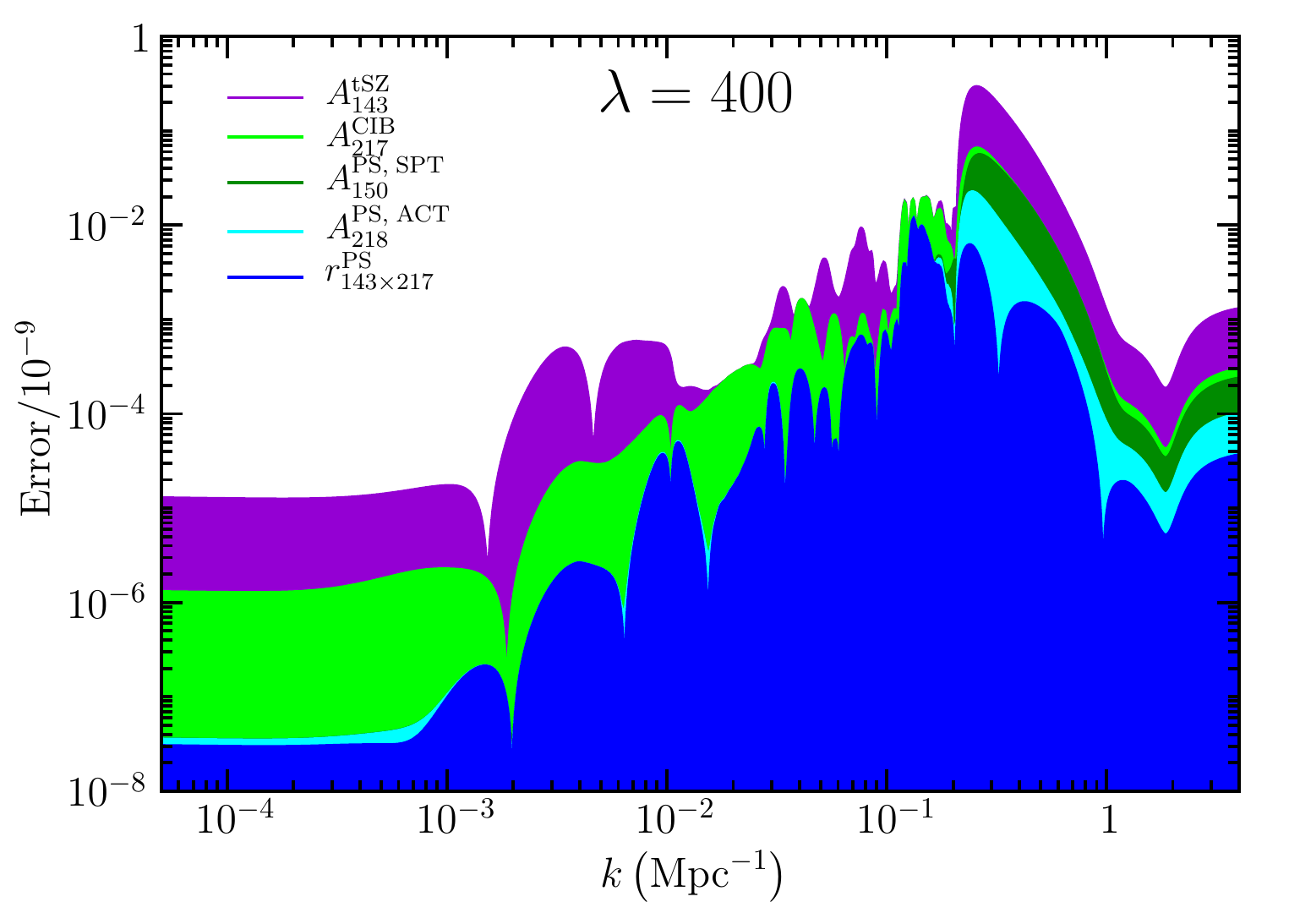}
\includegraphics*[angle=0,width=0.5\columnwidth,trim = 32mm 171mm 23mm
  15mm, clip]{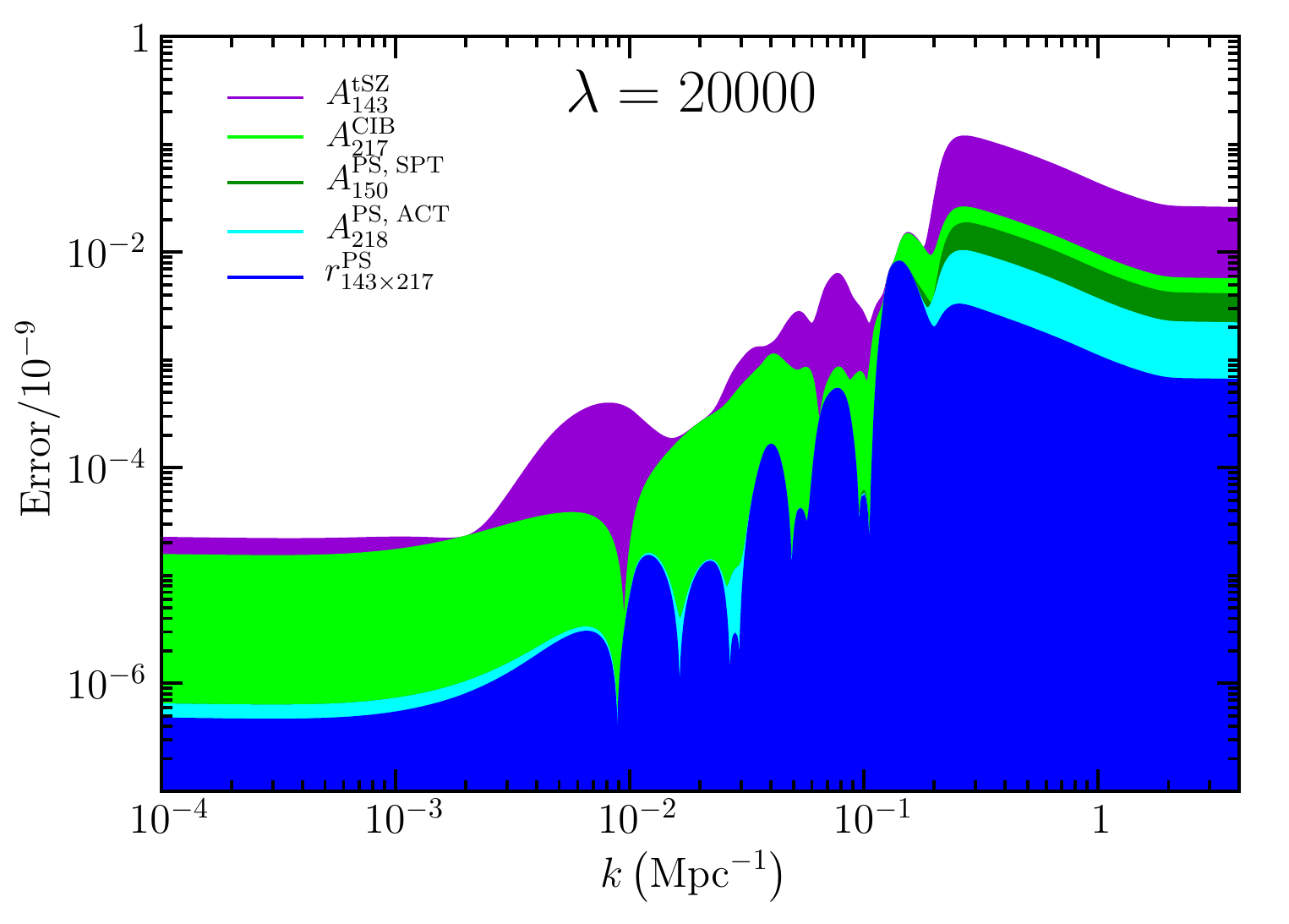}
  \caption{Contributions of some CMB foreground parameters to the
    square root of the diagonal elements of the matrix
    $\mathsf{\Sigma}_\mathrm{P}$ (\ref{sigmap}) for Planck, WMAP-9
    polarisation, ACT, SPT, WiggleZ, galaxy clustering, CFTHLenS and
    Lyman-$\alpha$ data (combination IV), with $\lambda =$ 400 and
    20000. The error contributions are added in quadrature.}
\label{fig:covparams1}
\end{figure}

\begin{figure}[tbh]
\includegraphics*[angle=0,width=0.5\columnwidth,trim = 32mm 171mm 23mm
  15mm, clip]{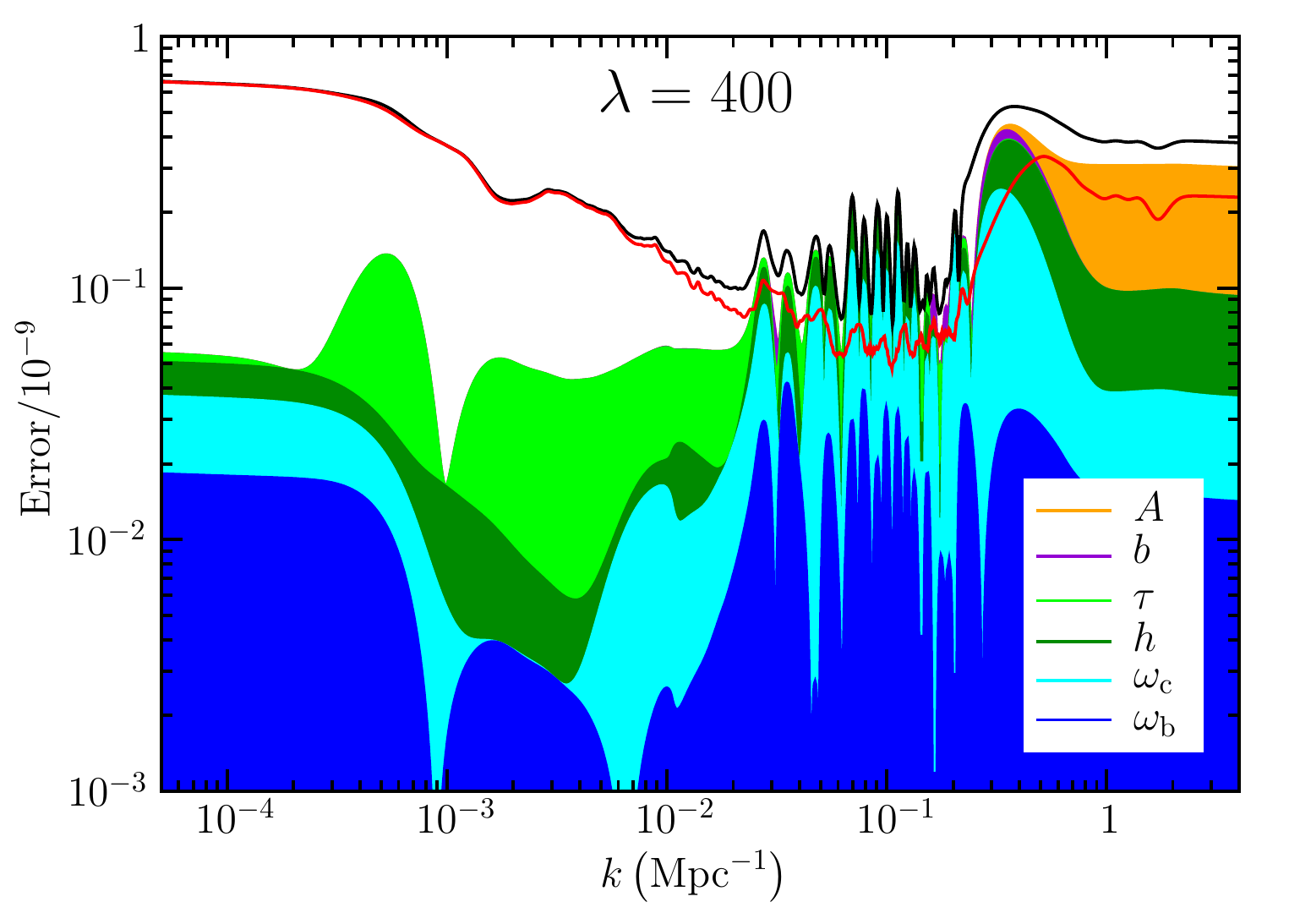}
\includegraphics*[angle=0,width=0.5\columnwidth,trim = 32mm 171mm 23mm
  15mm, clip]{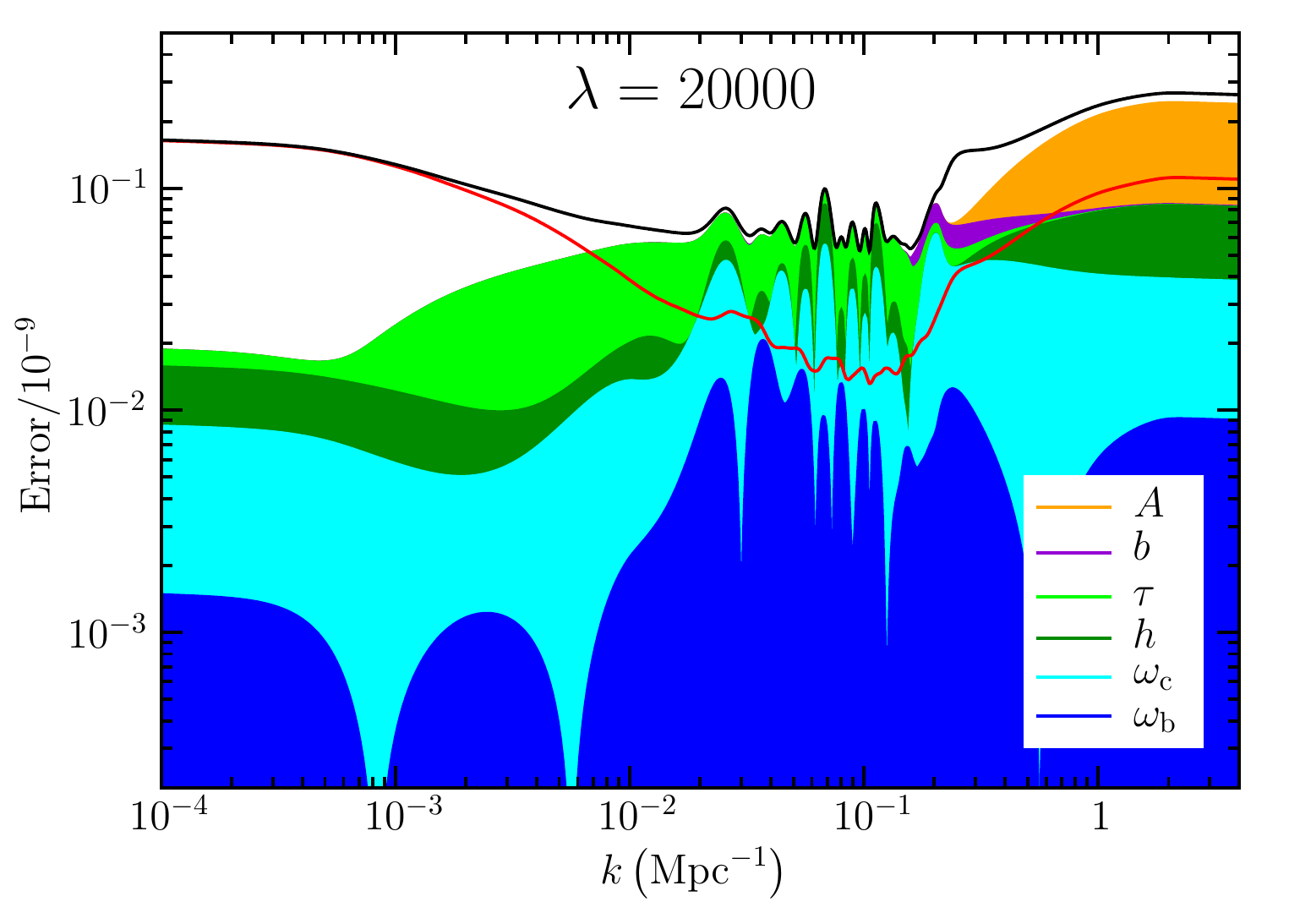}
  \caption{Contributions of different background parameters to the
    square root of the diagonal elements of the matrix
    $\mathsf{\Sigma}_\mathrm{P}$ (\ref{sigmap}) for data combination
    IV (Planck, WMAP-9 polarisation, ACT, SPT, WiggleZ, galaxy
    clustering, CFTHLenS, Lyman-$\alpha$) with $\lambda=$ 400 (left)
    and 20000 (right). The error contributions are added in
    quadrature. In both panels the red line is the square root of the
    diagonal elements of the matrix $\mathsf{\Sigma}_\mathrm{F}$
    (\ref{sigmaf}) and is included for comparison. The black line is
    the square root of the diagonal elements of the matrix
    $\mathsf{\Sigma}$ (\ref{sigmat}) and includes contributions to the
    total error from uncertainties in the background and nuisance
    parameters, as well as from noise in the data.}
\label{fig:covparams2}
\end{figure}

The effect of the foreground parameter uncertainties on the diagonal
elements of the matrix $\mathsf{\Sigma}_\mathrm{P}$ is shown in
Fig.\ref{fig:covparams1}.  The contribution of the foregrounds
$\mathrm{f}_\ell^I$ to the total TT angular power spectrum increases
with the multipole moment $\ell$. Hence the error due to uncertainties
in the foreground parameters is greatest at
$k\simeq 0.25\;\mathrm{Mpc}^{-1}$, which is approximately the smallest
scale probed by the CMB data.  As discussed in detail in
\cite{Hunt:2013bha}, the patterns of peaks on intermediate scales in
Fig.\ref{fig:covparams2} is due to the effects of uncertainties in the
background parameters propagating through the CMB acoustic peaks. The
Sachs-Wolfe plateau is more insensitive to the background parameters
and so the error is lower on large scales. The error on small scales
is dominated by the uncertainty in the Lyman-$\alpha$ calibration
parameter $A$. At $k\simeq 0.25\;\mathrm{Mpc}^{-1}$ the error due to
uncertainties in nuisance parameters is comparable with that from
background parameter uncertainties. On intermediate and small scales
the error due to uncertain background and nuisance parameter values is
greater than that due to noise in the data. We emphasise again that
our analysis \emph{assumes} the standard $\Lambda$CDM cosmology.

\subsection{Uncorrelated bandpowers \label{uncorr}}

\begin{figure}[tbh]
\includegraphics*[angle=0,width=0.5\columnwidth,trim = 32mm 171mm 23mm
  15mm, clip]{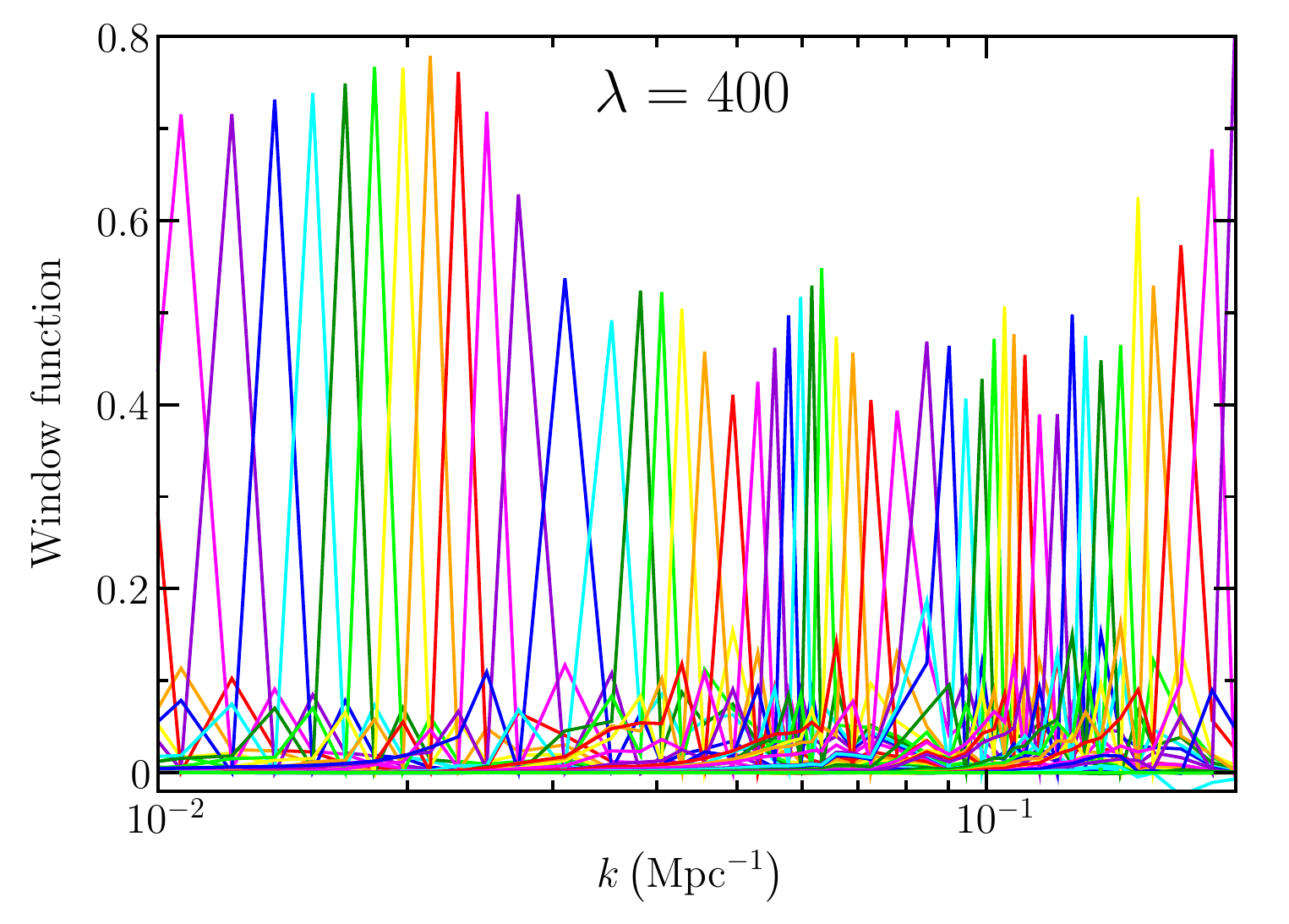}
\includegraphics*[angle=0,width=0.5\columnwidth,trim = 32mm 171mm 23mm
  15mm, clip]{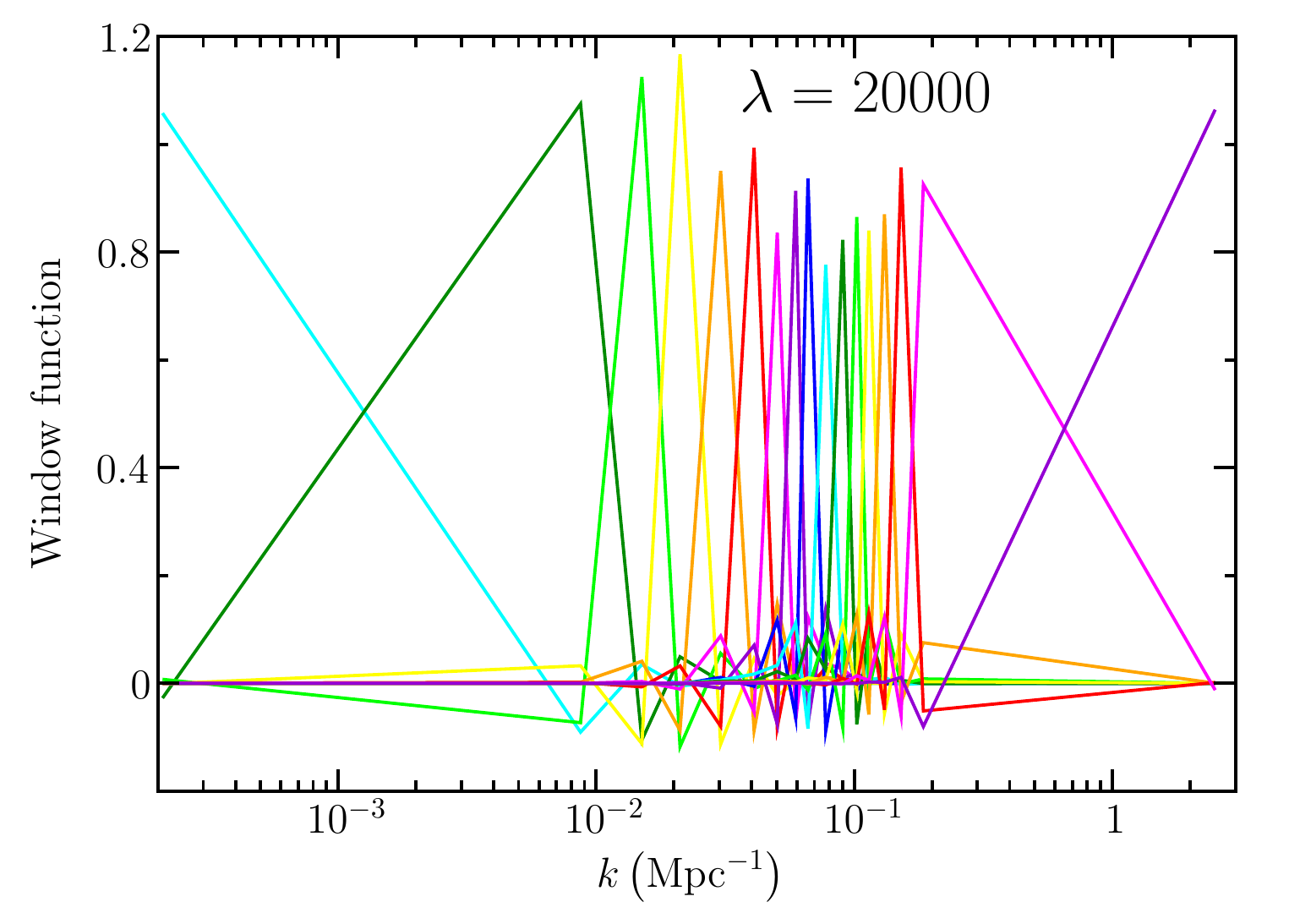}
  \caption{Bandpower window functions for $\lambda=$ 400, 20000 of
    reconstructions with data
    combination IV (Planck, WMAP-9 polarisation, ACT, SPT,
    WiggleZ, galaxy clustering, CFTHLenS, Lyman-$\alpha$).}
\label{fig:win}
\end{figure}

\begin{figure}[tbh]
\includegraphics*[angle=0,width=0.5\columnwidth,trim = 32mm 171mm 23mm
  15mm, clip]{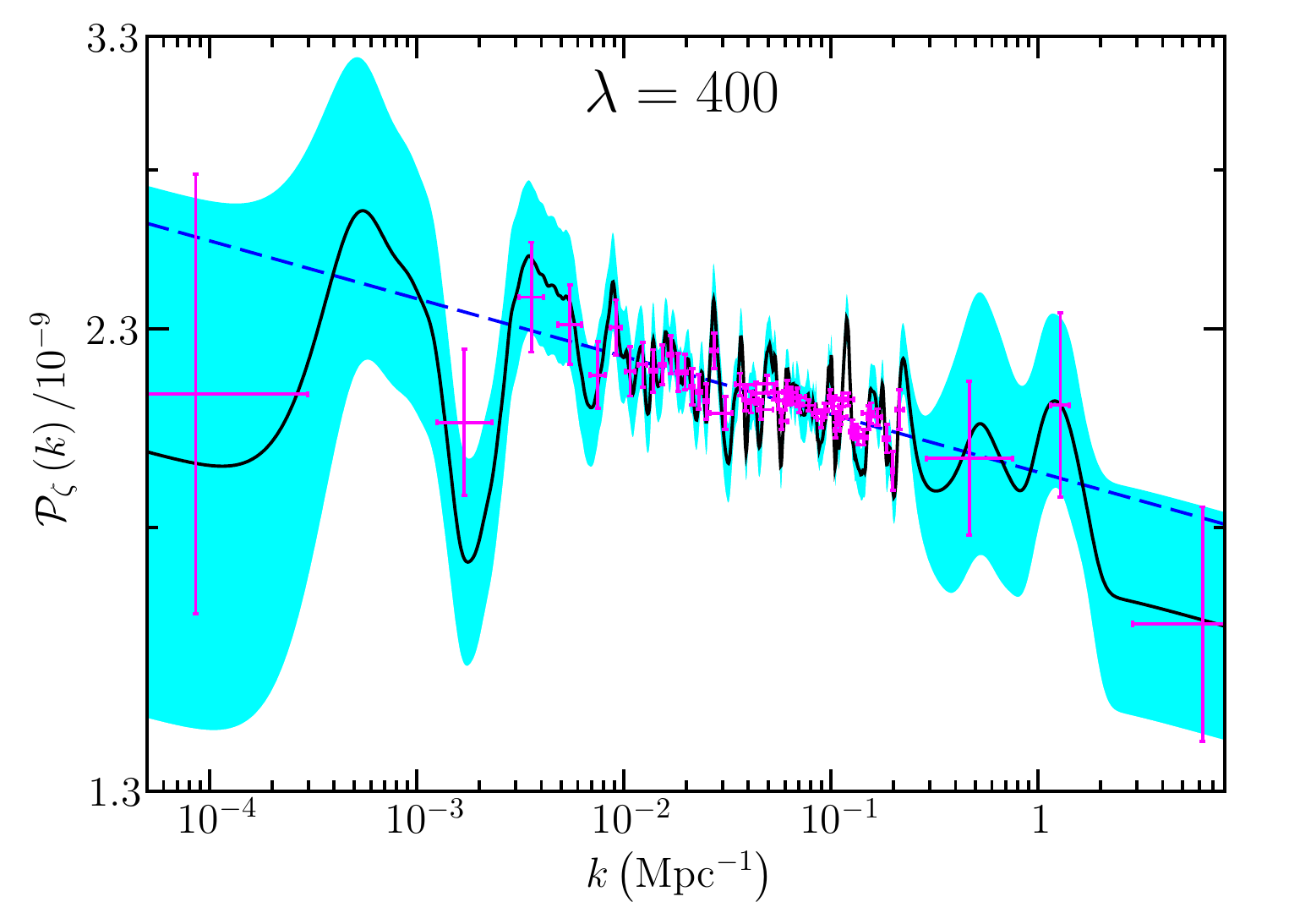}
\includegraphics*[angle=0,width=0.5\columnwidth,trim = 32mm 171mm 23mm
  15mm, clip]{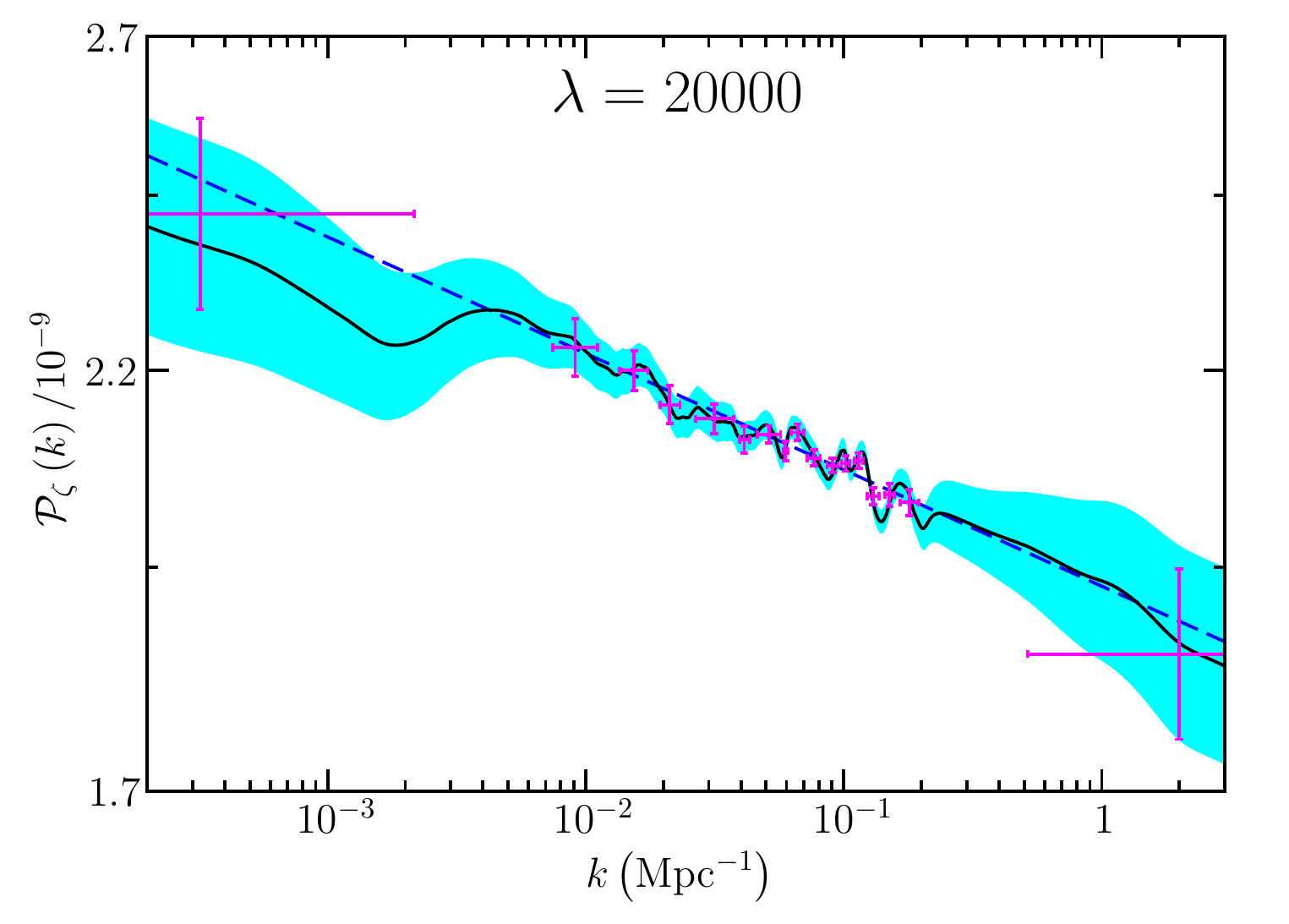}
  \caption{Decorrelated bandpowers for $\lambda=400$ and $20000$. The
    black line is the PPS recovered from data combination IV (Planck,
    WMAP-9 polarisation, ACT, SPT, WiggleZ, galaxy clustering,
    CFTHLenS, Lyman-$\alpha$). The light band is the $1\sigma$ error
    obtained from the square root of the diagonal elements of the
    frequentist covariance matrix $\mathsf{\Sigma}_\mathrm{F}$
    (\ref{sigmaf}). The vertical error bars are the $1\sigma$ errors
    given by the diagonal bandpower covariance matrix. The horizontal
    error bars indicate the locations of the 25th and 75th percentiles
    of the absolute value of the bandpower window functions.}
\label{fig:decorr}
\end{figure}

Our understanding of the recovered PPS is complicated by the
correlation between neighbouring PPS elements due to the smoothing
criterion. To overcome this we calculate \emph{uncorrelated}
bandpowers which represent the independent degrees of the freedom of
the reconstruction using the method of \cite{Hunt:2013bha}.
Correlated bandpowers (with a non-diagonal frequentist covariance
matrix $\mathsf{\Sigma}_\mathrm{N}$) are transformed into uncorrelated
bandpowers (with a diagonal covariance matrix) by multiplication with
a set of window functions. These are the rows of the Hermitian square
root of $\mathsf{\Sigma}_\mathrm{N}^{-1}$, normalised to sum to
unity. The effective number of free parameters of the reconstruction
is estimated by the quantity
$\nu_1\equiv\sum_{\mathbb{Z},i,a} W^{(\mathbb{Z})}_{ai}
M^{(\mathbb{Z})}_{ia}$.
Since $\nu_1=56.5$ for $\lambda=400$ and $\nu_1=16.85$ for
$\lambda=20000$ we choose $57$ bandpowers for $\lambda=400$ and $17$
bandpowers for $\lambda=20000$. The correlated bandpowers are chosen
so that the window functions are as well-behaved and non-negative as
possible. Figs.\ref{fig:win} and \ref{fig:decorr} show the window
functions and uncorrelated bandpowers. The window functions are lower
and less localised at the wavenumbers corresponding to the troughs of
the CMB angular power spectrum, where the resolution is reduced.

\subsection{Statistical significance of the features \label{statsig}}

\begin{figure}[tbh]
\includegraphics[width=0.5\columnwidth,trim = 32mm 171mm 23mm 15mm, clip]{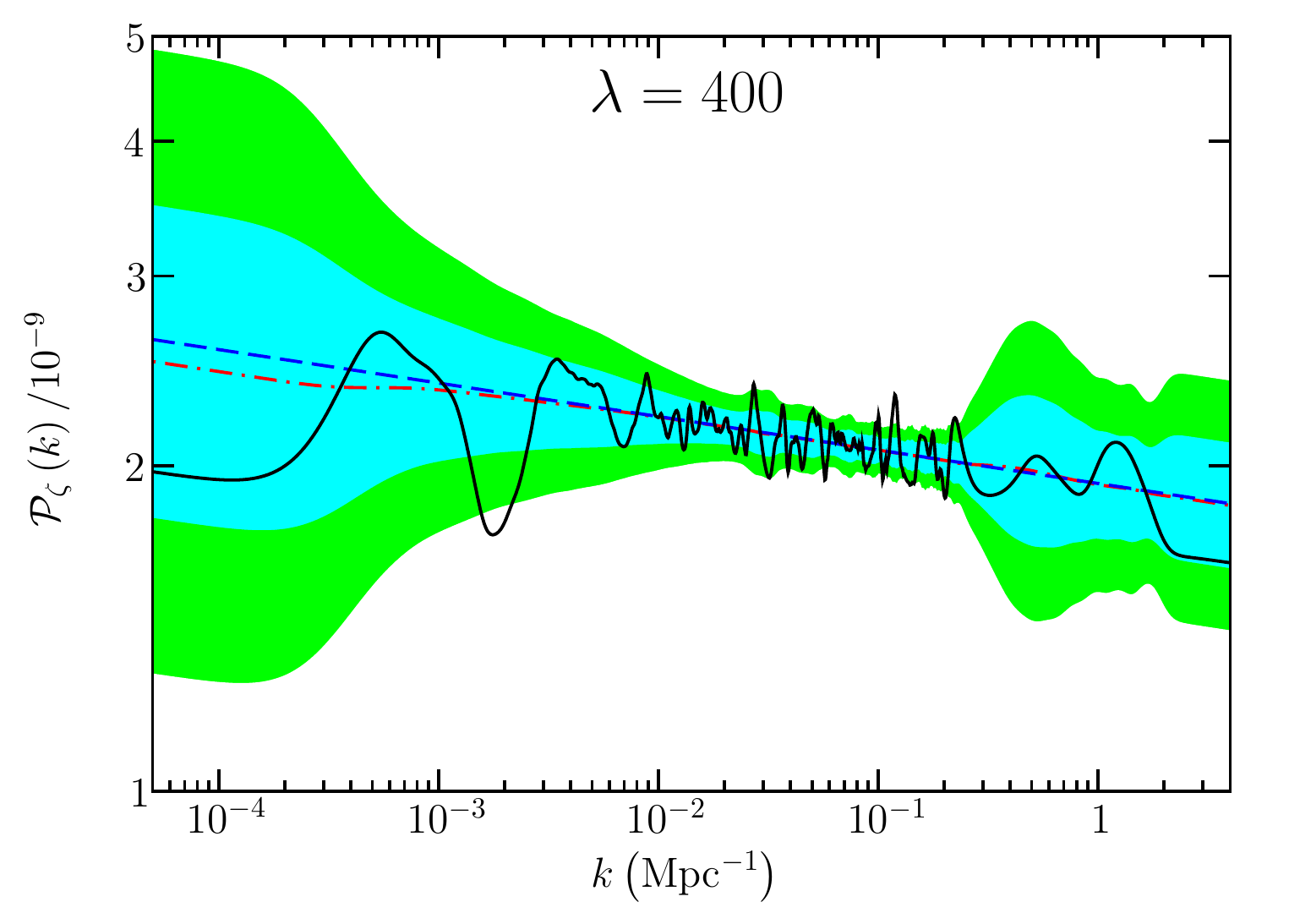}
\includegraphics[width=0.5\columnwidth,trim = 32mm 171mm 23mm 15mm, clip]{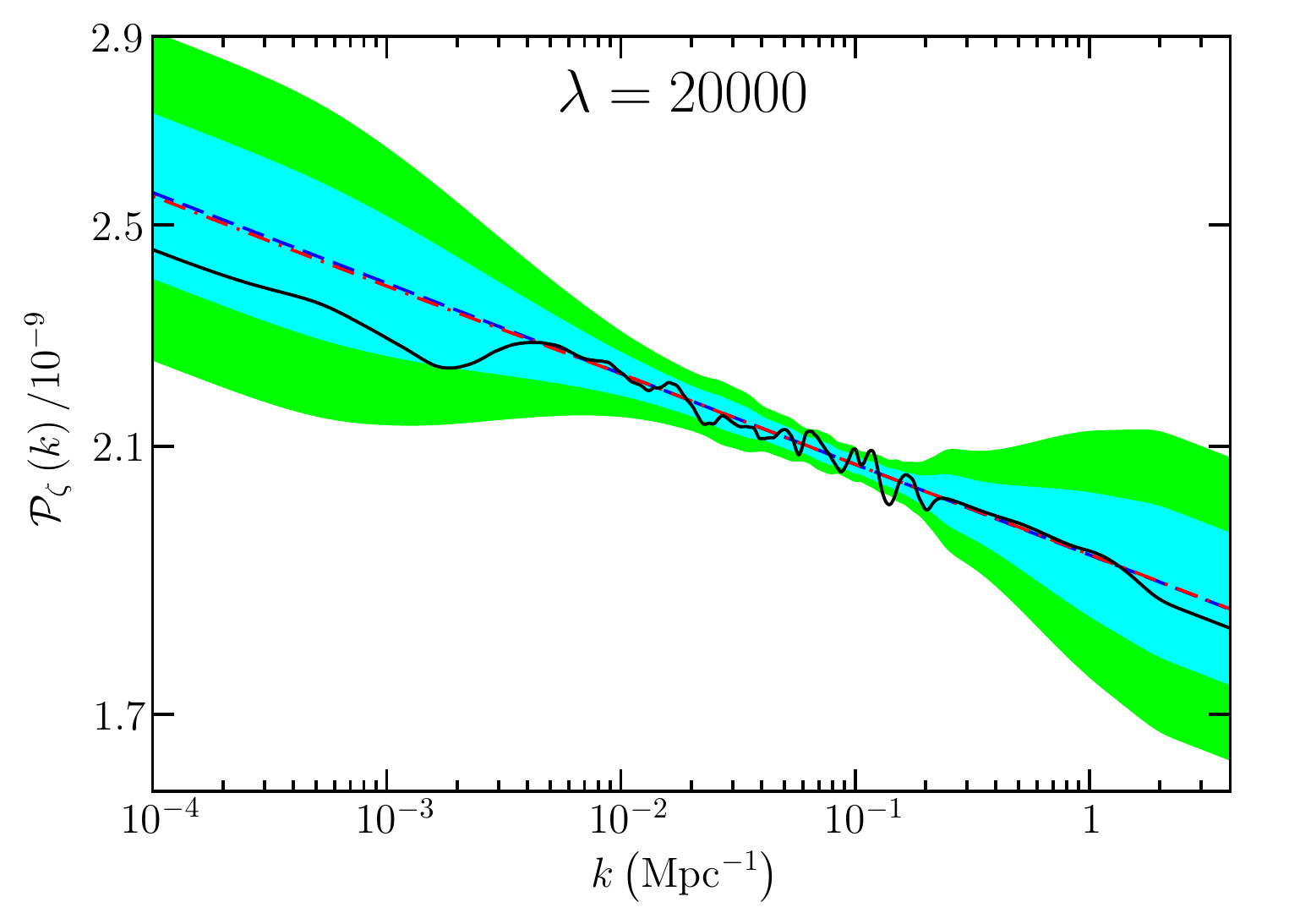}
\caption{Comparison of the PPS (full black line) recovered using data
  combination IV (Planck, WMAP-9 polarisation, ACT, SPT, WiggleZ,
  galaxy clustering, CFTHLenS, Lyman-$\alpha$) with the results of
  $10^6$ simulated reconstructions for $\lambda=400$ and $20000$ which
  were generated using a power-law PPS with $n_\mathrm{s}=0.969$
  (dashed blue line). The shaded bands indicate the $1\sigma$ and
  $2\sigma$ error estimate from Monte Carlo simulations, while the red
  dot-dashed line is the mean of the reconstructions.}
\label{fig:hypoth}
\end{figure}

We need to establish if the features in the PPS reconstructions are
consistent with noise-induced artifacts or if they represent genuine
departures of the true PPS from a power-law. We perform a hypothesis
test with the null hypothesis being that the true spectrum is the
best-fit power-law to data combination IV, which has
$n_\mathrm{s}=0.969$, and invert $10^6$ mock data realisations
generated using the null hypothesis PPS.  The $\ell<50$ CMB data
points were simulated by sampling a Wishart distribution as in
\cite{Hunt:2013bha}, while the other data points were drawn from
Gaussian distributions with the correct covariance matrices.  The
distribution of the results is compared to the reconstruction from the
real data in Fig.\ref{fig:hypoth}, which gives a visual indication of
the statistical significance of the features in the
reconstruction. The mean of the recovered spectra is biased low on
large scales with $\lambda=400$ due to the non-Gaussian CMB likelihood
function for low mulitipoles, as in \cite{Hunt:2013bha}.

To assess the evidence against the null hypothesis at a particular
wavenumber $k$ we use the \emph{local} test statistic
\begin{equation}
\label{t1stat}
\mathrm{T}\left(k\right)\equiv \sum_i \frac{(\hat{\mathrm{p}}_i-\mathrm{p}_i^\mathrm{PL})^2}{\sigma_i^2}\phi_i\left(k\right).
\end{equation}
Here $\sigma_i^2$ is the variance of
$\hat{\mathrm{p}}_i-\mathrm{p}_i^\mathrm{PL}$, the deviation of the
estimated PPS $\hat{\mathrm{p}}_i$ from the null hypothesis power-law
PPS $\mathrm{p}_i^\mathrm{PL}$ at wavenumber bin $i$. For a given
wavenumber we compute the $p$-value of $\mathrm{T}\left(k\right)$ (the
probability under the null hypothesis of exceeding the observed test
statistic value) using the distribution of $\mathrm{T}\left(k\right)$
in the $10^6$ simulated inversions\footnote{On small scales where the
  CMB likelihood is Gaussian, the $\mathrm{T}\left(k\right)$ statistic
  is $\chi^2$ distributed to a high degree of accuracy.}. The
wavenumbers with the lowest $p$-values are recorded in Tables
\ref{tab2} and \ref{tab3}.

\begin{table}[tbh]
\begin{center}
\footnotesize{
\begin{tabular}{c|ccccc}
\hline $k$/$\mathrm{Mpc}^{-1}$ & 0.00177 & 0.0272 & 0.0390 & 0.0573 & 0.101  
\\ \hline $p$-value & 0.0291 & 0.0219 & 0.0364 & 0.00674 & 0.00471 
\\ Stat.sig./$\sigma$ & 2.18 & 2.29 & 2.09 & 2.71 & 2.83 \\ \hline
\end{tabular} 
\vspace{0.25cm}
\vspace{0.25cm}
\begin{tabular}{c|ccccc}
\hline $k$/$\mathrm{Mpc}^{-1}$ & 0.105 & 0.119 & 0.140 & 0.202 & 0.223  
\\ \hline $p$-value & 0.0203 & $5.40\times 10^{-5}$ & 0.0474 & 0.0207 & 0.0304
\\ Stat.sig./$\sigma$ & 2.32 & 4.04 & 1.98 & 2.31 & 2.16 \\ \hline
\end{tabular}} 
\caption{\label{tab2} $\mathrm{T}\left(k\right)$ $p$-values of the
  highest significance features in the PPS recovered from data
  combination IV with $\lambda=400$. The $p$-values are also converted
  to equivalent two-sided Gaussian standard deviations to express the
  statistical significance.}
\end{center}  
\end{table}
\begin{table}[tbh]
\begin{center}
\footnotesize{
\begin{tabular}{c|ccccc}
\hline $k$/$\mathrm{Mpc}^{-1}$ & 0.0639 & 0.0873 & 0.0999 & 0.117 & 0.140  
\\ \hline $p$-value & 0.0902 & 0.140 & 0.100 & 0.0231 & 0.00398
\\ Stat.sig./$\sigma$ & 1.69 & 1.47 & 1.64 & 2.27 & 2.88 \\ \hline
\end{tabular}} 
\caption{\label{tab3} Same as Table \ref{tab2}, but for
  $\lambda=20000$.}
\end{center}   
\end{table}

For $\lambda=400$ the peak at $k\simeq 0.12\;\mathrm{Mpc}^{-1}$
represents a $4\sigma$ excursion, while the
$k\simeq 0.057\;\mathrm{Mpc}^{-1}$ dip and the
$k\simeq 0.10\;\mathrm{Mpc}^{-1}$ peak constitute $2.7\sigma$ and
$2.8\sigma$ deviations respectively. All the other features have less
than $2.4\sigma$ significance, including the dip at
$k\simeq 0.0018\;\mathrm{Mpc}^{-1}$ associated with the
$\ell\simeq 22$ power deficit and the dip at
$k\simeq 0.14\;\mathrm{Mpc}^{-1}$ where the unreliable 217 GHz Planck
spectrum is omitted. The statistical significance of the features in
the $\lambda=20000$ reconstruction is generally lower. The greatest
departure from a power-law is the $k\simeq 0.14\;\mathrm{Mpc}^{-1}$
dip at $2.9\sigma$, up from $2.0\sigma$ for $\lambda=400$.

While the $\mathrm{T}\left(k\right)$ statistic can be used to gauge
the significance of an individual feature, it must be remembered that
over a sufficiently large wavenumber interval the
$\mathrm{T}\left(k\right)$ $p$-value will be small at some $k$ purely
by chance even if the null hypothesis is true. This is an example of
the `look-elsewhere' effect \cite{Gross:2010qma,Vitells:2011da}, or
the problem of multiple comparisons, which is that the likelihood of a
false detection of an anomaly increases with the size of parameter
space searched. To account for this effect we use the \emph{global}
test statistic
\begin{equation}
\mathrm{T}_\mathrm{max}\equiv \max_k\, \mathrm{T}\left(k\right),
\end{equation}
equal to the maximum value of $\mathrm{T}\left(k\right)$ over the
wavenumber range of the recovered PPS. Clearly if there are
significant features \emph{anywhere} in the PPS,
$\mathrm{T}_\mathrm{max}$ will be greater than expected under the null
hypothesis. The $\mathrm{T}_\mathrm{max}$ $p$-value for the
\emph{most} significant feature at $k\simeq 0.12\;\mathrm{Mpc}^{-1}$,
again computed using the simulations, is 0.0239 for $\lambda=400$ and
0.172 for $\lambda=20000$.  This is equivalent to $2.26\sigma$ and
$1.37\sigma$ respectively, hence both reconstructions of the scalar
perturbations are statistically \emph{consistent} with a power-law and
there is no significant evidence presently for features in the PPS.

\section{Conclusions}

The generation of large-scale structure in the universe by growth of
initially small density fluctuations through gravitational instability
is akin to a scattering experiment at a high energy accelerator. The
`beam' here corresponds to the primordial perturbations, the `target'
to the (mainly dark) matter content of the universe, and the
`detector' to the universe as a whole, while the `signal' is the CMB
anisotropy or galaxy correlations. In contrast to the laboratory
situation where the only unknown is the physical interaction between
the beam and the target, in the cosmological context this is known to
be gravity. However all else is unknown or uncertain. We cannot
\emph{simultaneously} infer the properties of the target and the
detector with an unknown beam, hence there are `degeneracies' and
necessarily circularity in e.g. inferring cosmological parameters (the
`detector') or the nature of the dark matter (the `target') or the
spectrum of the density fluctuations (the `beam'). It is common in
particular for the spectrum to be taken to be a power-law and the dark
matter to be cold and collisionless, in determining the parameters of
the assumed $\Lambda$CDM cosmology.

A crucial consistency check is to reverse this procedure and attempt
to infer the PPS, as we have done following our method detailed
earlier \cite{Hunt:2013bha}, using Planck \cite{Ade:2013kta} and other
CMB and large-scale structure data sets. We find several features in
the spectrum, of which one has significance $\simeq 4 \sigma$ for
$\lambda=400$ ($\simeq 2.9 \sigma$ for $\lambda=20000$). This is
potentially of great interest as such features cannot be generated in
the standard slow-roll models of inflation driven by a scalar
field. However the feature is suspect (even though it is in the
supposedly clean 143x217 GHz spectrum) as it is associated with the
same multipole range $1700< \ell<1860$ as the 217x217 GHz
contamination. Hence we cannot claim that it is primordial in
origin. In addition its significance drops to $\sim 2 \sigma$ after we
account for the `look elsewhere' effect, hence there is presently
\emph{no} compelling evidence for a departure of the scalar
fluctuations from a power-law spectrum. 

Nevertheless we believe
that searches for spectral features are still the best direct probe of
inflation, especially given the lack of evidence for any
non-gaussianity in the CMB and the well recognised difficulties in
searches for the B-mode polarisation signal from inflationary
gravitational waves. In contrast to the latter signatures, the TT
signal is orders of magnitude higher, with systematics that can in
principle be better understood. Hence we intend to continue such
searches with further data releases from Planck and other CMB
experiments, as well as data from observational probes of large-scale
structure in the universe, which can be consistently analysed together
in our framework.

\section*{Acknowledgements}

We acknowledge use of the {\tt CAMB} and {\tt cosmoMC} codes and thank
the Planck team for making their data and analysis tools publicly
available. PH is grateful to the Niels Bohr Institute for hospitality
and SS acknowledges a DNRF Niels Bohr Professorship. We thank Pavel
Naselsky for very helpful discussions.

\newpage

\section{Data sets \label{dsets}}
We discuss the data sets used in our analysis; throughout we have
treated the data \emph{exactly} as recommended by the experimental
collaboration which provided it.

\subsection{Planck \label{planckdat}} 

The Planck temperature likelihood function $L_\text{Planck} $ is a
hybrid combination of a Gibbs sampler based Blackwell-Rao estimator
$L_\texttt{Comm}$ implemented in the \texttt{Commander} software code
for $2 \leq \ell \leq 49$, and a Gaussian pseudo-$C_\ell$
approximation $L_\texttt{CamSpec}$ for $50 \leq \ell \leq 2500$
computed by the \texttt{CamSpec} code \cite{Ade:2013kta}. Thus
$L_\text{Planck} = L_\texttt{Comm} + L_\texttt{CamSpec}$. The
\texttt{Commander} likelihood uses a low-resolution,
foreground-cleaned combination of the seven maps from 30 to 353 GHz,
while the \texttt{CamSpec} likelihood uses cross-spectra from the 100,
143 and 217 GHz channels. The multipole range for the $100\times100$
GHz and $143\times143$ GHz spectra is $50\leq\ell\leq1200$ and
$50\leq\ell\leq2000$ respectively, while the $217\times217$ GHz and
$143\times217$ GHz spectra both cover $500\leq\ell\leq2500$. The
\texttt{CamSpec} likelihood is \cite{camspec}
\begin{equation}
L_\mathrm{CamSpec}= \sum_{\ell \ell^\prime} \sum_{I I^\prime} \left(\frac{\mathrm{s}_\ell^\mathrm{TT} + \mathrm{f}_\ell^I}{\sigma_\ell^I} -\mathrm{d}_\ell^I \right)
\left(N_{\ell \ell^\prime}^{-1} \right)^{I I^\prime} \left(\frac{\mathrm{s}_{\ell^\prime}^\mathrm{TT} + \mathrm{f}_{\ell^\prime}^{I^\prime}} {\sigma_{\ell^\prime}^{I^\prime}} 
-\mathrm{d}_{\ell^\prime}^{I^\prime} \right),
\end{equation}
where the index $I$ labels the spectrum, i.e.
$I \in \left\{100
  \times100,\;143\times143,\;217\times217,\;143\times217\right\}$.
Here $\mathrm{s}_\ell^\mathrm{TT}$ is the theoretical temperature
angular power spectrum and $\mathrm{d}_\ell^I$ is the measured $I$th
cross-spectrum. The covariance matrices
$N_{\ell \ell^\prime}^{I I^\prime}$ incorporate the correlations
between the different spectra and are evaluated for a fixed fiducial
model. The $\mathrm{f}_\ell^I$ terms represent the unresolved
`foreground' which can include galactic point sources, clustered
sources in the cosmic infrared background (CIB), and the kinetic and
thermal Sunyaev-Zeldovich effects (kSZ and tSZ) from galaxy clusters,
as discussed in \cite{Ade:2013kta,Ade:2013zuv}. They are given by
\begin{eqnarray}
\mathrm{f}_\ell^{100 \times100} & = & A_{100}^\mathrm{PS} \tilde{\ell}^2 + A^\mathrm{kSZ} \mathrm{t}_\ell^\mathrm{kSZ} + c_1 A^\mathrm{tSZ} \mathrm{t}_\ell^\mathrm{tSZ}, \\
\mathrm{f}_\ell^{143 \times143} & = & A_{143}^\mathrm{PS} \tilde{\ell}^2 + A^\mathrm{kSZ} \mathrm{t}_\ell^\mathrm{kSZ} + c_2 A^\mathrm{tSZ} \mathrm{t}_\ell^\mathrm{tSZ}
 + c_3 A_{143}^\mathrm{CIB} \tilde{\ell}^{\gamma^\mathrm{CIB}} \nonumber \\ 
& & - 2 \left(c_2 c_3 A_{143}^\mathrm{CIB} A^\mathrm{tSZ} \right)^{1/2} \xi^{\mathrm{tSZ}\times\mathrm{CIB}} \mathrm{t}_\ell^{\mathrm{tSZ}\times\mathrm{CIB}}, \\
\mathrm{f}_\ell^{217 \times217} & = & A_{217}^\mathrm{PS} \tilde{\ell}^2 + A^\mathrm{kSZ} \mathrm{t}_\ell^\mathrm{kSZ} + c_4 A_{217}^\mathrm{CIB} \tilde{\ell}^{\gamma^\mathrm{CIB}}, \\
\mathrm{f}_\ell^{143 \times217} & = & r_{143\times217}^\mathrm{PS} \left(A_{143}^\mathrm{PS} A_{217}^\mathrm{PS}\right)^{1/2} \tilde{\ell}^2 + A^\mathrm{kSZ} \mathrm{t}_\ell^\mathrm{kSZ} \nonumber \\
& & + r_{143\times217}^\mathrm{CIB}\left(c_3 c_4 A_{143}^\mathrm{CIB} A_{217}^\mathrm{CIB} \right)^{1/2} \tilde{\ell}^{\gamma^\mathrm{CIB}} \nonumber \\
& & - \left(c_3 c_4 A_{217}^\mathrm{CIB} A^\mathrm{tSZ} \right)^{1/2} \xi^{\mathrm{tSZ}\times\mathrm{CIB}} \mathrm{t}_\ell^{\mathrm{tSZ}\times\mathrm{CIB}},
\end{eqnarray}
where $\tilde{\ell}\equiv \ell/3000$. Here
$\mathrm{t}_\ell^\mathrm{kSZ}$, $\mathrm{t}_\ell^\mathrm{tSZ}$ and
$\mathrm{t}_\ell^{\mathrm{tSZ}\times\mathrm{CIB}}$ are theoretical
`templates' for the kSZ and tSZ components, and for the tSZ and CIB
cross-correlation (which are fixed for the present analysis)
\cite{Ade:2013kta}. The constants $c_1$ to $c_4$ (all of order unity)
correct for the different bandpass responses of the Planck detectors,
while the remaining 11 parameters characterise the amplitudes and
cross-correlations of the various foreground components.

Beam and calibration errors are responsible for the following terms in
the likelihood:
\begin{equation}
\sigma_\ell^I = C_I\left(1 + \beta_1^{100\times100} \sum_i g_i^I E_{i\ell}^I\right)^{-1} .
\end{equation}
Here $C_I$ are the calibration factors for the different spectra, with
$C_{143\times143}=1$ and $C_{143\times217} = C_{217\times217}^{1/2}$,
so that only $C_{100\times100}$ and $C_{217\times217}$ are free
parameters. Uncertainties in the beam transfer functions are
parameterised by the beam error eigenmodes $E_{i\ell}^I$
\cite{Ade:2013dta}.  All of the beam eigenmode amplitudes apart from
$\beta_1^{100\times100}$ (the first of the $100 \times100$ beam
eigenmodes) are marginalised over analytically. This gives rise to the
second factor above where $g_i^I$ are the beam conditional means, with
$g_1^{100\times100}=1$.

\begin{figure}[tbh]
\includegraphics[angle=0,width=0.5\columnwidth,trim = 32mm 171mm 23mm
  15mm, clip]{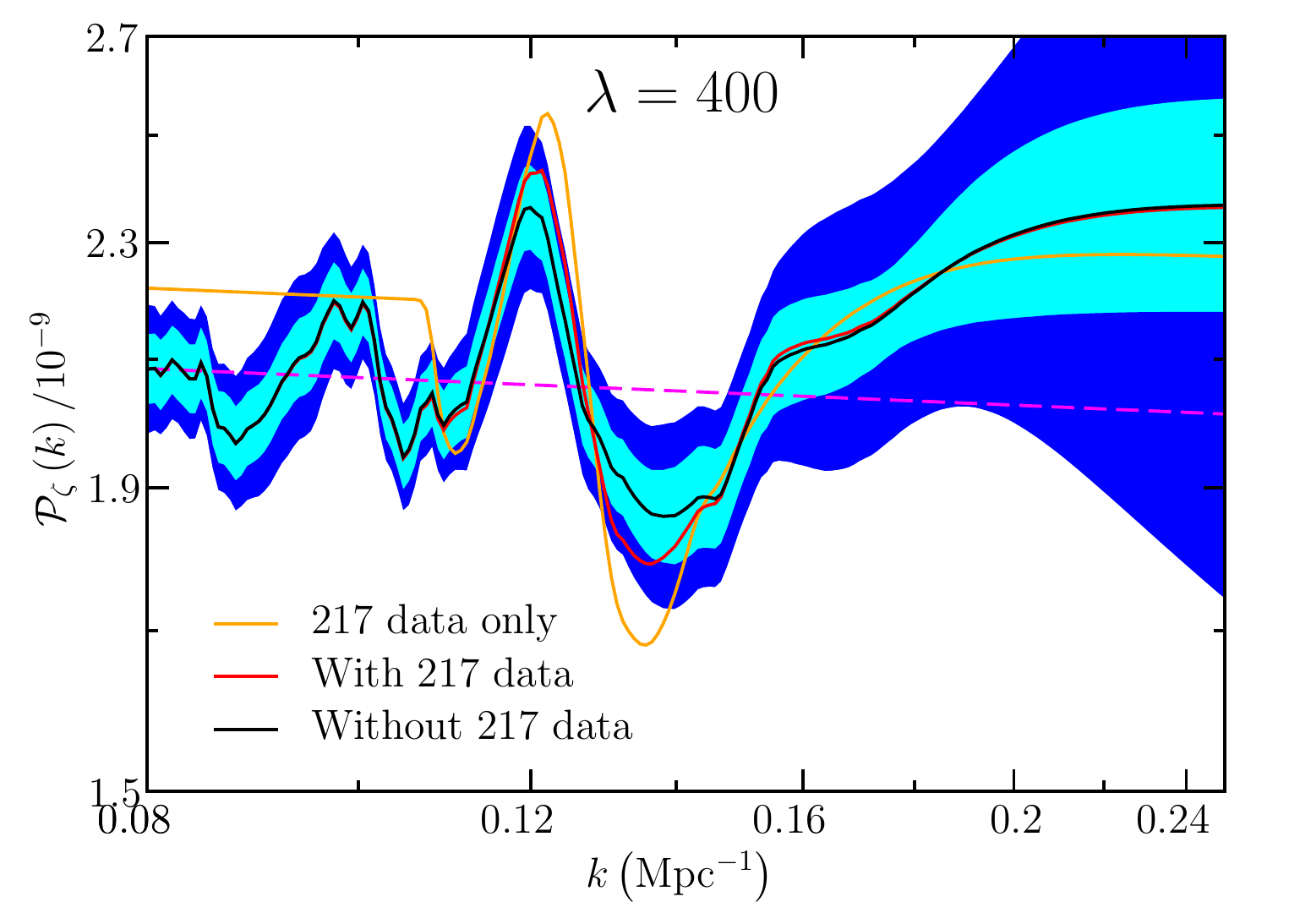}
\includegraphics[angle=0,width=0.5\columnwidth,trim = 32mm 171mm 23mm
  15mm, clip]{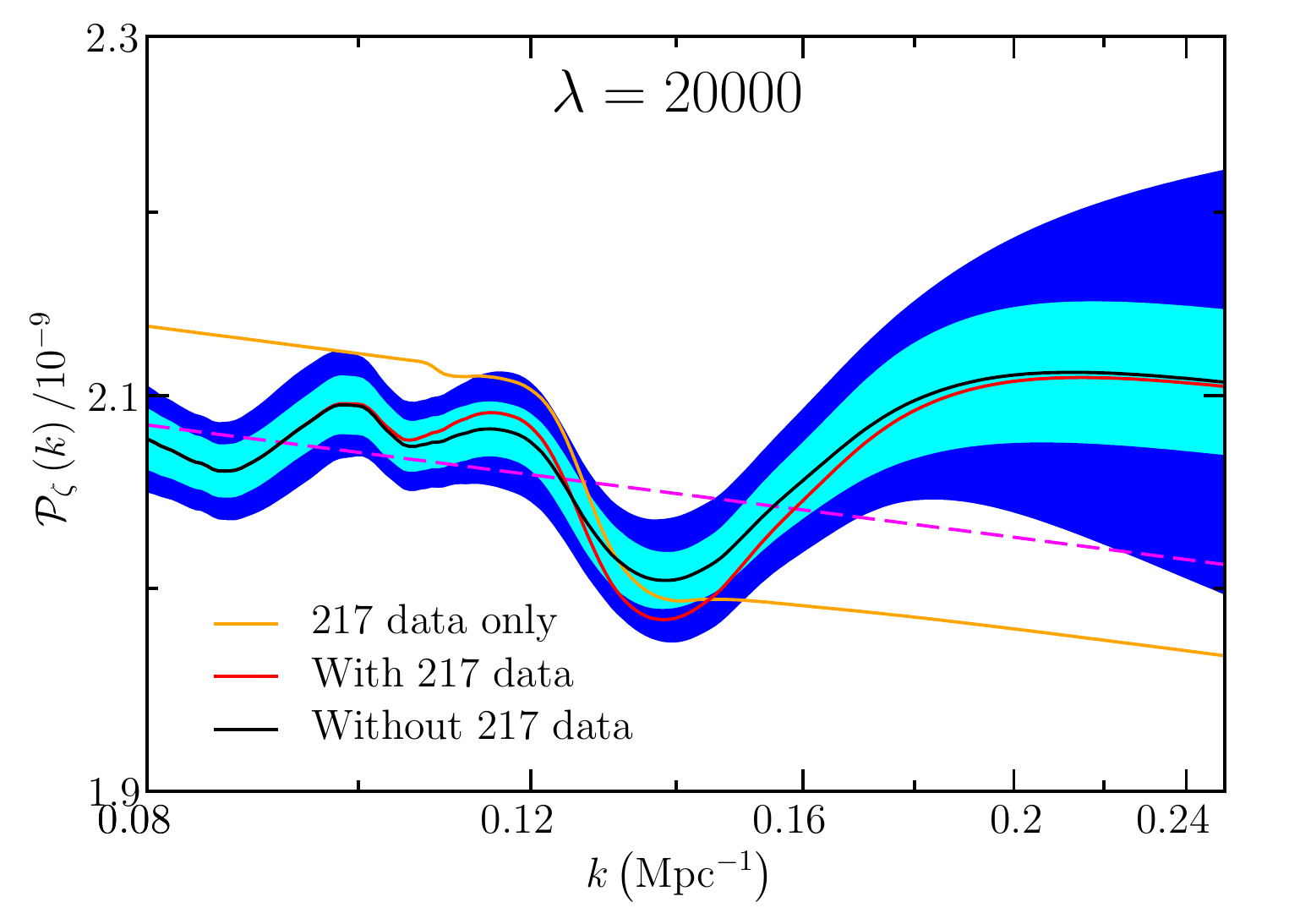}
  \caption{The PPS recovered from the Planck data with (red line) and
    without (black line) the 217 GHz data over the $1700<\ell<1860$
    multipoles, with $\lambda=400$ (left) and $\lambda=20000$
    (right). The orange line is the PPS recovered from the 217 GHz
    $1500<\ell<2060$ multipoles alone, while the dashed magenta line
    is the best-fit power-law with $n_\mathrm{s}=0.969$.  Removing the
    contaminated 217 GHz data does reduce the amplitude of the
    $k=0.14\;\mathrm{Mpc}^{-1}$ feature, but it still remains
    significant.}
\label{fig:exclude}
\end{figure}

Electromagnetic interference between the Planck satellite 4K
Joule-Thomson cryogenic cooler and the HFI bolometers produces
discrete `lines' in the power spectral density of the time-ordered
data, which manifest as features at certain multipoles in the measured
angular power spectrum \cite{Ade:2013kuy}. As first suggested in
\cite{Spergel:2013rxa} and later confirmed by the Planck team
\cite{Ade:2013zuv}, the correction for the 4K cooler lines was
imperfect, leading to an artifical dip at $\ell\simeq 1800$ in the
$217\times217$ GHz cross-spectrum. We therefore exclude the
$217\times217$ GHz data over the interval $1700<\ell<1860$ when
performing our reconstructions. Fig.\ref{fig:exclude} shows that the
effect of removing the data is to reduce the amplitude of the peak at
$k=0.12\;\mathrm{Mpc}^{-1}$ and the dip at $k=0.14\;\mathrm{Mpc}^{-1}$
in the recovered PPS.

Both the derivative $\partial L_\texttt{Comm}/\partial y_i$, which is
evaluated numerically, and the Hessian
$\partial^2 L_\texttt{Comm}/\partial y_i y_j$ are required in order to
obtain $\hat{\yB}$.  Now, for $\ell<50$ we use the expression
\begin{equation}
 \left(N^\mathrm{TT}\right)_{\ell\ell}^{-1}=
\frac{\left(2\ell+1\right) f_\mathrm{sky}^2}{2
\left(\mathrm{s}_\ell^\mathrm{TT}+\mathcal{N}_\ell^\mathrm{TT}\right)^2},
\end{equation}
for the diagonal elements of the inverse TT covariance matrix, where
$f_\mathrm{sky}=0.8$ is the effective fraction of the sky covered by
Planck. To estimate the TT noise spectrum we employ
\begin{equation}
\frac{1}{\mathcal{N}_\ell^\mathrm{TT}}=\sum_\nu \frac{1}{\left(\Delta_\nu^\mathrm{TT}
  \theta_\nu\right)^2}
\exp\left[-\frac{\ell\left(\ell+1\right)\theta_\nu}{8\ln2}\right],
\end{equation}
valid for an idealised CMB experiment with Gaussian beams and
isotropic Gaussian white noise. Here $\nu$ is the frequency of the
band, $\theta_\nu$ is the beam width and $\Delta_\nu^\mathrm{TT}$ is
the temperature noise per pixel. Following \cite{Hamann:2007sb} we
include only the 3 channels (100, 143 and 217 GHz) least affected by
foreground contamination and assume they have been perfectly
cleaned. Their specifications are listed in Table \ref{table}. The
subdominant off-diagonal elements of the covariance matrix are
neglected. The Hessian is then approximated by
$\partial^2 L_\texttt{Comm}/\partial y_i y_j \simeq
2\sum_{\ell\ell^\prime} \mathrm{p}_i W^\mathrm{TT}_{\ell
  i}\left(N^\mathrm{TT}\right)^{-1}_{\ell\ell^\prime}W^\mathrm{TT}_{\ell
  j} \mathrm{p}_j$,
which holds for a Wishart (or Gaussian) likelihood function.

\begin{table}[!h]
\begin{center}
\footnotesize{
\begin{tabular}{ccc}
\hline
 $\nu$/GHz & $\theta_\nu$ & $\Delta_\nu^\mathrm{TT}$/$\mu$K \\
\hline
 100 & $9.6^\prime$ &  8.2 \\
 143 & $7.0^\prime$ &  6.0 \\
 217 & $4.6^\prime$ & 13.1 \\
\hline
\end{tabular}}
\caption{Technical details for the 100, 143 and 217~GHz channels of
  the Planck HFI \cite{Tauber:2010du}.}
\label{table} 
\end{center} 
\end{table}

We do \emph{not} do any further processing of the maps such as
imposing a mask/apodization or `inpainting'. This may well be
necessary to remove possibly spurious features especially at high
$\ell$, but we consider that this is best done by the Planck
Collaboration themselves.

\subsection{ACT}

ACT measures CMB anisotropies at frequencies including 148 and 218 GHz
in two separate regions of the sky --- an equatorial strip (ACTe) and
a southern strip (ACTs) \cite{Das:2013zf,Dunkley:2013vu}. Hence the
total likelihood is
$L_\mathrm{ACT}\equiv L_\mathrm{ACTe} + L_\mathrm{ACTs}$ where
\begin{equation}
L_\mathrm{ACTe/s}= \sum_{b b^\prime} \sum_{I I^\prime} \sum_{\alpha
  \alpha^\prime}\left(\frac{\mathrm{s}_b^I + \mathrm{f}_b^I}{\sigma^I}
-\mathrm{d}_b^{I\alpha} \right) \left(N_{b b^\prime}^{-1} \right)^{I
  I^\prime}_{\alpha \alpha^\prime}
\left(\frac{\mathrm{s}_{b^\prime}^{I^\prime} +
  \mathrm{f}_{b^\prime}^{I^\prime}}{\sigma^{I^\prime}}
-\mathrm{d}_{b^\prime}^{I^\prime\alpha^\prime} \right).
\end{equation}
Here $I \in \left\{148\times148,\;148\times218,\;218\times218\right\}$
labels the three cross-frequency pairs used in the likelihood. ACT
measured bandpowers (labelled $b$) in bands of multiple $\ell$, with
window functions $W_{b\ell}^I$, thus
$\mathrm{s}_b^I = \sum_\ell W_{b\ell}^I \mathrm{s}_\ell^\mathrm{TT}$.
The $148\times148$ GHZ bandpowers lie in the range
$1000\leq \ell \leq 3250$, while the $148\times218$ and $218\times218$
GHz bandpowers cover $1500\leq \ell \leq 3250$. We do not use higher
multipole bandpowers as secondary CMB anisotropies dominate over the
PPS-dependent primary anisotropies on such small scales. ACTe (ACTs)
data was collected over two (three) seasons of observations, hence the
bandpower measurements $\mathrm{d}_b^{I\alpha}$ are labelled both by
pairs of seasons (the index $\alpha$) and pairs of frequencies (the
index $I$). For ACTe there are three different season pairs for the
$148\times148$ and $218\times218$ GHz bandpowers, and four season
pairs for the $148\times218$ GHz bandpowers, leading to a total of
$10$ bandpower sets. For ACTs there are $21$ bandpower sets as there
are six season pairs for the $148\times148$ and $218\times218$ GHz
bandpowers, and nine season pairs for the $148\times218$ GHz
bandpowers.

The ACT foreground model is the same as the one used for Planck,
therefore
\begin{eqnarray}
\mathrm{f}_\ell^{148 \times148} & = & A_{148}^\mathrm{PS,\;ACT} \tilde{\ell}^2 + A^\mathrm{kSZ} \mathrm{t}_\ell^\mathrm{kSZ} + c_5 A^\mathrm{tSZ} \mathrm{t}_\ell^\mathrm{tSZ} 
 + c_6^2 A_{143}^\mathrm{CIB} \tilde{\ell}^{\gamma^\mathrm{CIB}} \nonumber \\
& &  \mbox{} + c_7^2 A_\mathrm{dust}^\mathrm{ACTe} \tilde{\ell}^{-0.7} 
- 2 c_6 \left(c_5 A_{143}^\mathrm{CIB} A^\mathrm{tSZ} \right)^{1/2} \xi^{\mathrm{tSZ}\times\mathrm{CIB}} \mathrm{t}_\ell^{\mathrm{tSZ}\times\mathrm{CIB}}, \\
\mathrm{f}_\ell^{148 \times218} & = & r_\mathrm{150\times220}^\mathrm{PS}\left(A_{148}^\mathrm{PS,\;ACT} A_{218}^\mathrm{PS,\;ACT}\right)^{1/2} \tilde{\ell}^2 + A^\mathrm{kSZ} \mathrm{t}_\ell^\mathrm{kSZ} \nonumber \\ 
& & \mbox{} + c_6 c_8 r_{143\times217}^\mathrm{CIB}\left(A_{143}^\mathrm{CIB} A_{217}^\mathrm{CIB} \right)^{1/2} \tilde{\ell}^{\gamma^\mathrm{CIB}} 
 + c_7 c_9 A_\mathrm{dust}^\mathrm{ACTe} \tilde{\ell}^{-0.7} \nonumber \\
& & \mbox{} - c_8 \left(c_5 A_{217}^\mathrm{CIB} A^\mathrm{tSZ} \right)^{1/2} \xi^{\mathrm{tSZ}\times\mathrm{CIB}} \mathrm{t}_\ell^{\mathrm{tSZ}\times\mathrm{CIB}}, \\
\mathrm{f}_\ell^{218 \times218} & = & A_{218}^\mathrm{PS,\;ACT} \tilde{\ell}^2 + A^\mathrm{kSZ} \mathrm{t}_\ell^\mathrm{kSZ} + c_8^2 A_{217}^\mathrm{CIB} \tilde{\ell}^{\gamma^\mathrm{CIB}}
 + c_9^2 A_\mathrm{dust}^\mathrm{ACTe} \tilde{\ell}^{-0.7}. 
\end{eqnarray}
The extra terms proportional to $\ell^{-0.7}$ represent a residual
`Galactic cirrus' contribution \cite{Ade:2013zuv}. The constants $c_4$
to $c_9$ correspond to scalings of the foreground parameters between
the Planck and ACT effective frequencies. The calibration factors are
$\sigma^{148 \times148} = \left(y_{148}^\mathrm{ACTe/s}\right)^2$,
$\sigma^{148 \times218} = y_{148}^\mathrm{ACTe/s}
y_{217}^\mathrm{ACTe/s} $
and $\sigma^{218 \times218} = \left(y_{217}^\mathrm{ACTe/s}\right)^2$
where $y_{148}^\mathrm{ACTe/s}$ and $y_{218}^\mathrm{ACTe/s}$ are the
map-level calibration parameters for ACTe and ACTs.  Beam errors are
included in the covariance matrices
$N_{b b^\prime \alpha \alpha^\prime}^{I I^\prime}$ \cite{Ade:2013zuv}.

\subsection{SPT}

SPT mapped the CMB anisotropies at 95, 150 and 220 GHz. Following the
Planck team only the data reported in \cite{Reichardt:2011yv} is
used. The SPT likelihood is
\begin{equation}
L_\mathrm{SPT} = \sum_{b b^\prime} \sum_{I I^\prime} \left(\frac{\mathrm{s}_b^I + \mathrm{f}_b^I}{\sigma^I} -\mathrm{d}_b^{I} \right)
\left(N_{b b^\prime}^{-1} \right)^{I I^\prime} \left(\frac{\mathrm{s}_{b^\prime}^{I^\prime} + \mathrm{f}_{b^\prime}^{I^\prime}}{\sigma^{I^\prime}} 
-\mathrm{d}_{b^\prime}^{I^\prime} \right),
\end{equation}
where
$I \in
\left\{95\times95,\;95\times150,\;95\times220,\;150\times150,\;150\times220,\;220\times220\right\}$
labels the cross-spectra. We use the bandpowers
$\mathrm{s}_b^I = \sum_\ell W_{b\ell}^I \mathrm{s}_\ell^\mathrm{TT}$
in the range $2000 \leq \ell \leq 3250$. The foreground bandpowers are
$\mathrm{f}_b^I = \sum_\ell W_{b\ell}^I \mathrm{f}_\ell^I$ where
\begin{eqnarray}
\mathrm{f}_\ell^{95 \times95} & = & A_{95}^\mathrm{PS,\;SPT} \tilde{\ell}^2 + A^\mathrm{kSZ} \mathrm{t}_\ell^\mathrm{kSZ} + c_{10} A^\mathrm{tSZ} \mathrm{t}_\ell^\mathrm{tSZ} 
 + c_{11} \tilde{\ell}^{-1.2},  \\
\mathrm{f}_\ell^{95 \times150} & = & r_{95\times150}^\mathrm{PS} \left( A_{95}^\mathrm{PS,\;SPT} A_{150}^\mathrm{PS,\;SPT} \right)^{1/2}\tilde{\ell}^2 
+ A^\mathrm{kSZ} \mathrm{t}_\ell^\mathrm{kSZ} + c_{12} A^\mathrm{tSZ} \mathrm{t}_\ell^\mathrm{tSZ} \nonumber \\
& & \mbox{} + \left(c_{11} c_{13} \right)^{1/2} \tilde{\ell}^{-1.2} - c_{14} \left(c_{10} A_{143}^\mathrm{CIB} A^\mathrm{tSZ} \right)^{1/2} \xi^{\mathrm{tSZ}\times\mathrm{CIB}} 
\mathrm{t}_\ell^{\mathrm{tSZ}\times\mathrm{CIB}}, \\
\mathrm{f}_\ell^{95 \times220} & = & r_{95\times220}^\mathrm{PS} \left( A_{95}^\mathrm{PS,\;SPT} A_{220}^\mathrm{PS,\;SPT} \right)^{1/2}\tilde{\ell}^2 
+ A^\mathrm{kSZ} \mathrm{t}_\ell^\mathrm{kSZ}  \nonumber \\
& & \mbox{} + \left(c_{11} c_{15} \right)^{1/2} \tilde{\ell}^{-1.2} - c_{16} \left(c_{10} A_{143}^\mathrm{CIB} A^\mathrm{tSZ} \right)^{1/2} \xi^{\mathrm{tSZ}\times\mathrm{CIB}} 
\mathrm{t}_\ell^{\mathrm{tSZ}\times\mathrm{CIB}}, \\
\mathrm{f}_\ell^{150 \times150} & = & A_{150}^\mathrm{PS,\;SPT} \tilde{\ell}^2 + A^\mathrm{kSZ} \mathrm{t}_\ell^\mathrm{kSZ} + c_{17} A^\mathrm{tSZ} \mathrm{t}_\ell^\mathrm{tSZ}
+ c_{14}^2 A_{143}^\mathrm{CIB} \tilde{\ell}^{\gamma^\mathrm{CIB}} \nonumber \\
& & \mbox{} + c_{13} \tilde{\ell}^{-1.2} - 2 c_{14} \left(c_{17} A_{143}^\mathrm{CIB} A^\mathrm{tSZ} \right)^{1/2} \xi^{\mathrm{tSZ}\times\mathrm{CIB}}
\mathrm{t}_\ell^{\mathrm{tSZ}\times\mathrm{CIB}}, \\
\mathrm{f}_\ell^{150 \times220} & = & r_{150\times220}^\mathrm{PS} \left( A_{150}^\mathrm{PS,\;SPT} A_{220}^\mathrm{PS,\;SPT} \right)^{1/2}\tilde{\ell}^2 
+ A^\mathrm{kSZ} \mathrm{t}_\ell^\mathrm{kSZ}  \nonumber \\
& & + c_{14} c_{16} r_{143\times217}^\mathrm{CIB}\left(A_{143}^\mathrm{CIB} A_{217}^\mathrm{CIB} \right)^{1/2} \tilde{\ell}^{\gamma^\mathrm{CIB}}
+ \left(c_{13} c_{15} \right)^{1/2} \tilde{\ell}^{-1.2} \nonumber \\
& & \mbox{}- c_{16} \left(c_{17} A_{217}^\mathrm{CIB} A^\mathrm{tSZ} \right)^{1/2} \xi^{\mathrm{tSZ}\times\mathrm{CIB}}
\mathrm{t}_\ell^{\mathrm{tSZ}\times\mathrm{CIB}}, \\ 
\mathrm{f}_\ell^{220 \times220} & = & A_{220}^\mathrm{PS,\;SPT} \tilde{\ell}^2 + A^\mathrm{kSZ} \mathrm{t}_\ell^\mathrm{kSZ} + + c_{16}^2 A_{217}^\mathrm{CIB} \tilde{\ell}^{\gamma^\mathrm{CIB}} 
 + c_{15} \tilde{\ell}^{-1.2}.  
\end{eqnarray}
The $\ell^{-1.2}$ terms originate from Galactic dust emission
\cite{Ade:2013zuv}. The calibration uncertainties are related to the
map-level calibration parameters $y_{95}^\mathrm{SPT}$,
$y_{150}^\mathrm{SPT}$ and $y_{220}^\mathrm{SPT}$ through
$\sigma^{95 \times95} = \left(y_{95}^\mathrm{SPT}\right)^2$,
$\sigma^{95 \times150} = y_{95}^\mathrm{SPT} y_{150}^\mathrm{SPT}$,
$\sigma^{95 \times220} = y_{95}^\mathrm{SPT} y_{220}^\mathrm{SPT} $,
$\sigma^{150 \times150} = \left(y_{150}^\mathrm{SPT}\right)^2$,
$\sigma^{150 \times220} = y_{150}^\mathrm{SPT} y_{220}^\mathrm{SPT} $
and $\sigma^{220 \times220} = \left(y_{220}^\mathrm{SPT}\right)^2$
\cite{Ade:2013zuv}.

\subsection{WiggleZ}

The WiggleZ galaxy redshift survey measured the galaxy power spectrum
$\mathcal{P}_\mathrm{gal}\left(k\right)$ using the photometric
redshift estimates of $1.7\times 10^5$ galaxies over seven regions of
the sky in four redshift bins centred at
$z = \left\{0.22, 0.41, 0.60, 0.78\right\}$, with a total volume of
$\sim 1\;\mathrm{Gpc}^3$. In what follows the index $I$ labels the
redshift bin and the index $r$ labels the sky region.  To obtain the
theoretical galaxy power spectrum at each redshift we use the `N-body
simulation calibrated without damping' method recommended by the
WiggleZ team \cite{Parkinson:2012vd}, so that
\begin{equation}
\mathcal{P}_\mathrm{gal}^I\left(k\right) \equiv b^2 \frac{\mathcal{P}_\mathrm{hf}^{\mathrm{it},I}\left(k\right)}{\mathcal{P}_\zeta^\mathrm{it}\left(k\right)}
\frac{\mathcal{P}_\mathrm{poly}^{\mathrm{fid},I}\left(k\right)}{\mathcal{P}_\mathrm{hf}^{\mathrm{fid},I}\left(k\right)}
\mathcal{P_\zeta}\left(k\right).
\label{pgal}
\end{equation}
Here $\mathcal{P}_\zeta^\mathrm{it}\left(k\right)$ is a
power-law fit in the wavenumber range
$0.01 < k < 0.3\;\mathrm{Mpc}^{-1}$ to the PPS recovered at each
iteration of the Newton-Raphson minimisation, and
$\mathcal{P}_\mathrm{hf}^{\mathrm{it},I}$ is the \texttt{Halofit}
\cite{Smith:2002dz} fitting formula for the nonlinear matter power
spectrum at redshift $I$ corresponding to
$\mathcal{P}_\zeta^\mathrm{it}\left(k\right)$. Hence on small
scales where nonlinear effects are negligible,
$\mathcal{P}_\mathrm{hf}^{\mathrm{it},I}\left(k\right)/\mathcal{P}_\zeta^\mathrm{it}\left(k\right)$
equals $T^2\left(k\right)$, the square of the linear matter transfer
function. The power-law fit is used because it is unclear how to apply
the \texttt{Halofit} formula to matter power spectra with localised
features. The factor
$\mathcal{P}_\mathrm{poly}^{\mathrm{fid},I}\left(k\right)/\mathcal{P}_\mathrm{hf}^{\mathrm{fid},I}\left(k\right)$
accounts for additional nonlinear and redshift space distortion
effects specific to WiggleZ, as determined from the \texttt{GiggleZ}
N-body simulations. The quantity
$\mathcal{P}_\mathrm{hf}^{\mathrm{fid},I}$ is the \texttt{Halofit}
nonlinear matter power spectrum at redshift $I$ for the
\texttt{GiggleZ} fiducial cosmological model, and
$\mathcal{P}_\mathrm{poly}^{\mathrm{fid},I}$ is a fifth-order
polynomial fit to the \texttt{GiggleZ} power spectrum at redshift $I$.

The galaxy power spectrum $\mathcal{P}_\mathrm{gal}^I\left(k\right)$
is related to $\mathrm{d}_a^{Ir}$, the $a^\mathrm{th}$ power spectrum
measurement in the $r^\mathrm{th}$ region at the $I^\mathrm{th}$
redshift, by a convolution with a window function
$W_a^{Ir}\left(k\right)$ that depends on the WiggleZ survey
geometry. As in our previous paper \cite{Hunt:2013bha} we transform
the window function into
$\mathcal{W}_a^{Ir}\left(k\right)\equiv \left(\gamma^I\right)^{-2}
W_a^{Ir}\left(\gamma^I k\right)$
to account for the fact that the mapping from redshift space to real
space depends on the assumed cosmological model. Here the
`Alcock-Paczynski scaling factor' $\gamma^I$ is
\begin{equation}
\gamma^I \equiv
\left[\frac{\left(D_\mathrm{A}^I\right)^2 H_\mathrm{fid}^I}{\left(D_\mathrm{A,fid}^I\right)^2 H^I}\right]^{1/3},
\label{scale}
\end{equation}
where $D_\mathrm{A}^I$ is the angular diameter distance and $H^I$ is
the Hubble parameter, both at the $I^\mathrm{th}$ redshift, and the
subscript `fid' refers to the quantities for the fiducial model
assumed by the WiggleZ team. The WiggleZ data points are then
\begin{equation}
\mathrm{d}_a^{Ir} = \int^{\infty}_0
\mathcal{W}_a^{Ir}\left(k\right)\mathcal{P}_\mathrm{gal}^I\left(k\right)\,\mathrm{d}k + \mathrm{n}_a^{Ir},
\end{equation}
where $\mathrm{n}_a^{Ir}$ is an additive noise term. Using \ref{basis} 
and \ref{pgal} gives $s_a^{Ir} = \sum_i
W_{ai}^{Ir}\mathrm{p}_i$. The likelihood function has the Gaussian
form
\begin{equation}
L_\mathrm{WiggleZ}= \sum_{a a^\prime} \sum_{Ir} \left(\mathrm{s}_a^{Ir} -\mathrm{d}_a^{Ir} \right)
\left(N_{a a^\prime}^{-1} \right)^{Ir} \left(\mathrm{s}_{a^\prime}^{Ir} -\mathrm{d}_{a^\prime}^{Ir} \right).
\end{equation}
Note that the different redshift bins and the sky regions are uncorrelated.

\subsection{Galaxy clusters \label{clusterdat}}

The variance of the matter density contrast smoothed over a scale $R$
using the top-hat filter
$F\left(x\right)=3\left(\sin x -x \cos x\right)/x^3$ is
\begin{equation}
\sigma^2_R=\frac{1}{2\pi^2}\int F^2\left(kR\right) \mathcal{P}_\mathrm{m}\left(k\right) k^2 \mathrm{d}k.
\label{sig8}
\end{equation}
Observations of galaxy cluster abundance, when fitted to semi-analytic
predictions for the halo mass function, constrain the combination
$\sigma_8 \Omega_\mathrm{m}^q$ where $q\simeq 0.4$
\cite{Allen:2011zs,Weinberg:2012es}.  In \cite{Ade:2013lmv} the Planck
collaboration compiled the results of 5 recent galaxy cluster
experiments and presented them in their table 2 as constraints on the
quantity
$\Sigma_8=\sigma_8 \left(\Omega_\mathrm{m}/0.27\right)^{0.3}$.

The Chandra Cluster Cosmology Project (CCCP) used X-ray observations
of 49 nearby ($z<0.2$) and 37 distant ($0.4<z<0.9$) galaxy clusters to
obtain $\Sigma_8=0.784\pm0.027$ \cite{Vikhlinin:2008ym}. The clusters
were first detected by the ROSAT satellite and then reobserved with
the Chandra satellite. From 10810 clusters of the optically selected
SDSS MaxBCG catalogue, which lie in the range $0.1<z<0.3$, Rozo
\emph{et al.} found $\Sigma_8=0.806\pm0.033$ \cite{Rozo:2009jj}. In
the likeihood analyses of the CCCP and MaxBCG data,
$\omega_\mathrm{b}$ and $n_\mathrm{s}$ were held fixed at values
consistent with the WMAP5 results. In the MaxBCG analysis the hubble
parameter was set to $h=0.7$, while a prior on $h$ derived from Hubble
Space Telescope (HST) observations was applied in the CCCP analysis.

A collection of 15 SZ clusters in the range $0.2<z<1.4$ detected with
the Sunyaev-Zeldovich (SZ) effect by ACT with optical follow-up
observations gave $\Sigma_8=0.848\pm0.032$
\cite{Hasselfield:2013wf}. This measurement neglects the uncertainty
in the SZ scaling relation parameters, which were held fixed at values
taken from a certain gas pressure profile model. The Planck
collaboration found $\Sigma_8=0.764\pm0.025$ using 189 high
signal-to-noise clusters from the Planck SZ catalogue with redshifts
up to $z=1$, when the hydrostatic mass bias was allowed to vary
between zero and $30\%$. If the bias was fixed at the best-fit value
from numerical simulations of $20\%$ then $\Sigma_8=0.78\pm0.01$
\cite{Ade:2013lmv}. A study of 698 clusters at redshift $z<0.5$ from
the REFLEX II X-ray catalogue using X-ray luminosity as a mass proxy
derived $\Sigma_8=0.80\pm0.03$ \cite{Bohringer:2014ooa}. It held most
of the other cosmological parameters fixed at values consistent with
the WMAP9 and Planck CMB results. Using 100 SZ clusters in the range
$0.3<z<1.4$ identified by SPT (of which 63 had optical velocity
dispersion measurements and 16 had X-ray observations from either
Chandra or the XMM-Newton satellite), \cite{Bocquet:2014nta} reported
$\Sigma_8=0.809\pm0.036$.  Both the ACT and SPT
analyses employed priors from Big Bang Nucleosynthesis (BBN) and HST
data, while Planck used BBN and Baryonic Acoustic Oscillations (BAO)
constraints instead.

The scatter in the $\Sigma_8$ measurements is greater than the quoted
errors, which indicates the presence of unknown systematic errors. We
summarise the measurements as $\Sigma_8=0.797\pm0.050$ and use this in
our work. The likelihood function for galaxy clusters is
\begin{equation}
L_\mathrm{GC}= \frac{\left(\mathrm{s} -\mathrm{d}\right)^2}{\sigma^2}.
\end{equation}
Here $d=\left(0.27/\Omega_\mathrm{m}\right)^{0.6}\Sigma_8^2$,
$\sigma$ is the uncertainty in $d$ and $s=\sum_i W_i \mathrm{p}_i$, 
where $W_i$ is derived from Eq.(\ref{sig8}).  

\subsection{CFHTLenS}

The Canada-France-Hawaii Telescope Lensing Survey (CFHTLenS) covers an
area of 154 square degrees in five optical bands.  The two-point
cosmic shear correlation functions $\xi_+\left(\vartheta\right)$ and
$\xi_-\left(\vartheta\right)$ were estimated from the ellipticity and
photometric redshift measurements of 4.2 million galaxies in the
redshift range $0.2<z<1.3$. Two-point shear statistics are related to
the convergence power spectrum $\mathcal{P}_\kappa\left(\ell\right)$,
which is given by a weighted integral of the matter power spectrum
$\mathcal{P}_\mathrm{m}\left(k,z\right)$ along the line of sight
\cite{Refregier:2003ct,Hoekstra:2008db}:
\begin{equation}
\mathcal{P}_\kappa\left(\ell\right)=\frac{9\Omega_\mathrm{m}^2 H_0^4}{4c^4} \int_0^{\chi_\mathrm{H}}
\frac{g^2 \left(\chi\right)}{a^2\left(\chi\right)} \mathcal{P}_\mathrm{m}\left[\frac{\ell}{D_A\left(\chi\right)},z\left(\chi\right) \right]\mathrm{d}\chi.
\end{equation}
Here  
\begin{equation}
\chi\left(z\right)=c \int_0^z \frac{\mathrm{d}z^\prime}{H\left(z^\prime\right)}
\end{equation}
is the radial comoving distance of a source at redshift $z$,
$\chi_\mathrm{H}$ denotes the horizon distance and
$a\left(\chi\right)$ is the scale factor at a distance $\chi$. The
comoving angular diameter distance $D_A\left(\chi\right)$ out to a
distance $\chi$ depends on the curvature of the universe:
\begin{equation}
D_A\left(\chi\right) = \left\{ \begin{array}{lll} 
      c H_0^{-1} \Omega_K^{-1/2} \sinh{\left(\Omega_K^{1/2} c^{-1}H_0 \chi \right) } \;\;\;\;\;\;
      & \mbox{for} \;\; & \Omega_K>0 \\
      \chi & \mbox{for} & \Omega_K=0 \\
      c H_0^{-1} \left|\Omega_K\right|^{-1/2} \sin{\left( \left|\Omega_K\right|^{1/2}c^{-1}H_0 \chi \right)} & \mbox{for} &
      \Omega_K<0 \; .
    \end{array}
  \right.
\end{equation}
The lensing efficiency function $g\left(\chi\right)$ is defined as
\begin{equation}
g\left(\chi\right)=\int_\chi^{\chi_\mathrm{H}} \rho\left[z\left(\chi^\prime\right)\right] \frac{\mathrm{d}z}{\mathrm{d}\chi^\prime} 
\frac{D_A\left(\chi^\prime-\chi\right)}{D_A\left(\chi^\prime\right)} \mathrm{d}\chi^\prime,
\end{equation}
where $\rho\left(z\right)$ is the redshift distribution of the source
galaxies normalised to unity,
\begin{equation}
\int_0^\infty \rho\left(z\right) \mathrm{d}z=1.
\end{equation}
The shear correlation functions are Hankel transforms of the
convergence power spectrum,
\begin{equation}
\xi_{+/-}\left(\vartheta\right) = \frac{1}{2\pi} \int_0^\infty J_{0/4}\left(\ell \vartheta\right)\mathcal{P}_\kappa\left(\ell\right) \ell\, \mathrm{d}\ell,
\end{equation}
where $\xi_+$ and $\xi_-$ correspond to $J_0$ and $J_4$ respectively,
Bessel functions of the first kind of order $0$ and $4$. Using the
substitution $k=\ell/D_A\left(\chi^\prime\right)$ this can be
rewritten as
\begin{equation}
\xi_{+/-}\left(\vartheta\right) =  \int^{\infty}_0
  \mathcal{K}_{+/-}\left(\bm{\theta},\vartheta,k\right)\mathcal{P_\zeta}.
  \left(k\right)\,\mathrm{d}k.
\end{equation}
Here the integral kernels are
\begin{equation}
\mathcal{K}_{+/-}\left(\bm{\theta},\vartheta,k\right)=\frac{9\Omega_\mathrm{m}^2 H_0^4}{8\pi c^4} \int_0^{\chi_\mathrm{H}}
\frac{D_A^2 \left(\chi\right) g^2 \left(\chi\right)}{a^2\left(\chi\right)} J_{0/4}\left[k D_A\left(\chi\right) \vartheta\right]     
\frac{\mathcal{P}_\mathrm{hf}^\mathrm{it}\left[k,z\left(\chi\right)\right]}{\mathcal{P}_\zeta^\mathrm{it}\left(k\right)}
k\,\mathrm{d}\chi,
\end{equation}
where $\mathcal{P}_\zeta^\mathrm{it}\left(k\right)$ is a
power-law fit in the wavenumber range
$0.01 < k < 0.3\;\mathrm{Mpc}^{-1}$ to the PPS recovered at each
iteration of the Newton-Raphson minimisation, and
$\mathcal{P}_\mathrm{hf}^\mathrm{it}\left(k,z\right)$ is the
\texttt{Halofit} nonlinear matter power spectrum corresponding to
$\mathcal{P}_\zeta^\mathrm{it}\left(k\right)$. We exclude
angular scales for which nonlinear evolution alters the shear
correlation functions by more than $20\%$. Thus $\xi_+$ and $\xi_-$
data points are retained for $\vartheta > 12$ arc min and
$\vartheta > 53$ arc min respectively, as $\xi_-$ is more sensitive to
nonlinear effects than $\xi_+$.

Denoting the observed values and the theoretical predictions of
$\xi_{+/-}\left(\vartheta_a\right)$ by $d_a^\mu$ and $s_a^\mu$
respectively, the likelihood function is
\begin{equation}
L_\mathrm{CFHTLenS}= \sum_{a a^\prime} \sum_{\mu \nu} \left(\mathrm{s}_a^\mu -\mathrm{d}_a^\mu \right)
\left(N_{a a^\prime}^{-1} \right)^{\mu\nu} \left(\mathrm{s}_{a^\prime}^\nu -\mathrm{d}_{a^\prime}^\nu \right).
\end{equation}
Here $a$ and $a^\prime$ label the angular scale while $\mu$ and $\nu$
stand for the `$+$' and `$-$' components.  The covariance matrix
$N_{aa^\prime}^{\mu\nu}$ was calculated for a fiducial model.

\subsection{Lyman-$\alpha$ data}

VHS \cite{Viel:2004bf} used Lyman-$\alpha$ forest observations by
Croft \cite{Croft:2000hs} and LUQAS (Large Sample of UVES QSO
Absorption Spectra) \cite{Kim:2003qt} to estimate the linear matter
power spectrum on scales
$0.3 \;h/\mathrm{Mpc}\lesssim k \lesssim 3 \;h/\mathrm{Mpc}$. The
LUQAS sample consists of 27 high-resolution quasar spectra taken by
the UVES spectrograph on the Very Large Telescope. The Croft sample
comprises 30 high-resolution and 23 low-resolution spectra obtained
using the HIRES and LRIS instruments of the Keck observatory. The mean
absorption redshift of the Croft and LUQAS samples is
$z_\mathrm{Croft}\simeq2.72$ and $z_\mathrm{LUQAS}\simeq2.25$.  VHS
employed the so-called `effective bias' method \cite{Gnedin:2001wg}
calibrated by a suite of hydrodynamical simulations to infer the
matter power spectrum from the transmitted flux power spectrum of the
two datasets. The VHS results were subsequently incorporated into the
CosmoMC module \texttt{lya.f90} \cite{Lesgourgues:2007te}.

The Lyman-$\alpha$ likelihood is $L_{\mathrm{Ly}\alpha} \equiv L_\mathrm{Croft} +
L_\mathrm{LUQAS}$ where
\begin{equation}
L_\mathrm{Croft/LUQAS}= \sum_{a} \frac{\left(\mathrm{s}_a /Q_\Omega^\mathrm{Croft/LUQAS} - A \,\mathrm{d}_a \right)^2}{\sigma_a^2}.
\end{equation}
Here $\mathrm{s}_a$ and $\mathrm{d}_a$ are the theoretical and
measured matter power spectrum data points respectively, $\sigma_a^2$
is the variance of the uncorrelated measurement errors and
$A=1\pm 0.29$ is the overall calibration error of the effective bias
method.  The latter originates mainly from uncertainties in the
numerical simulations, the effective optical depth, the mean
temperature of the intergalactic medium and the slope of the
temperature-density relation. The factor
\begin{equation}
Q_\Omega^\mathrm{Croft/LUQAS}=\left[\frac{2.4}{1+1.4 \,\Omega_\mathrm{m}^{0.6}\left(z_\mathrm{Croft/LUQAS}\right)}\right]^2
\end{equation}
accounts for the dependence of the inferred matter power spectrum on
the matter density $\Omega_\mathrm{m}$ at the redshift of the
Lyman-$\alpha$ data.

\subsection{WMAP9 polarisation}

The pixel-based WMAP9 polarisation likelihood $L_\mathrm{WP}$ covers
$\ell\leq 23$ and uses the WMAP9 polarisation maps at 33, 41 and 61
GHz.  The Planck team updated the temperature map used in constructing
the likelihood to the Planck Commander map. Our handling of this data
set is the same as in our previous paper \cite{Hunt:2013bha}.

\section{Error analysis \label{error}}

When generalised to include perturbative nonlinear effects Eq.(\ref{int1}) becomes
\begin{eqnarray}
  \mathrm{d}_a^{(\mathbb{Z})} & = &  \mathrm{c}_a^{(\mathbb{Z})}\left(\bm{\theta}\right)
+ \int^{\infty}_0
  \mathcal{K}_a^{(\mathbb{Z})}\left(\bm{\theta},k\right)\mathcal{P_\zeta}
  \left(k\right)\,\mathrm{d}k \nonumber \\
& & + \int^{\infty}_0 \int^{\infty}_0
  \mathcal{K}_a^{(\mathbb{Z})}\left(\bm{\theta},k_1, k_2 \right)\mathcal{P_\zeta}
  \left(k_1\right) \mathcal{P_\zeta}\left(k_2\right)\,\mathrm{d}k_1 \, \mathrm{d}k_2 + \mathrm{n}_a^{(\mathbb{Z})},
\label{int2}
\end{eqnarray}
which is valid in the mildly nonlinear regime. Using the expansion
Eq.(\ref{basis}) gives
\begin{equation}
\mathrm{d}_a^{(\mathbb{Z})}=\mathrm{c}_a^{(\mathbb{Z})}\left(\bm{\theta}\right)
+\sum_i W_{ai}^{(\mathbb{Z})}\left(\bm{\theta}\right)\,\mathrm{p}_i
+\frac{1}{2} \sum_{ij} S_{aij}^{(\mathbb{Z})}\left(\bm{\theta}\right)\,\mathrm{p}_i\,\mathrm{p}_j+\mathrm{n}_a^{(\mathbb{Z})},
\end{equation}
where
\begin{equation}
\label{relerror} 
 S_{aij}^{(\mathbb{Z})}\left(\bm{\theta}\right)= \int^{k_{i+1}}_{k_i} \int^{k_{j+1}}_{k_j}
 \mathcal{K}_a^{(\mathbb{Z})}\left(\bm{\theta},k_1,k_2\right)
\, \mathrm{d}k_1\,\mathrm{d}k_2.
\end{equation}  
We emphasise that the additional nonlinear terms are \emph{not} used
elsewhere in this paper but are included here for completeness.

Any deconvolution method for recovering the PPS defines a transfer
function $\mathcal{T}$ which gives the relationship of the estimate
$\hat{\yB}$ to the true PPS $\yB_\mathrm{tru}$. It depends on the true
background parameters $\bm{\theta}_\mathrm{tru}$, the estimated
background parameters $\hat{\bm{\theta}}$ and the noise in the data
$\nB$ so that
\begin{equation}
 \hat{\yB}\left(\dB,\hat{\bm{\theta}}\right) =
 \bm{\mathcal{T}}\left(\yB_\mathrm{tru},\bm{\theta}_\mathrm{tru},
 \hat{\bm{\theta}},\nB\right).
\label{transfer}
\end{equation}
Performing a Taylor expansion of $\mathcal{T}$ about a fiducial PPS
$\yB_\mathrm{fid}$ close to $\yB_\mathrm{tru}$ yields
\begin{eqnarray}
  \hat{\mathrm{y}}_i\left(\dB,\hat{\bm{\theta}}\right) & = &
  \mathcal{T}_i\left(\yB_\mathrm{fid},\bm{\theta}_\mathrm{tru},\bm{\theta}_\mathrm{tru},
  \mathbf{0}\right) + \sum_{j}R_{ij}\,\Delta
  \mathrm{y}_j+\frac{1}{2}\sum_{j,k}Y_{ijk}\,\Delta
  \mathrm{y}_j\,\Delta \mathrm{y}_k \nonumber\\ & &
  +\sum_{\mathbb{Z},a}
  M_{ia}^{(\mathbb{Z})}\,\mathrm{n}_a^{(\mathbb{Z})}+\sum_\alpha      
  M_{i\alpha}\,\mathrm{u}_\alpha+\sum_{\mathbb{Z},j,a}Z_{ija}^{(\mathbb{Z})}\,\Delta
  \mathrm{y}_j
  \,\mathrm{n}_a^{(\mathbb{Z})}+\sum_{j,\alpha}Z_{ij\alpha}\,\Delta
  \mathrm{y}_j\,\mathrm{u}_\alpha \nonumber\\ & &
  +\frac{1}{2}\sum_{\mathbb{Z},\mathbb{Z'},a,b}X_{iab}^{(\mathbb{Z}\mathbb{Z'})}\,
  \mathrm{n}_a^{(\mathbb{Z})}\,\mathrm{n}_b^{(\mathbb{Z'})} +
  \sum_{\mathbb{Z},a,\alpha}X_{ia\alpha}^{(\mathbb{Z})}\,\mathrm{n}_a^{(\mathbb{Z})}\,
  \mathrm{u}_\alpha +
  \frac{1}{2}\sum_{\alpha,\beta}X_{i\alpha\beta}\,\mathrm{u}_\alpha\,\mathrm{u}_\beta
  +\ldots
\label{recerr}
\end{eqnarray}
Here $\Delta \mathrm{y}_i\equiv \mathrm{y}_{\mathrm{tru}|i}-\mathrm{y}_{\mathrm{fid}|i}$, 
\begin{eqnarray}
  M_{ia}^{(\mathbb{Z})}\equiv\left.\frac{\partial\hat{\mathrm{y}}_i}{\partial
    \mathrm{d}_{a}^{(\mathbb{Z})}}\right|_{\hat{\dB}_\mathrm{fid},\bms{\theta}_\mathrm{tru}},
  \qquad
  M_{i\alpha}\equiv\left.\frac{\partial\hat{\mathrm{y}}_i}{\partial
    \theta_\alpha}\right|_{\hat{\dB}_\mathrm{fid},\bms{\theta}_\mathrm{tru}},
  \qquad\qquad\qquad\qquad \nonumber
  \\ X_{iab}^{(\mathbb{Z}\mathbb{Z'})} \equiv
  \left.\frac{\partial^{2}\hat{\mathrm{y}}_i}{\partial
    \mathrm{d}_{a}^{(\mathbb{Z})}\partial
    \mathrm{d}_{b}^{(\mathbb{Z'})}}\right|_{\hat{\dB}_\mathrm{fid},\bms{\theta}_\mathrm{tru}},
  \qquad X_{ia\alpha}^{(\mathbb{Z})} \equiv
  \left.\frac{\partial^{2}\hat{\mathrm{y}}_i}{\partial
    \mathrm{d}_{a}^{(\mathbb{Z})}\partial
    \theta_\alpha}\right|_{\hat{\dB}_\mathrm{fid},\bms{\theta}_\mathrm{tru}},  
  \qquad X_{i\alpha\beta} \equiv
  \left.\frac{\partial^{2}\hat{\mathrm{y}}_i}{\partial
    \theta_\alpha\partial
    \theta_\beta}\right|_{\hat{\dB}_\mathrm{fid},\bms{\theta}_\mathrm{tru}},\nonumber 
\end{eqnarray}
where $\hat{\dB}_\mathrm{fid}$ denotes collectively the datasets
estimated from $\yB_\mathrm{fid}$, i.e.\
\begin{equation}
\hat{\mathrm{d}}_{\mathrm{fid}|a}^{(\mathbb{Z})}=\mathrm{c}_a^{(\mathbb{Z})}\left(\bm{\theta}_\mathrm{tru}\right)
+\sum_i W_{ai}^{(\mathbb{Z})}\left(\bm{\theta}_\mathrm{tru}\right)\,\mathrm{p}_{\mathrm{fid}|i}
+\frac{1}{2} \sum_{ij} S_{aij}^{(\mathbb{Z})}\left(\bm{\theta}_\mathrm{tru}\right)\,\mathrm{p}_{\mathrm{fid}|i}\,\mathrm{p}_{\mathrm{fid}|j},
\end{equation}
and
\begin{eqnarray}
R_{ij} & \equiv & \sum_{\mathbb{Z},a}M_{ia}^{(\mathbb{Z})}W_{aj}^{(\mathbb{Z})}
  \left(\bm{\theta}_\mathrm{tru}\right) \mathrm{p}_j + \sum_{\mathbb{Z},a,k} M_{ia}^{(\mathbb{Z})} S_{ajk}^{(\mathbb{Z})}
  \left(\bm{\theta}_\mathrm{tru}\right) \mathrm{p}_j \, \mathrm{p}_k, \\
  Y_{ijk} & \equiv &\sum_{\mathbb{Z},\mathbb{Z'},a,b}
  X_{iab}^{(\mathbb{Z}\mathbb{Z'})}W_{aj}^{(\mathbb{Z})}
  \left(\bm{\theta}_\mathrm{tru}\right) W_{bk}^{(\mathbb{Z'})}\left(\bm{\theta}_\mathrm{tru}\right) \mathrm{p}_j \, \mathrm{p}_k,
+ \sum_{\mathbb{Z},a} M_{ia}^{(\mathbb{Z})} S_{ajk}^{(\mathbb{Z})}
  \left(\bm{\theta}_\mathrm{tru}\right) \mathrm{p}_j \, \mathrm{p}_k, \nonumber \\
& & + \sum_{\mathbb{Z},\mathbb{Z'},a,b,l}
  X_{iab}^{(\mathbb{Z}\mathbb{Z'})}S_{ajl}^{(\mathbb{Z})}
  \left(\bm{\theta}_\mathrm{tru}\right) W_{bk}^{(\mathbb{Z'})}\left(\bm{\theta}_\mathrm{tru}\right) \mathrm{p}_j \, \mathrm{p}_k \, \mathrm{p}_l
+\delta_{jk} \sum_{\mathbb{Z},a} M_{ia}^{(\mathbb{Z})} W_{aj}^{(\mathbb{Z})}
  \left(\bm{\theta}_\mathrm{tru}\right) \mathrm{p}_k, \nonumber \\
& &  + \sum_{\mathbb{Z},\mathbb{Z'},a,b,l}
  X_{iab}^{(\mathbb{Z}\mathbb{Z'})}W_{aj}^{(\mathbb{Z})}
  \left(\bm{\theta}_\mathrm{tru}\right) S_{bkl}^{(\mathbb{Z'})}\left(\bm{\theta}_\mathrm{tru}\right) \mathrm{p}_j \, \mathrm{p}_k \, \mathrm{p}_l
+\delta_{jk} \sum_{\mathbb{Z},a,l} M_{ia}^{(\mathbb{Z})} S_{ajl}^{(\mathbb{Z})}
  \left(\bm{\theta}_\mathrm{tru}\right) \mathrm{p}_k\, \mathrm{p}_l, \nonumber \\
& & +\sum_{\mathbb{Z},\mathbb{Z'},a,b,l,m}
  X_{iab}^{(\mathbb{Z}\mathbb{Z'})}S_{ajl}^{(\mathbb{Z})}
  \left(\bm{\theta}_\mathrm{tru}\right) S_{bkm}^{(\mathbb{Z'})}\left(\bm{\theta}_\mathrm{tru}\right) \mathrm{p}_j \, \mathrm{p}_k \, \mathrm{p}_j \, \mathrm{p}_m, 
\end{eqnarray}
\begin{eqnarray}
 Z_{ija}^{(\mathbb{Z})}& \equiv &
  \sum_{\mathbb{Z'},b}X_{iab}^{(\mathbb{Z}\mathbb{Z'})}W_{bj}^{(\mathbb{Z'})}
  \left(\bm{\theta}_\mathrm{tru}\right)\mathrm{p}_j
+\sum_{\mathbb{Z'},b,k}X_{iab}^{(\mathbb{Z}\mathbb{Z'})}S_{bjk}^{(\mathbb{Z'})}
  \left(\bm{\theta}_\mathrm{tru}\right)\mathrm{p}_j \, \mathrm{p}_k, \\
  Z_{ij\alpha}& \equiv & \sum_{\mathbb{Z},a}X_{ia\alpha}^{(\mathbb{Z})}W_{aj}^{(\mathbb{Z})}
  \left(\bm{\theta}_\mathrm{tru}\right) \mathrm{p}_j
+\sum_{\mathbb{Z},a,k}X_{ia\alpha}^{(\mathbb{Z})}S_{ajk}^{(\mathbb{Z})}
  \left(\bm{\theta}_\mathrm{tru}\right) \mathrm{p}_j \, \mathrm{p}_k.  
\label{matrices}
\end{eqnarray}  
These matrices characterise the inversion and are discussed in more
detail in \cite{Hunt:2013bha}.  In particular the sensitivity matrices 
$M_{ia}^{(\mathbb{Z})}$ give the dependence of the estimated PPS 
on the data points $d_a^{(\mathbb{Z})}$. They control the manner in which 
noise in the data produces artifacts in the recovered PPS.
The first-order resolution matrix $\mathsf{R}$ describes the linear mapping from
$\yB_\mathrm{tru}$ to $\hat{\yB}$. The closer $\mathsf{R}$ is to the
identity matrix $\mathsf{I}$, the better the resolution and the lower
the bias of the inversion method. For Tikhonov regularisation
$\sum_j R_{ij}\simeq 1$ for all $i$ so that the estimated PPS is
correctly scaled. The second-order resolution matrix $\mathsf{Y}$
details a quadratic mapping of $\yB_\mathrm{tru}$ to $\hat{\yB}$ and
should vanish in order to minimise the bias. For Tikhonov
regularisation analytic expressions for these inversion matrices can
be derived as in \cite{Hunt:2013bha}:
\begin{equation}
M_{ia}^{(\mathbb{Z})} = -\sum_{j} A_{ij}^{-1} B_{ja}^{(\mathbb{Z})},\qquad M_{i\alpha} =
-\sum_{j} A_{ij}^{-1} B_{j\alpha},
\end{equation}
\begin{equation}
X_{iab}^{(\mathbb{Z}\mathbb{Z'})} = -\sum_{j,k,l} A_{ij}^{-1} C_{jkl} M_{ka}^{(\mathbb{Z})}
M_{lb}^{(\mathbb{Z'})}- \sum_{j,k} A_{ij}^{-1} E_{jka}^{(\mathbb{Z})}
M_{kb}^{(\mathbb{Z'})}-\sum_{j,k} A_{ij}^{-1} E_{jkb}^{(\mathbb{Z'})}
M_{ka}^{(\mathbb{Z})}-\sum_{j} A_{ij}^{-1} D_{jab}^{(\mathbb{Z}\mathbb{Z'})},
\end{equation}
\begin{equation}
X_{ia\alpha}^{(\mathbb{Z})} = -\sum_{j,k,l} A_{ij}^{-1} C_{jkl} M_{ka}^{(\mathbb{Z})}
M_{l\alpha}- \sum_{j,k} A_{ij}^{-1} E_{jka}^{(\mathbb{Z})}
M_{k\alpha}-\sum_{j,k} A_{ij}^{-1} E_{jk\alpha} M_{ka}^{(\mathbb{Z})}-\sum_{j}
A_{ij}^{-1} D_{ja\alpha}^{(\mathbb{Z})}.
\end{equation}
\begin{equation}
X_{i\alpha\beta} = -\sum_{j,k,l} A_{ij}^{-1} C_{jkl} M_{k\alpha}
M_{l\beta}- \sum_{j,k} A_{ij}^{-1} E_{jk\alpha} M_{k\beta}-\sum_{j,k}
A_{ij}^{-1} E_{jk\beta} M_{k\alpha}-\sum_{j} A_{ij}^{-1}
D_{j\alpha\beta},
\end{equation}
\begin{equation}
 \begin{array}{ccc}
   A_{ij} \equiv \left.\frac{\displaystyle\partial^{2}Q}
   {\displaystyle\partial \mathrm{y}_i\partial
     \mathrm{y}_j}\right|_{\hat{\yB}_\mathrm{fid},\hat{\dB}_\mathrm{fid},\bms{\theta}_\mathrm{tru}}, &
   B_{ia}^{(\mathbb{Z})}\equiv\left.\frac{\displaystyle\partial^{2}Q} 
   {\displaystyle\partial \mathrm{y}_i\partial
  \mathrm{d}_a^{(\mathbb{Z})}}\right|_{\hat{\yB}_\mathrm{fid},\hat{\dB_\mathrm{fid}},\bms{\theta}_\mathrm{tru}}, &
   B_{i\alpha} \equiv \left.\frac{\displaystyle\partial^{2}Q}
   {\displaystyle\partial \mathrm{y}_i\partial
     \theta_\alpha}\right|_{\hat{\yB}_\mathrm{fid},\hat{\dB}_\mathrm{fid},\bms{\theta}_\mathrm{tru}}
   \\ C_{ijk}\equiv\left.\frac{\displaystyle\partial^{3}Q}
      {\displaystyle\partial \mathrm{y}_{i}\partial \mathrm{y}_j\partial
        \mathrm{y}_k}\right|_{\hat{\yB}_\mathrm{fid},\hat{\dB}_\mathrm{fid},\bms{\theta}_\mathrm{tru}}, &  
      D_{iab}^{(\mathbb{Z}\mathbb{Z'})} \equiv \left.\frac{\displaystyle\partial^{3}Q} 
      {\displaystyle\partial \mathrm{y}_{i}\partial \mathrm{d}_{a}^{(\mathbb{Z})}\partial
        \mathrm{d}_{b}^{(\mathbb{Z'})}}\right|_{\hat{\yB}_\mathrm{fid},\hat{\dB}_\mathrm{fid},\bms{\theta}_\mathrm{tru}},
      & D_{ia\alpha}^{(\mathbb{Z})} \equiv
      \left.\frac{\displaystyle\partial^{3}Q}{\displaystyle\partial
        \mathrm{y}_{i}\partial \mathrm{d}_{a}^{(\mathbb{Z})}\partial \theta_\alpha}
      \right|_{\hat{\pB}_\mathrm{fid},\hat{\dB}_\mathrm{fid},\bms{\theta}_\mathrm{tru}}
      \\ D_{i\alpha\beta}\equiv\left.\frac{\displaystyle\partial^{3}Q}
         {\displaystyle\partial \mathrm{y}_{i}\partial \theta_\alpha\partial
           \theta_\beta}\right|_{\hat{\yB}_\mathrm{fid},\hat{\dB}_\mathrm{fid},\bms{\theta}_\mathrm{tru}},
         & E_{ija}^{(\mathbb{Z})} \equiv
         \left.\frac{\displaystyle\partial^{3}Q}{\displaystyle\partial
           \mathrm{y}_{i}\partial \mathrm{y}_{j}\partial \mathrm{d}_{a}^{(\mathbb{Z})}}
         \right|_{\hat{\yB}_\mathrm{fid},\hat{\dB}_\mathrm{fid},\bms{\theta}_\mathrm{tru}}, &
         E_{ij\alpha} \equiv
         \left.\frac{\displaystyle\partial^{3}Q}{\displaystyle\partial
           \mathrm{y}_{i}\partial \mathrm{y}_{j}\partial \theta_\alpha}
         \right|_{\hat{\yB}_\mathrm{fid},\hat{\dB}_\mathrm{fid},\bms{\theta}_\mathrm{tru}}. \\
 \end{array}
\end{equation}
Here the derivatives are evaluated at $\hat{\yB}_\mathrm{fid}\equiv\hat{\yB}\left(\hat{\dB}_\mathrm{fid},\bm{\theta}_\mathrm{tru}\right)$.
 
\begin{figure}[t!]
\includegraphics[width=0.5\columnwidth,trim = 32mm 171mm 23mm 15mm, clip]{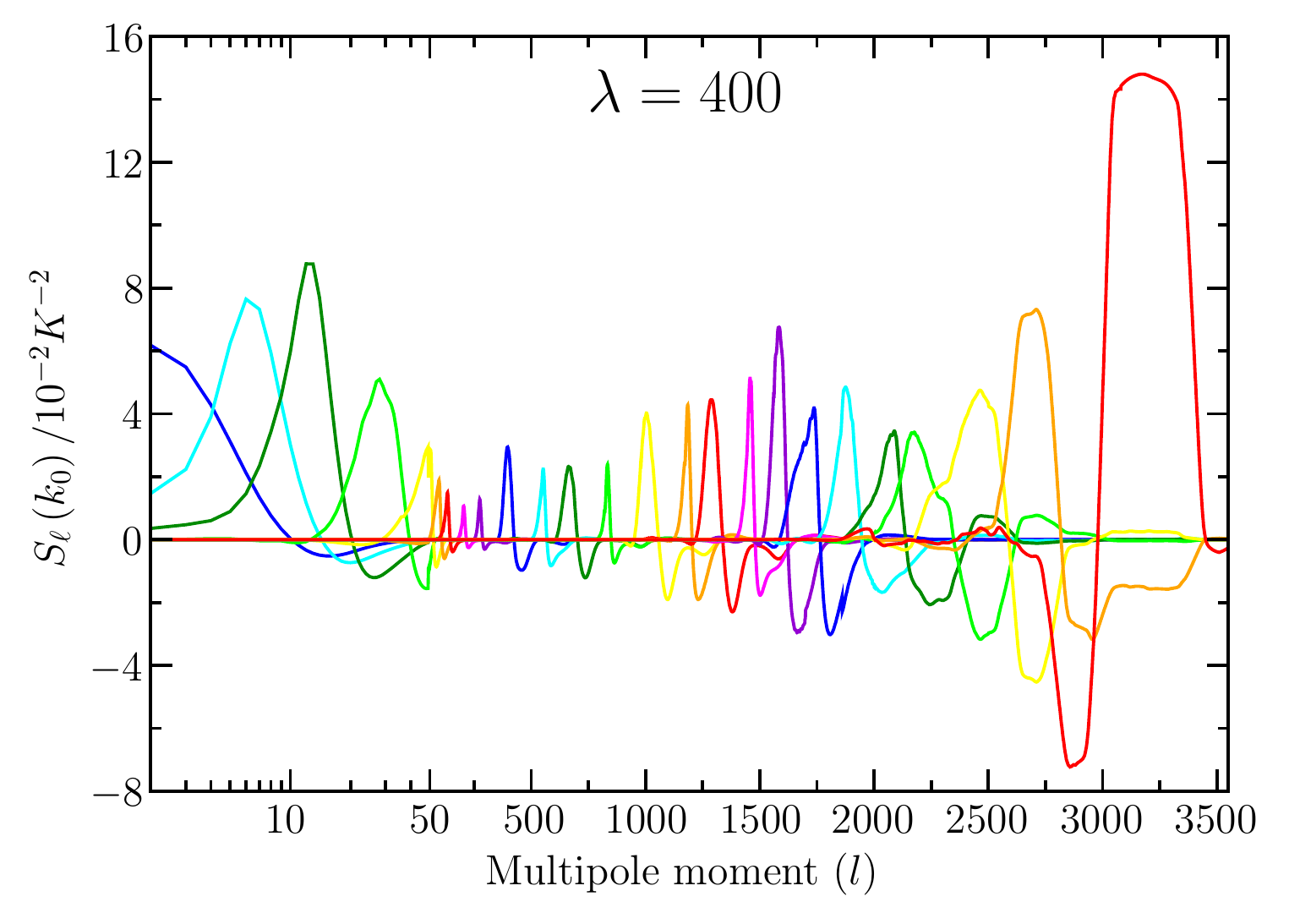}
\includegraphics[width=0.5\columnwidth,trim = 32mm 171mm 23mm 15mm, clip]{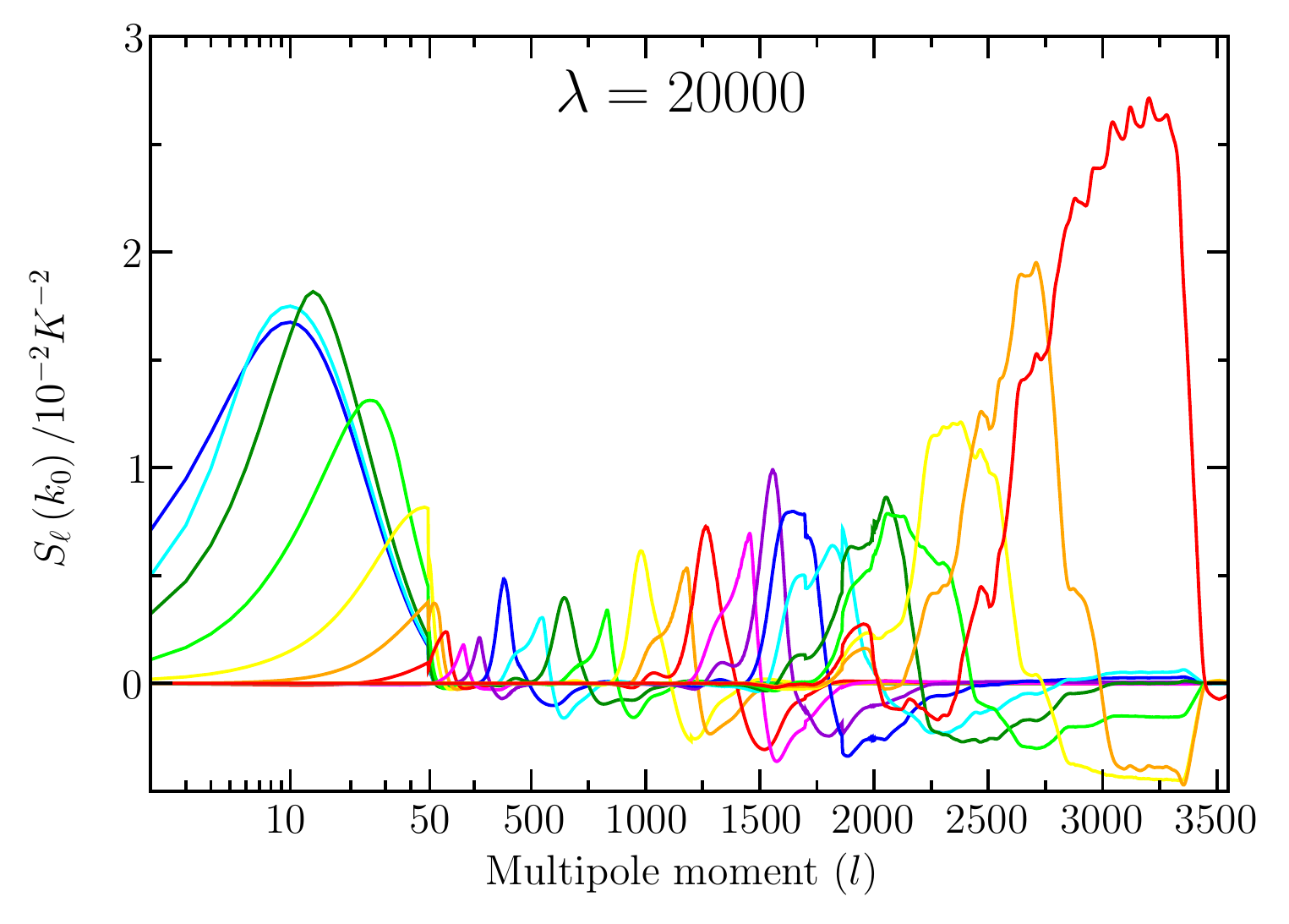}
\caption{Left: TT sensitivity kernels $S_\ell\left(k_0\right)$ for the
    Planck, WMAP-9 polarisation, ACT and SPT data (combination II) with
    $\lambda=400$. The kernels are shown for $k_0=10^{-4}$,
    $5\times10^{-4}$, $10^{-3}$, $2.5\times10^{-3}$, $5\times10^{-3}$,
    $7.5\times10^{-3}$, 0.01, 0.015, 0.02, 0.03, 0.04, 0.05, 0.06,
    0.075, 0.085, 0.095, 0.105, 0.115, 0.125, 0.138, 0.15, 0.162, 0.18,
    0.2, 0.3$\;\mathrm{Mpc}^{-1}$. Note that the horizontal axis changes 
    from a logarithmic to a linear scale at $\ell=50$. Right: Same as 
    left panel but for $\lambda=20000$.}
\label{fig:sens} 
\end{figure}

Since the CMB data points depend on the underlying TT spectrum $\mathrm{s}_\ell^\mathrm{TT}$ we introduce the TT sensitivity kernels
\begin{equation}
S_\ell\left(k_0\right) \equiv \sum_i \frac{\partial\hat{\mathrm{y}}_i}{\partial
    \mathrm{s}_\ell^\mathrm{TT}} \phi_i\left(k_0\right) =  \sum_{\mathbb{Z},a,i}  M_{ia}^{(\mathbb{Z})}  \frac{\partial \mathrm{d}_a^{(\mathbb{Z})}}{\partial
    \mathrm{s}_\ell^\mathrm{TT}} \phi_i\left(k_0\right).
\end{equation}
The TT sensitivity kernels for some selected $k_0$ values are shown in
Fig.\ref{fig:sens}.  To a first approximation the amplitude of the
sensitivity kernels varies inversely with the height of the TT
spectrum because the PPS is almost scale-invariant. Thus the
$k_0=0.015\;\mathrm{Mpc}^{-1}$ kernel, which corresponds to the first
acoustic peak, is the smallest. The kernels are particularly well
localised for the scales $0.01\leq k_0 \leq 0.115\;\mathrm{Mpc}^{-1}$,
below which the signal-to-noise ratio of the Planck data falls
sharply. The ACT and SPT bandpowers, which sample the TT spectrum less
densely, lead to more irregular sensitivity kernels. The kernels
associated with the CMB acoustic peaks are narrower than those at the
troughs. For $\lambda=20000$ the kernels are broader and less
localised than for $\lambda=400$.

The functions
\begin{eqnarray}
R\left(k_0,k\right) & \equiv & \sum_{i,j} R_{ij} \phi_i\left(k_0\right)\phi_j\left(k\right), \\
Y\left(k_0,k_1,k_2\right) & \equiv & \sum_{i,j,k} Y_{ijk} \phi_i\left(k_0\right)\phi_j\left(k_1\right)\phi_k\left(k_2\right),
\end{eqnarray}
known as resolution kernels are more suitable for plotting than the
resolution matrices.  The first-order kernel $R\left(k_0,k\right)$
describes the extent to which the estimated PPS is a smoothed version
of the true PPS. For fixed $k_0$ it is a sharply peaked function of
$k$ (ideally centred at $k=k_0$) which represents the wavenumber range
over which the true PPS is smoothed.
 
\begin{figure}[t!]
\includegraphics[angle=0,width=0.5\columnwidth,trim = 32mm 171mm 23mm
  15mm, clip]{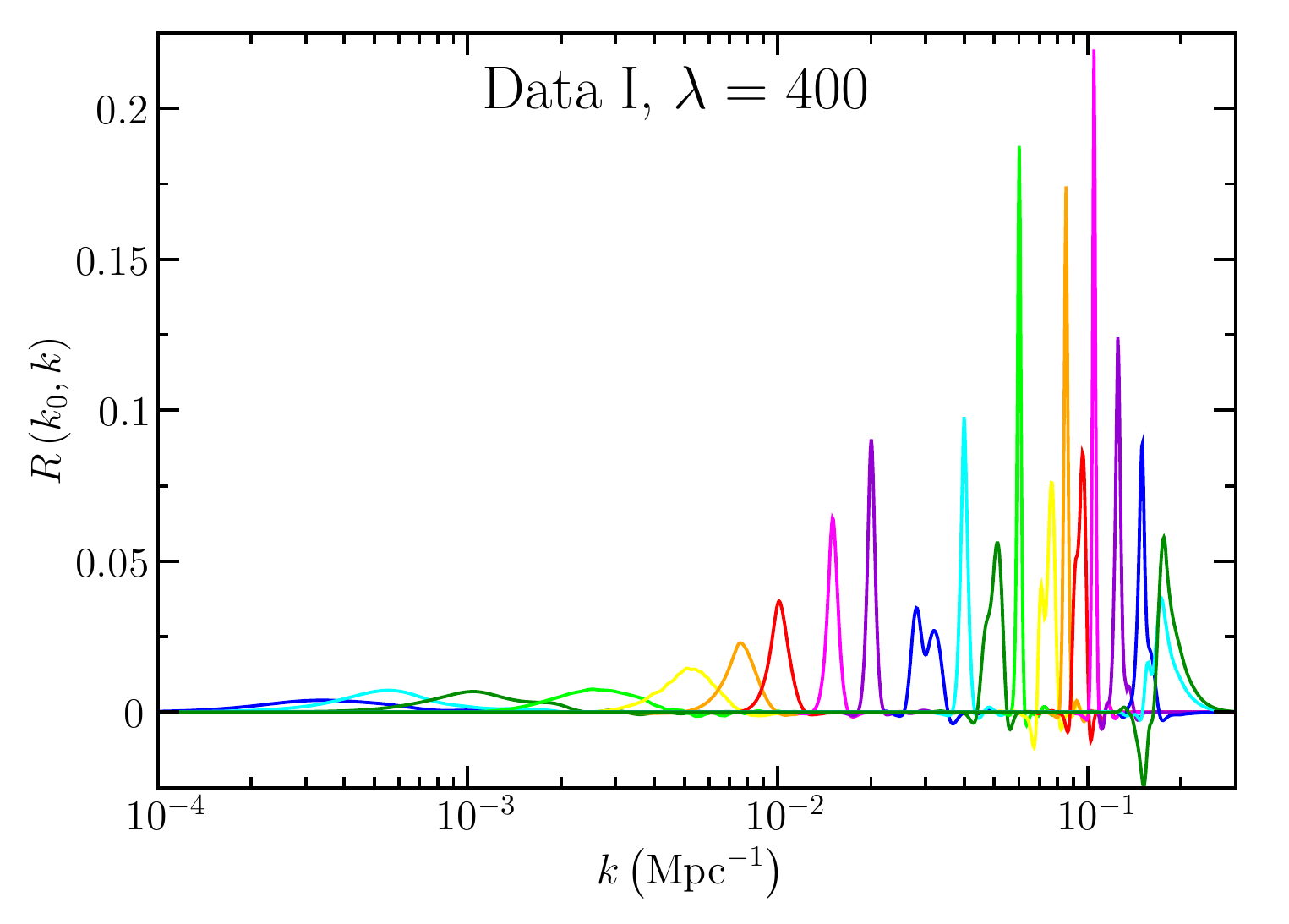}
\includegraphics[angle=0,width=0.5\columnwidth,trim = 32mm 171mm 23mm
  15mm, clip]{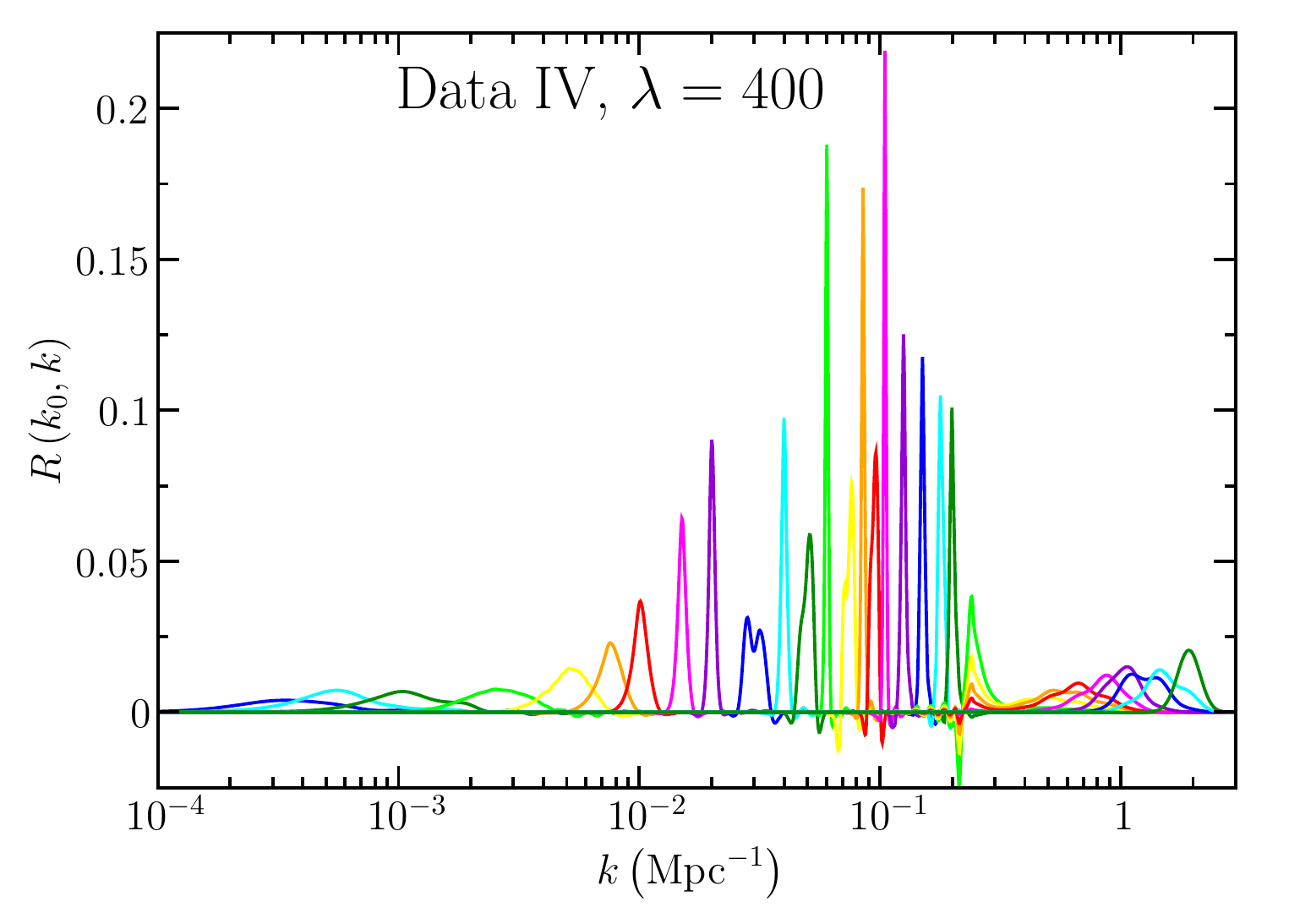}
\includegraphics[angle=0,width=0.5\columnwidth,trim = 32mm 171mm 23mm
  15mm, clip]{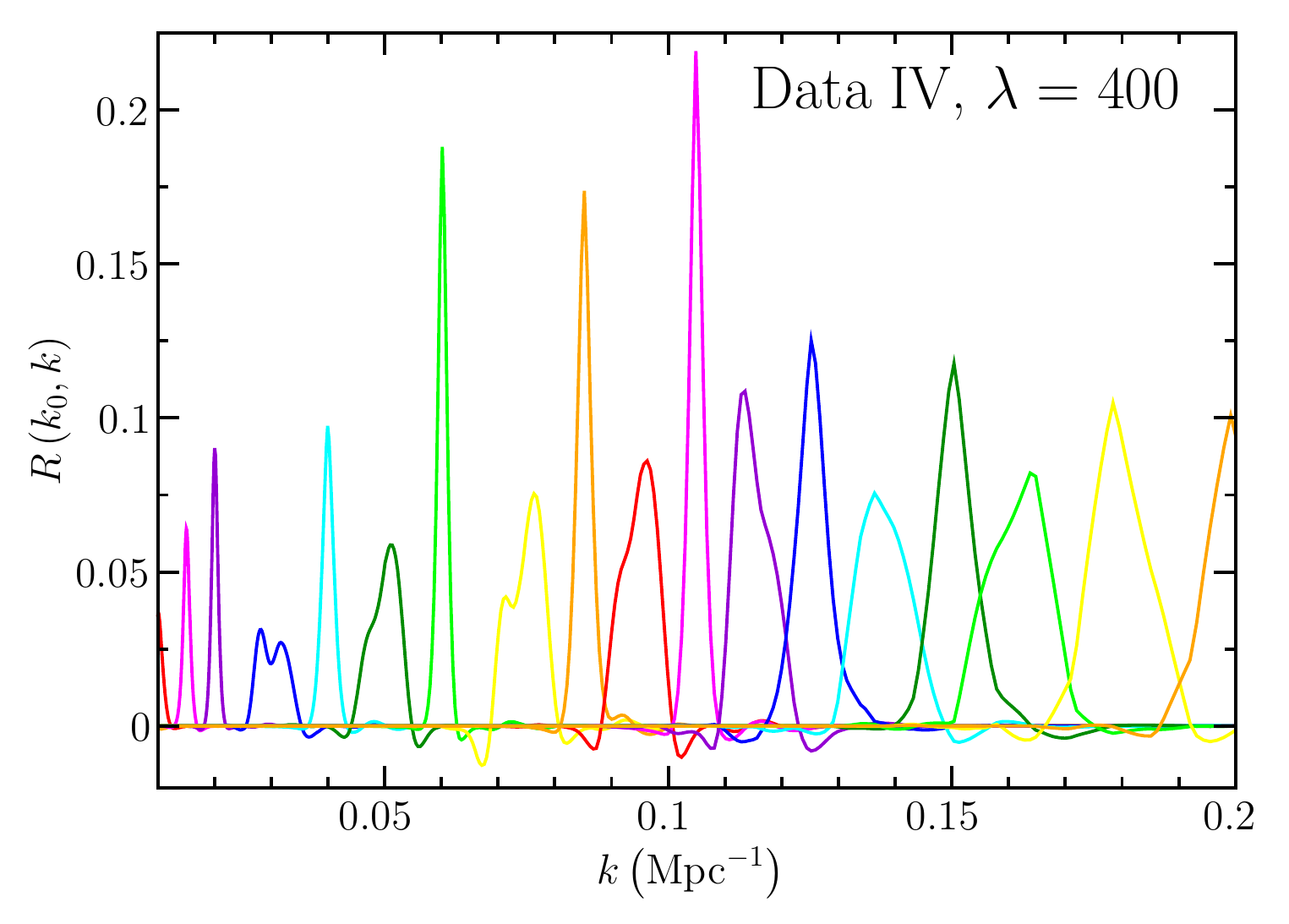}
\includegraphics[angle=0,width=0.5\columnwidth,trim = 32mm 171mm 23mm
  15mm, clip]{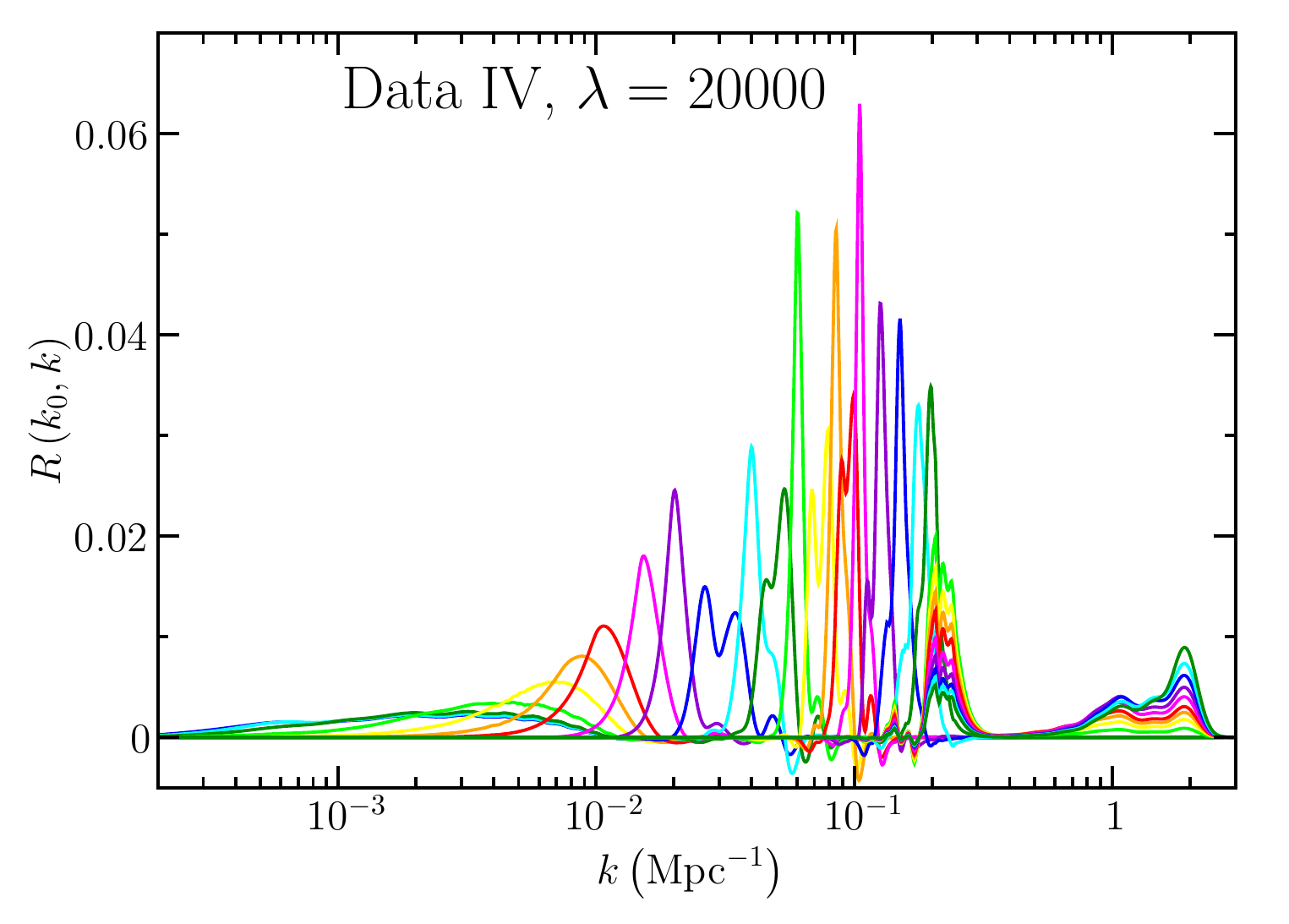}
  \caption{Top left: First-order resolution kernels $R(k_0,k)$ for the
    Planck and WMAP-9 polarisation data (combination I) with
    $\lambda=400$. The kernels are shown for $k_0=10^{-4}$,
    $5\times10^{-4}$, $10^{-3}$, $2.5\times10^{-3}$, $5\times10^{-3}$,
    $7.5\times10^{-3}$, 0.01, 0.015, 0.02, 0.03, 0.04, 0.05, 0.06,
    0.075, 0.085, 0.095, 0.105, 0.125, 0.15, 0.18,
    0.2$\;\mathrm{Mpc}^{-1}$.  Top right: Resolution kernels for the
    Planck, WMAP-9 polarisation, ACT, SPT, WiggleZ, galaxy clustering,
    CFTHLenS and Lyman-$\alpha$ data (combination IV) with
    $\lambda=400$. The $k_0$ values are the same as in the top left
    panel, plus $k_0=0.3$, 0.4, 0.5, 0.6, 0.8, 1, 1.25, 1.5,
    2.0$\;\mathrm{Mpc}^{-1}$.  Bottom left: Same as the top right
    panel, but for $k_0=0.01$, 0.015, 0.02, 0.03, 0.04, 0.05, 0.06,
    0.075, 0.085, 0.095, 0.105, 0.115, 0.125, 0.138, 0.15, 0.162,
    0.18, 0.2$\;\mathrm{Mpc}^{-1}$. Bottom right: Same as the top
    right panel, but with $\lambda=20000$.}
\label{fig:resol1}
\end{figure}

Features in the true PPS $\mathcal{P}_\zeta\left(k\right)$ much
broader than the resolution kernel $R\left(k_0,k\right)$ are recovered
well by the estimated PPS
$\hat{\mathcal{P}}_\zeta\left(k_0\right)$, while features much   
narrower are smoothed out. The resolution kernels for some chosen
values of the target wavenumber $k_0$ are displayed in
Fig.\,\ref{fig:resol1}. Since the resolution kernels depend on both
the integral kernels $\mathcal{K}_a^{(\mathbb{Z})}$ and the error in
the data, the greatest resolution is attained on intermediate scales
where the cosmic variance and noise in the data is minimised. A clear
pattern is that the resolution is better at wavenumbers corresponding
to the 7 peaks of the CMB TT spectrum observed by Planck than the
troughs. This is because the TT integral kernels are narrower at the
acoustic peaks than the troughs. There is a loss of resolution at
$k\simeq 0.4\; \mathrm{Mpc}^{-1}$ between the lower wavenumbers
covered by the CMB and WiggleZ datasets and the higher wavenumbers
covered by the VHS Lyman-$\alpha$ data. Although this gap is spanned
by the CFHTLenS weak lensing data it has comparatively poor resolution
as it is sparser and noiser than the other datasets.

\begin{figure}[tbh]
\includegraphics[width=0.5\columnwidth,trim = 32mm 171mm 23mm 15mm, clip]{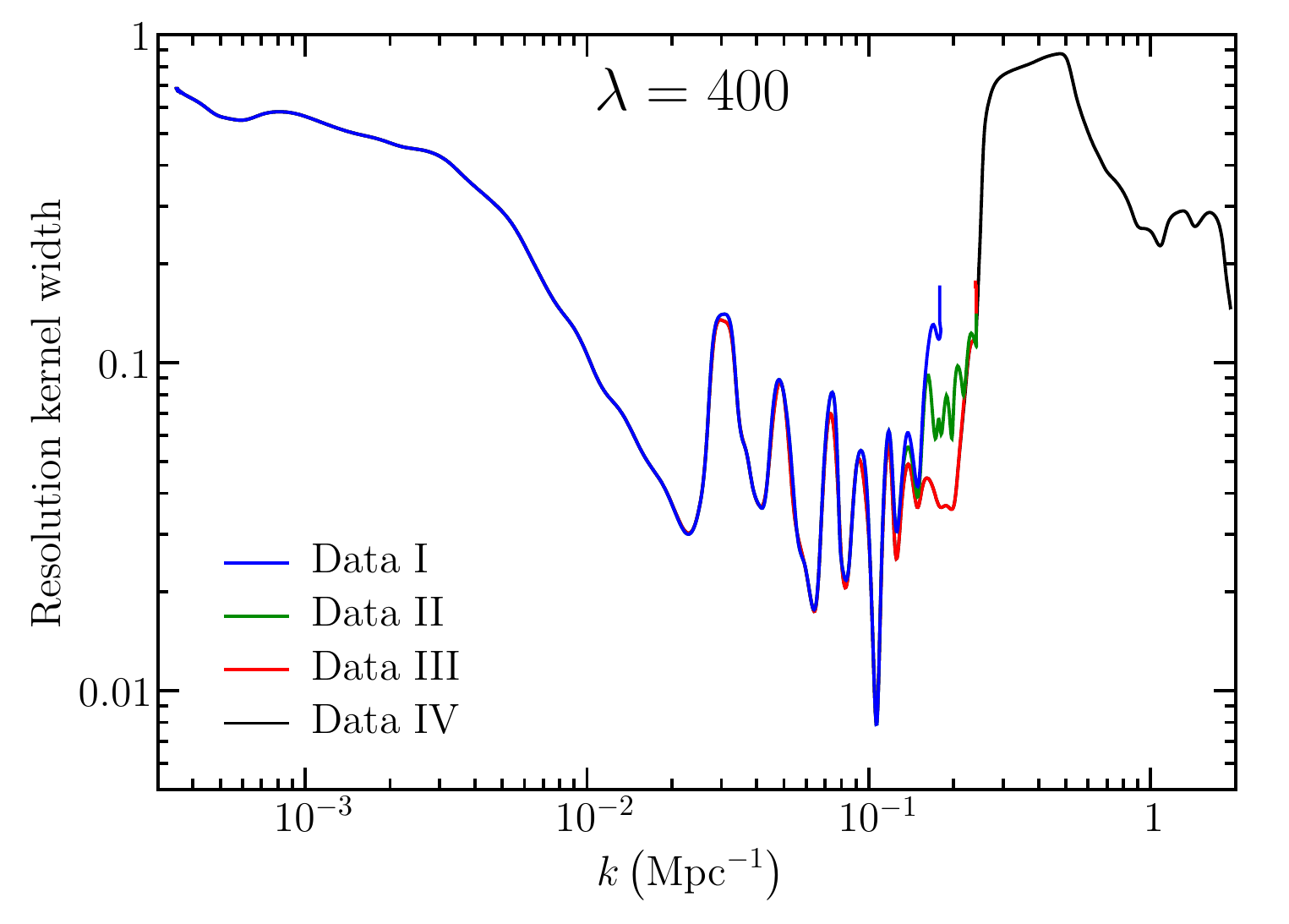}
\includegraphics[width=0.5\columnwidth,trim = 32mm 171mm 23mm 15mm, clip]{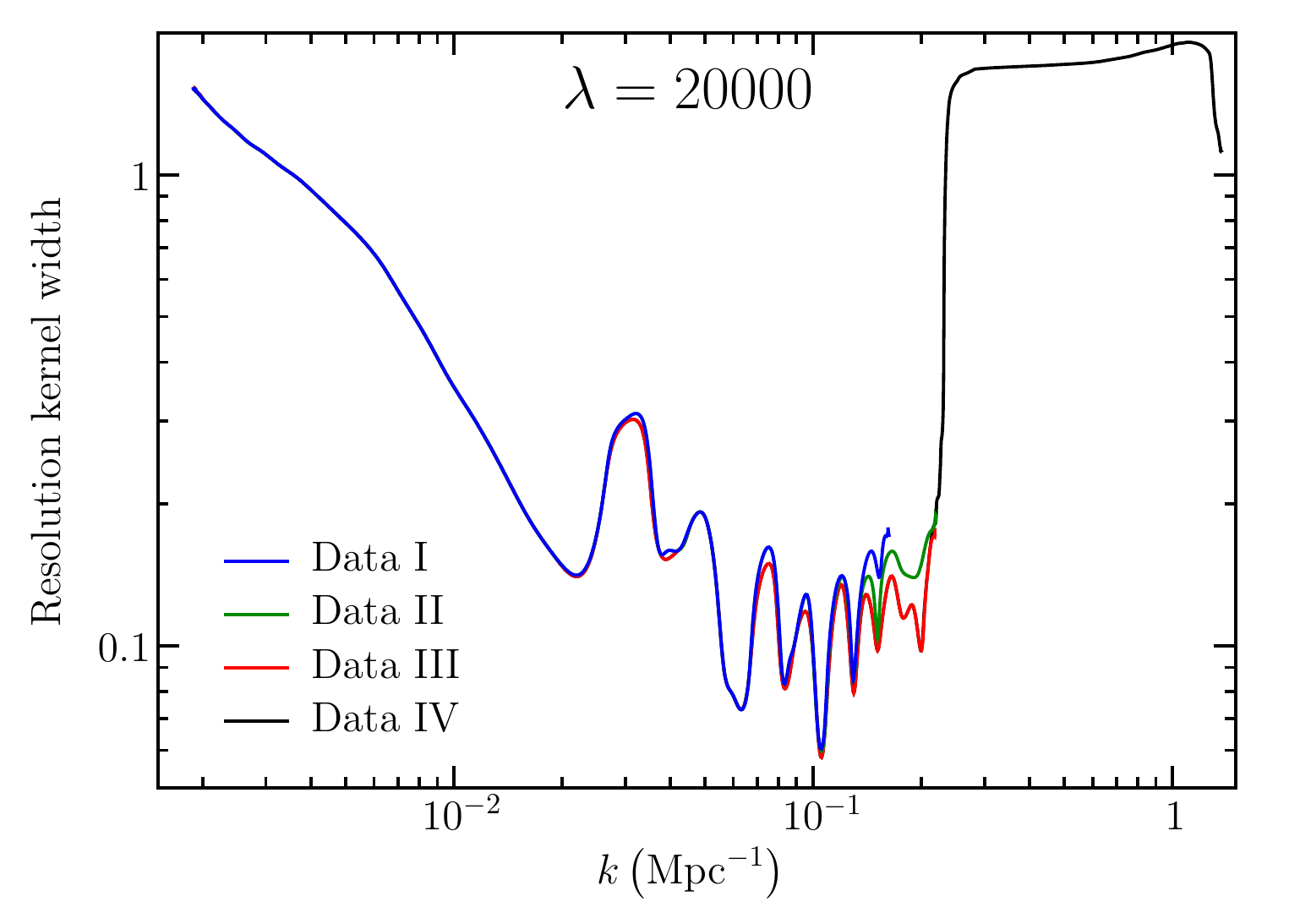}
\caption{Left: resolution kernel width plotted against the kernel
  location with $\lambda=400$ for the four data set
  combinations. Right: resolution kernel width plotted against the
  kernel location with $\lambda=20000$ for the four data set
  combinations. The influence of the CMB acoustic peaks on the
  resolution is clearly apparent.}
\label{fig:resol2} 
\end{figure}

To measure the resolution we use the width of the resolution kernels,
taken to be the quantity $\ln\left(k_{75}/k_{25}\right)$, the
logarithmic wavenumber interval between the 25th and 75th percentiles
of the area underneath the absolute value of the resolution kernel. In
Fig.\,\ref{fig:resol2} the kernel width is plotted against the
location of the kernel, defined as the wavenumber of the 50th
percentile $k_{50}$. The greater resolution at the acoustic peaks and
the increase in resolution caused by the addition of extra datasets
can clearly be seen.

Using Eq.(\ref{recerr}) the total frequentist
covariance matrix $\mathsf{\Sigma}$ of the estimated PPS
is 
\begin{eqnarray}
\label{sigmat}  
\mathsf{\Sigma} &\equiv& \langle\left(\hat{\yB} -
\langle \hat{\yB}\rangle\right) 
\left(\hat{\yB}-\langle \hat{\yB}\rangle\right)^\mathrm{T}\rangle=
\mathsf{\Sigma}_\mathrm{F}+\mathsf{\Sigma}_\mathrm{P}+\ldots,\\
\label{sigmaf}
\mathsf{\Sigma}_{\mathrm{F}|ij} &\equiv&
\sum_{\mathbb{Z},a,b} M^{(\mathbb{Z})}_{ia} N^{(\mathbb{Z})}_{ab} M^{(\mathbb{Z})}_{jb}, \\
\label{sigmap}
\mathsf{\Sigma}_{\mathrm{P}|ij} &\equiv& M_{i\alpha} U_{\alpha\beta} M_{j\beta},
\end{eqnarray}
where the angled brackets denote the average over an ensemble of
spectra estimated from repeated identical independent measurements of
the data and the background parameters.  Thus to first order
$\mathsf{\Sigma}$ is the sum of $\mathsf{\Sigma}_\mathrm{F}$ which
arises from the data noise and $\mathsf{\Sigma}_\mathrm{P}$ which
arises from errors in the background parameters.

Tikhonov regularisation has a natural Bayesian interpretation as a
two-stage hierarchical Bayes model with a hyperparameter
$\tilde{\lambda}$, as discussed in \cite{Hunt:2013bha}. The maximum
\emph{a posteriori} estimate that maximises the posterior distribution
of the PPS given the data $P\left(\yB|\dB\right)$ coincides with
$\hat{\yB}$ when the prior distributions of $\tilde{\lambda}$ and
$\bm{\theta}$ are
$P\left(\tilde{\lambda}\right) =
\delta\left(\tilde{\lambda}-\lambda\right)$
and
$P\left(\bm{\theta}\right)=\delta\left(\bm{\theta}-\hat{\bm{\theta}}\right)$. The
Bayesian covariance matrix $\mathsf{\Pi}$ which describes the shape of
$P\left(\yB|\dB\right)$ is then given by
\begin{equation}
\mathsf{\Pi}^{-1}_{ij}\equiv\left.\frac{1}{2}
\frac{\displaystyle\partial^{2}Q\left(\yB,\dB,\hat{\bm{\theta}},\lambda\right)}
       {\displaystyle\partial \mathrm{y}_i\partial \mathrm{y}_j}\right|_{\hat{\yB}}.
\label{sigb}
\end{equation}

As the regularisation parameter $\lambda$ decreases, each element of
the recovered PPS is effectively dependent on fewer data
points. Mathematically, each row of the sensitivity matrices
$M_{ia}^{(\mathbb{Z})}$ becomes a narrower and more sharply peaked
function of the index $a$. This means that the resolution kernels also
become narrower, but the noise artifact term
$\sum_{\mathbb{Z},a}
M_{ia}^{(\mathbb{Z})}\,\mathrm{n}_a^{(\mathbb{Z})}$
in Eq.(\ref{recerr}) becomes more significant as the noise vectors
$\mathrm{n}_a^{(\mathbb{Z})}$ are less averaged out.  Thus there is an
unavoidable trade-off between the resolution and the variance of the
recovered PPS.

\begin{figure}[h!]
\includegraphics[width=0.5\columnwidth,trim = 32mm 171mm 23mm 15mm, clip]{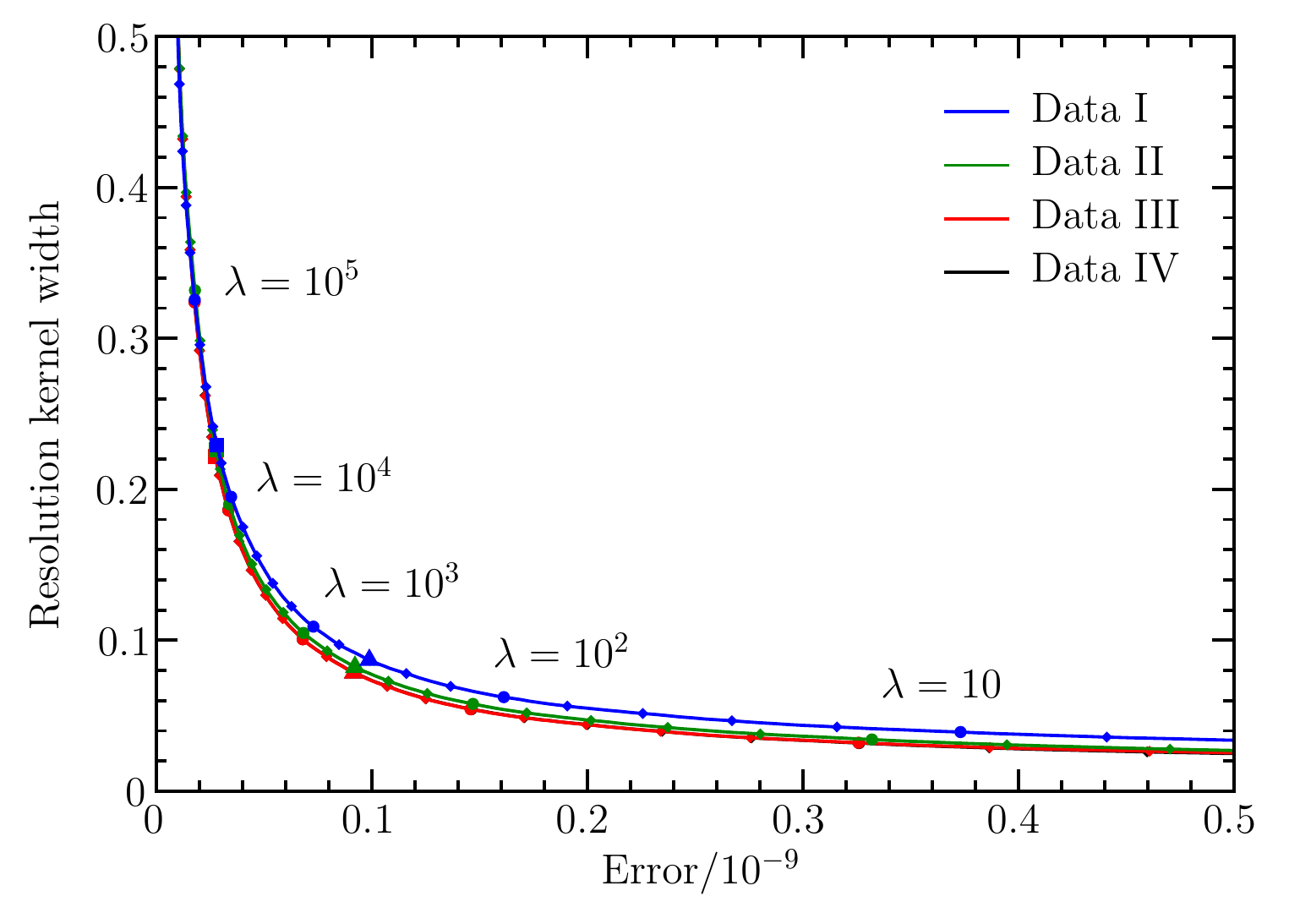}
\includegraphics[width=0.5\columnwidth,trim = 32mm 171mm 23mm 15mm, clip]{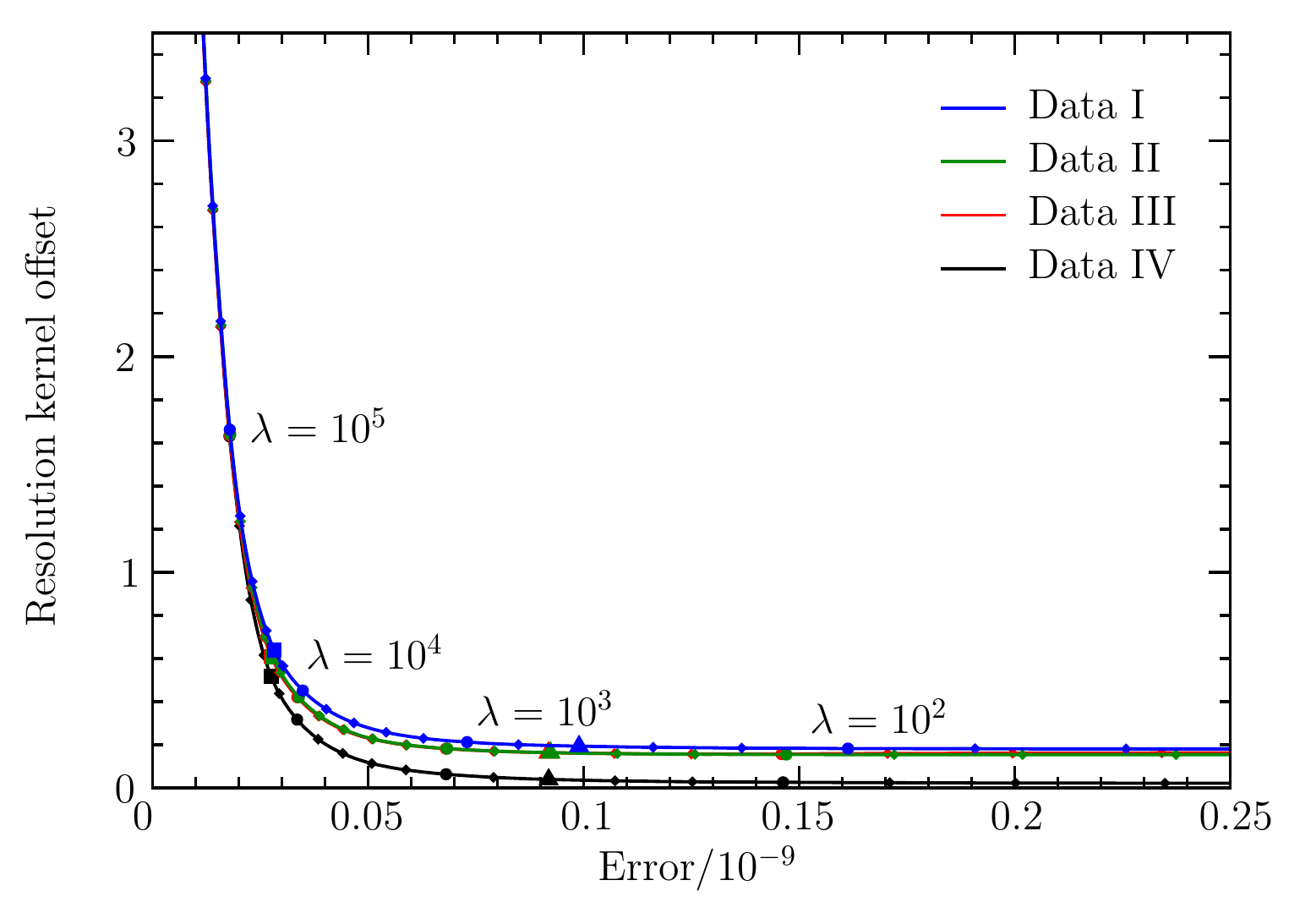}
\caption{Trade-off curve of the width of the resolution kernel versus
  error (left panel) and of its offset versus error (right panel), as
  a function of the regularisation parameter $\lambda$ for the four
  data set combinations. In both panels the triangles and squares
  correspond to $\lambda=$400 and 20000 respectively, and the diamonds
  mark logarithmically equal intervals of $\lambda$.}
\label{fig:tradeoff} 
\end{figure}

This can be seen in Fig.\,\ref{fig:tradeoff} in which two measures of
resolution quality (kernel width and offset) are plotted against the
reconstruction error. The error is computed from the frequentist
covariance matrix $\mathsf{\Sigma}_\mathrm{F}$ averaged over
$5\times 10^{-3} \leq k \leq 0.25\; \mathrm{Mpc}^{-1}$. The kernel
width is defined as in Fig.\,\ref{fig:resol2}, but now averaged over
the kernels $R\left(k_0,k\right)$ with
$5\times 10^{-3} \leq k_0 \leq 0.25\; \mathrm{Mpc}^{-1}$.  The offset
of the kernel $R\left(k_0,k\right)$ is defined as the quantity
$\left|k_{50}-k_0\right|/k_0$. The offset is averaged over the kernels
$R\left(k_0,k\right)$ with
$3.5\times 10^{-4} \leq k_0 \leq 1.9\; \mathrm{Mpc}^{-1}$.
 
The trade-off curves in Fig.\,\ref{fig:tradeoff} are shaped like the
letter L. The corner of the curves represents the optimum compromise
between resolution and variance. The regularisation parameter values
$\lambda=400$ and $\lambda=20000$ used in our analysis are close to
the corner of the curves and are thus satisfactory according to this
criterion.

\end{document}